\documentclass[a4paper,11pt]{article}
\pdfoutput=1 

\usepackage{jheppub} 

\usepackage{graphicx}
\usepackage{dcolumn}
\usepackage{bm}

\usepackage{slashed}
\usepackage{multirow}

\usepackage{subfigure}
\usepackage{float}
\usepackage{amsmath}
\usepackage{bbm}

\providecommand{\tabularnewline}{\\}

\newcounter{YJC}

\begin{document}

\title{Searching for the neutral triple gauge couplings in the process $\mu^+\mu^-\to \gamma \nu\bar{\nu}$ at muon colliders}

\author[a,b, c]{Wei Xie,}
\author[c,d,1]{Ji-Chong Yang,\note{ Corresponding author.}}

\affiliation[a]{College of Mathematics and Physics, China Three Gorges University, No. 8 University Road, Yichang 443002, China}
\affiliation[b]{Center for Astronomy and Space Sciences, China Three Gorges University,  No. 8 University Road, Yichang 443002, China}
\affiliation[c]{Department of Physics, Liaoning Normal University, No. 850 Huanghe Road, Dalian 116029, China}
\affiliation[d]{Center for Theoretical and Experimental High Energy Physics, Liaoning Normal University, No. 850 Huanghe Road, Dalian 116029, China}
\emailAdd{xiewei@ctgu.edu.cn}
\emailAdd{yangjichong@lnnu.edu.cn}

\abstract{
We investigate the sensitivity of future high-energy muon colliders
to neutral triple gauge couplings (nTGCs) through the process $\mu^{+}\mu^{-}\to\gamma\nu\bar{\nu}$
within the Standard Model Effective Field Theory (SMEFT) framework.
Extending beyond previous studies, we consider a set of 14 dimension-8
operators, including both Higgs-related and pure gauge structures.
By computing the cross sections and performing Monte Carlo simulations
at multiple center-of-mass energies (3-30 TeV), we demonstrate that
the annihilation process dominates over vector boson fusion (VBF)
at TeV scales. We also explore the impact of beam polarization and
show that the $(-+)$ polarization enhances sensitivity to several
operators. After the study of the event selection strategies, we show
that muon colliders can impose stronger expected constraints on nTGCs
operators than current LHC bounds, with two of the pure gauge operators
yielding the most stringent expected constraints. We also evaluate
the contribution of CP-violating pure gauge operators to the electron
electric dipole moment (EDM), finding that the expected constraints
from muon colliders are stronger than those from EDM measurements.
}

\maketitle
\flushbottom

\section{Introduction }

Over the past few decades, the Standard Model (SM) of particle physics
has achieved remarkable success, with the majority of experimental
measurements showing excellent agreement with its theoretical predictions.
The Large Hadron Collider (LHC) has revolutionized high-energy particle
physics, most notably through the discovery of the Higgs boson. However,
equally important is the striking absence of clear evidence for new
physics (NP) up to approximately the TeV scale. The SM is widely regarded
as an effective low-energy theory, incomplete in describing the full
picture of fundamental interactions. It fails to account for several
observed phenomena, such as the nature of dark matter, the matter-antimatter
asymmetry in the universe, and the origin of neutrino masses. As a
result, searching for NP beyond the Standard Model (BSM) has become
one of the central objectives of current and upcoming high-energy
collider experiments, which aim to probe a higher range of energies
with greater precision in order to uncover potential signatures of
new particles or interactions.

The Standard Model Effective Field Theory (SMEFT)~\cite{weinberg,SMEFTReview1,SMEFTReview2,SMEFTReview3}
extends the SM by incorporating higher-dimensional effective operators
constructed from the known SM fields, in addition to the renormalizable
(dimension-4) SM Lagrangian. This framework offers a systematic and
model independent approach to exploring potential NP beyond the SM.
The higher-dimensional operators in SMEFT, characterized by dimension
$d>4$, are suppressed by inverse powers of a high-energy ultraviolet
(UV) cutoff scale $\Lambda$, which is assumed to be much larger than
the electroweak Higgs vacuum expectation value (VEV). This scale $\Lambda$
is typically associated with the mass scale of the unknown NP dynamics.
Among these operators, the dimension-5 operators are the leading corrections
and dimension-6 operators appear next in the expansion and have been
widely investigated in the context of collider phenomenology~\cite{Brivio:2017vri,Ellis:2014dva,Ellis:2014jta,He:2015spf,Ge:2016tmm,deBlas:2016ojx,Durieux:2017rsg,Murphy:2017omb,Ellis:2018gqa,Remmen:2019cyz,Koksal:2019oqt,Ellis:2020unq}.

A notable evolution is occurring in the phenomenological landscape
as dimension-8 operators receive rising attention from both theorists
and experimentalists~\cite{Ellis:2017edi,Ellis:2018cos,Alday:2014qfa}.
One example is neutral triple gauge couplings (nTGCs)~\cite{Gounaris:1999kf,Gounaris:2000dn,Degrande:2013kka,Senol:2018cks,Senol:2022snc,Ellis:2024omd},
such as $ZZ\gamma$ and $Z\gamma\gamma$. The nTGCs are absent in
the dimension-6 operators and the leading contributions arise from
dimension-8 operators within the SMEFT approach. These operators generally
involve the Higgs doublet $H$, and the induced nTGCs vanish in the
limit as the Higgs vacuum expectation value $\langle H\rangle\to0$.
This implies that the origin of these operators is intrinsically connected
to the mechanism of spontaneous electroweak symmetry breaking. Consequently,
probing nTGCs not only offers insight into potential NP at dimension-8
but also provides a novel avenue for exploring the dynamics of the
Higgs boson and electroweak symmetry breaking. Phenomenological studies
of these dimension-8 operators have been conducted at future $e^{+}e^{-}$
colliders including CEPC, FCC-ee, ILC and CLIC via the process $e^{+}e^{-}\to Z\gamma$
and $e^{+}e^{-}\to ZZ$(with $Z\to\ell^{+}\ell^{-},\nu\bar{\nu}$,$q\bar{q}$)~\cite{Ellis:2019zex,Ellis:2020ljj,Liu:2024tcz,Fu:2021mub,Yang:2021kyy,Guo:2024qyx,Jahedi:2022duc,Jahedi:2023myu,Ellis:2025jgt},
and at LHC and future hadron colliders via the process $pp(q\bar{q})\to Z\gamma$~\cite{Ellis:2022zdw,Ellis:2023ucy}.
$\mu^{-}\mu^{+}\rightarrow Z\gamma\rightarrow\nu\bar{\nu}\gamma$
process at the future muon collider with a center-of-mass (c.m.) energy
of 3 TeV has also been investigated to examine nTGCs~\cite{Senol:2022snc}. 

To probe energy scales beyond the reach of the LHC, a compelling objective
for the next-generation energy frontier collider is to reach, and
surpass, the 10 TeV scale. Multi-TeV muon colliders~\cite{muoncollider1,muoncollider2,muoncollider4,muoncollider6,muoncollider7,muoncollider8,Costantini:2020stv,AlAli:2021let,Han:2020uid}
have been proposed for directly probing the energy frontier, combining
unprecedented collision energies with a clean leptonic environment.
Muon mass (roughly 200 times that of the electron) results in much
less synchrotron radiation, making it feasible to achieve multi-TeV
collisions in a reasonably compact circular accelerator. Moreover,
muon colliders are among the most power-efficient designs in terms
of luminosity per unit of energy consumed, making them especially
attractive in the context of global efforts toward more sustainable
and energy-efficient research infrastructure. Therefore, the muon
collider stands out as an excellent platform for probing NP beyond
the electroweak scale and for conducting precise measurements of the
nTGCs.

While some previous nTGCs studies focused on a single dimension-8
operator~\cite{Fu:2021mub,Yang:2021kyy,Guo:2024qyx,Ellis:2019zex},
or set of four operators~\cite{Senol:2022snc,Senol:2018cks}, newly
constructed nTGCs operators have been taken into account recently~\cite{Ellis:2020ljj,Ellis:2025jgt,Liu:2024tcz,Ellis:2022zdw,Ellis:2023ucy},
by including CP-violating (CPV) operators and pure gauge operators.
Based on the operators considered in these references, we consider
a set of 14 nTGCs operators, including 10 Higgs-related operators
and 4 pure gauge operators. In most cases, the sensitivity to NP is
driven by interference between the SM and NP amplitudes. Since CPV
operators typically do not interfere with the SM at leading order,
their signals are generally suppressed. However, at high energies,
the overlap of phase space between the SM and NP becomes smaller,
and the interference plays a less significant role. This opens up
the possibility that CPV effects may become more visible. Therefore,
it is necessary to carefully examine the sensitivity to both CP-conserving
(CPC) and CPV operators in the high-energy regime.

In this work, we utilize the process $\mu^{+}\mu^{-}\to\gamma\nu\bar{\nu}$
to study nTGCs induced by 14 dimension-8 operators at muon colliders.
To preserve symmetry, an operator contributing to
the nTGCs vertices will at the same time also contribute to other
vertices, which includes the $WW\gamma$ coupling. Since this coupling
is a part of the dimension-8 nTGC operators, the vector boson fusion
(VBF) contribution from this coupling is also a NP effect. Consequently,
we take into account all possible vertices generated by the considered
nTGC operators. These nTGCs-induced processes can be classified into
the VBF process and the annihilation process $\mu^{+}\mu^{-}\to Z\gamma$,
with the subsequent decay $Z\to\nu\bar{\nu}$. While previous studies
of nTGCs mainly focus on the annihilation process on electron-positron
or muon colliders~\cite{Senol:2022snc,Durieux:2017rsg,Ellis:2020ljj,Ellis:2025jgt,Fu:2021mub,Jahedi:2022duc,Jahedi:2023myu,Liu:2024tcz},
it is generally believed that at muon colliders, VBF processes will
overtake annihilation in cross section once the c.m. energy surpasses
a few TeV, due to the logarithmic energy enhancement characteristic
of VBF topologies. The process $\mu^{+}\mu^{-}\to\gamma\nu\bar{\nu}$
can proceed via $W^{+}W^{-}\to\gamma$ fusion, where the outgoing
neutrinos lead to missing energy in the final state. This topology
is especially promising for probing higher-dimensional operators,
particularly at high energies. Moreover, since the $WW\gamma$ vertex
is present in the SM, interference effects between SM and NP amplitudes
may enhance the sensitivity to certain operators. The annihilation
process $\mu^{+}\mu^{-}\to Z\gamma\to\nu\bar{\nu}\gamma$ has a reconstructible
photon and missing energy, with a resonant $Z$ peak of the invariant
mass spectrum of missing neutrinos, providing clear signal-background
separation. In order to disentangle operator contributions and identifying
the full nTGCs structure, we will compare the VBF and annihilation
channels at muon colliders. Studying both channels allows coverage
over a wide energy range and helps identify energy-dependent deviations
from the SM.

A major advantage of lepton colliders is the ability to control the
polarization of the initial-state leptons. Beam polarization can enhance
sensitivity to NP effects. It enables the construction of observables
sensitive to specific operator structures, helps disentangle interference
between SM and NP contributions, and reduces background contamination.
Recent studies have demonstrated that beam polarization can significantly
tighten expected constraints on nTGCs in processes such as $e^{+}e^{-}\rightarrow Z\gamma$
and $e^{+}e^{-}\rightarrow ZZ$~\cite{Jahedi:2023myu,Guo:2024qyx}.
Therefore, it is important to investigate the role of initial-state
polarization in probing nTGCs at future muon colliders.

The electron electric dipole moment (EDM) is a highly
sensitive probe of CP violation and BSM physics~\cite{Alarcon:2022ero,Pospelov:2025vzj,Brod:2022bww}.
In the SMEFT framework, CPV effects arise from higher-dimensional
operators, with dimension-8 terms potentially contributing at the
one-loop level~\cite{Panico:2018hal,Ardu:2025rqy}. Owing to its
exceptional experimental precision, the EDM measurement provides stringent
constraints on such operators, offering a complementary approach to
collider searches for nTGCs. In this work, as a complement to collider
analysis, we perform a preliminary study of the contribution of CPV
nTGCs operators to the electron EDM and derive the corresponding constraints
on the two pure gauge dimension-8 operators.

This paper is organized as follows. In Section~\ref{sec:formalism-nTGC},
the set of 14 nTGCs operators are introduced and classified. Then
the comparison of contributions of VBF and annihilation process is
discussed, and the beam polarization effects on the cross sections
at muon colliders are explored. The partial wave unitarity bounds
are also presented. In Section~\ref{sec:Numerical}, the numerical
results for both the unpolarized and polarized cases are presented.
In Section~\ref{sec:EDM}, we study how CPV nTGCs operators contribute
to the electron EDM and how the EDM bound constrains nTGCs coefficients.
Finally, we conclude our research in Section~\ref{sec:Summary}.

\section{\label{sec:formalism-nTGC}The Contribution of nTGCs at Muon
Colliders}

\subsection{Neutral Triple Gauge Couplings}

The self-interactions of gauge bosons, arising from the non-Abelian
$SU(2)_{L}\times U(1)_{Y}$ gauge symmetry in the electroweak sector
of the SM, are of significant interest as they offer powerful means
to test the theory's predictions at the TeV energy scale. The nTGCs,
involving the photon and Z boson ($Z\gamma\gamma$ and $Z\gamma Z$),
do not receive contributions from SM and dimension-6 effective operators
at tree level. Therefore, any observed nTGCs effect can serve as a
critical indicator of NP from dimension-8 effective operators. In
the framework of SMEFT, the Lagrangian involving operators of dimension-8
can be written as~\cite{Degrande:2013kka}
\begin{equation}
\mathcal{L}_{\mathrm{nTGCs}}=\mathcal{L}_{\mathrm{SM}}+\frac{C_{i}}{\Lambda^{4}}\mathcal{O}_{i},
\end{equation}
where $C_{i}$ are the dimensionless Wilson coefficients which act
as parameters to probe NP effect, and $\Lambda$ is the cut-off scale. 

Recently, a set of seven CPC dimension-8 nTGCs operators have been
formulated based on a renormalizable model involving heavy vector-like
fermions~\cite{Ellis:2024omd}, which generate the operators via
one-loop corrections. Also a new set of CPV dimension-8 nTGCs operators
have been constructed by establishing a proper SMEFT basis~\cite{Ellis:2023ucy},
ensuring consistency with electroweak gauge symmetry. We include the
relevant operators from these newly constructed operator sets and
consider a set of 14 nTGCs operators, including 10 Higgs-related operators
and 4 pure gauge operators, as listed in Table~\ref{tab:List-of-nTGCs}.
Dimension-6 SMEFT operators can contribute to nTGCs
at the loop level. The loop contributions are suppressed by $1/\Lambda^{2}$,
which can overcome the $1/\Lambda^{4}$ suppression of dimension-8
operators if $\Lambda$ is large, where dimension-6 operators may
dominate nTGCs at high scales ($\Lambda\gtrsim10~\mathrm{TeV}$)~\cite{Degrande:2013kka}.
In this work, we employ the ``one operator at a time'' approach
and do not consider the loop contributions from dimension-6 operators.
The primary goal of this work is to investigate the expected constraints
on dimension-8 operators at future muon colliders and to compare them
with the relevant experimental results~\cite{CMS:2020gtj,ATLAS:2025ply}
as well as phenomenological studies~\cite{Senol:2022snc}. 

\begin{table}[tp]
\centering
\begin{tabular}{|c|c|c|c|c|c|c|c|c|c|c|}
\hline 
Operator & $\mathcal{O}_{\tilde{W}W}$ & $\mathcal{O}_{\tilde{W}W}^{\prime}$ & $\mathcal{O}_{\tilde{B}B}$ & $\mathcal{O}_{\tilde{B}B}^{\prime}$ & $\mathcal{O}_{\tilde{B}W}$ & $\mathcal{O}_{\tilde{B}W}^{\prime}$ & $\mathcal{O}_{\tilde{W}B}$ & $\mathcal{O}_{BW}$ & $\mathcal{O}_{WW}$ & $\mathcal{O}_{BB}$\tabularnewline
\hline 
\hline 
Coefficient & $C_{1}$ & $C_{2}$ & $C_{3}$ & $C_{4}$ & $C_{5}$ & $C_{6}$ & $C_{7}$ & $C_{8}$ & $C_{9}$ & $C_{10}$\tabularnewline
\hline 
CP & CPC & CPC & CPC & CPC & CPC & CPC & CPC & CPV & CPV & CPV\tabularnewline
\hline 
\end{tabular}

\medskip{}

\begin{tabular}{|c|c|c|c|c|}
\hline 
Operator & $\mathcal{O}_{G+}$ & $\mathcal{O}_{G-}$ & $\widetilde{\mathcal{O}}_{G+}$ & $\widetilde{\mathcal{O}}_{G-}$\tabularnewline
\hline 
\hline 
Coefficient & $C_{11}$ & $C_{12}$ & $C_{13}$ & $C_{14}$\tabularnewline
\hline 
CP & CPC & CPC & CPV & CPV\tabularnewline
\hline 
\end{tabular}

\caption{\protect\label{tab:List-of-nTGCs}List of nTGCs operators and their
corresponding coefficients considered in this work. CPC stands for
CP conservation. CPV stands for CP violation. The upper panel of the
table corresponds to Higgs-related operators, and the lower panel
of the table corresponds to pure gauge operators.}
\end{table}

Among the 10 nTGCs operators with Higgs doublets, there are 7 CPC
operators:

\begin{equation}
\ensuremath{\mathcal{O}_{\tilde{W}W}=\mathrm{i}H^{\dagger}\tilde{W}_{\mu\nu}W^{\nu\rho}\{D_{\rho},D_{\mu}\}H+\mathrm{h.c.},}
\end{equation}
\begin{equation}
\ensuremath{\mathcal{O}_{\tilde{W}W}^{\prime}=\mathrm{i}H^{\dagger}\tilde{W}_{\mu\nu}(D_{\rho}W^{\nu\rho})D_{\mu}H+\mathrm{h.c.},}
\end{equation}
\begin{equation}
\mathcal{O}_{\tilde{B}B}=iH^{\dagger}\tilde{B}_{\mu\nu}B^{\nu\rho}\{D_{\rho},D_{\mu}\}H+\mathrm{h.c.},
\end{equation}
\begin{equation}
\mathcal{O}_{\tilde{B}B}^{\prime}=\mathrm{i}H^{\dagger}\tilde{B}_{\mu\nu}(D_{\rho}B^{\nu\rho})D_{\mu}H+\mathrm{h.c.},
\end{equation}
\begin{equation}
\mathcal{O}_{\tilde{B}W}=\mathrm{i}H^{\dagger}\tilde{B}_{\mu\nu}W^{\nu\rho}\{D_{\rho},D^{\mu}\}H+\mathrm{h.c.},
\end{equation}
\begin{equation}
\mathcal{O}_{\tilde{B}W}^{\prime}=\mathrm{i}H^{\dagger}\tilde{B}_{\mu\nu}(D_{\rho}W^{\nu\rho})D^{\mu}H+\mathrm{h.c.},
\end{equation}
\begin{equation}
\mathcal{O}_{\tilde{W}B}=\mathrm{i}H^{\dagger}\tilde{W}_{\mu\nu}B^{\nu\rho}\{D_{\rho},D^{\mu}\}H+\mathrm{h.c.},
\end{equation}
and 3 CPV operators:
\begin{equation}
\mathcal{O}_{BW}=iH^{\dagger}B_{\mu\nu}W^{\mu\rho}\left\{ D_{\rho},D^{\nu}\right\} H+\mathrm{h.c.},
\end{equation}
\begin{equation}
\mathcal{O}_{WW}=iH^{\dagger}W_{\mu\nu}W^{\mu\rho}\{D_{\rho},D^{\nu}\}H+\mathrm{h.c.},
\end{equation}
\begin{equation}
\mathcal{O}_{BB}=iH^{\dagger}B_{\mu\nu}B^{\mu\rho}\left\{ D_{\rho},D^{\nu}\right\} H+\mathrm{h.c.}.
\end{equation}

There are 4 nTGCs operators constructed from pure gauge fields~\cite{Ellis:2020ljj,Ellis:2023ucy,Liu:2024tcz},
including 2 CPC operators:
\begin{equation}
g\mathcal{O}_{G+}=\widetilde{B}_{\mu\nu}W^{a\mu\rho}(D_{\rho}D_{\lambda}W^{a\nu\lambda}+D^{\nu}D^{\lambda}W_{\lambda\rho}^{a}),
\end{equation}
\begin{equation}
g\mathcal{O}_{G-}=\widetilde{B}_{\mu\nu}W^{a\mu\rho}(D_{\rho}D_{\lambda}W^{a\nu\lambda}-D^{\nu}D^{\lambda}W_{\lambda\rho}^{a}),
\end{equation}
and 2 CPV operators:
\begin{equation}
g\widetilde{\mathcal{O}}_{G+}=B_{\mu\nu}W^{a\mu\rho}(D_{\rho}D_{\lambda}W^{a\nu\lambda}+D^{\nu}D^{\lambda}W_{\lambda\rho}^{a}),
\end{equation}
\begin{equation}
g\widetilde{\mathcal{O}}_{G-}=B_{\mu\nu}W^{a\mu\rho}(D_{\rho}D_{\lambda}W^{a\nu\lambda}-D^{\nu}D^{\lambda}W_{\lambda\rho}^{a}).
\end{equation}
In the above equations $D_{\mu}\equiv\partial_{\mu}-igW_{\mu}^{i}\sigma^{i}-i(g^{\prime}/2)B_{\mu}Y$
is the conventional covariant derivative, $H$ is the SM Higgs doublet,
and $\tilde{B}_{\mu\nu}\equiv\epsilon_{\mu\nu\alpha\beta}B^{\alpha\beta}$,
$\tilde{W}_{\mu\nu}\equiv\epsilon_{\mu\nu\alpha\beta}W^{\alpha\beta}$,
$W_{\mu\nu}\equiv W_{\mu\nu}^{a}\sigma^{a}/2$ where $\sigma^{a}$
are Pauli matrices. 

The matrix element for a given process can be formally written as:
\begin{equation}
|\mathcal{M}_{\mathrm{tot}}|^{2}=|\mathcal{M}_{\mathrm{SM}}|^{2}+2\mathcal{R}\left(\mathcal{M}_{\mathrm{SM}}\mathcal{M}_{\mathrm{dim-8}}^{*}\right)+|\mathcal{M}_{\mathrm{dim-8}}|^{2}.
\end{equation}
For CPC operators, contribution to the total cross section arises
from both the pure NP term $|\mathcal{M}_{\mathrm{dim-8}}|^{2}$ and
the interference term $2\mathcal{R}\left(\mathcal{M}_{\mathrm{SM}}\mathcal{M}_{\mathrm{dim-8}}^{*}\right)$.
The cross section shows asymmetric variation with $C_{i}/\Lambda^{4}$.
For CPV dimension-8 operators, interference term integrates to zero
over phase space. Therefore, we have no interference cross section
for CPV operators. The total cross section shows symmetric variation
with $C_{i}/\Lambda^{4}$ in this case.

Experimental efforts to search for these nTGCs have been conducted
at various colliders, but no signals have been observed so far. Both
LEP and Tevatron have contributed to these searches, yet the most
stringent constraints currently come from the ATLAS and CMS experiments
at the LHC. Specifically, ATLAS has recently presented their limits
on the couplings $C_{\tilde{B}W}/\Lambda^{4}$, $C_{WW}/\Lambda^{4}$,
$C_{BW}/\Lambda^{4}$, $C_{BB}/\Lambda^{4}$, $C_{G+}/\Lambda^{4}$
and $C_{G-}/\Lambda^{4}$ using the $pp\to Z\gamma\to\ell^{+}\ell^{-}\gamma$
channel at a c.m. energy of $\sqrt{s}=13\ \mathrm{TeV}$ with an integrated
luminosity of $140\ \mathrm{fb}^{-1}$ at the LHC~\cite{ATLAS:2025ply}.
For the $C_{WW}/\Lambda^{4}$ coupling, the tightest bound comes from
CMS via the $pp\to ZZ\to4\ell$ process, using data collected at $\sqrt{s}=13\ \mathrm{TeV}$
with an integrated luminosity of $137\ \mathrm{fb}^{-1}$~\cite{CMS:2020gtj}.
The projected 95\% confidence level constraints on dimension-8 nTGCs
operators from both ATLAS and CMS are summarized in Table~\ref{tab:LHC-constraint}.

\begin{table}[tp]
\centering
\begin{tabular}{c|c|c|c}
\hline 
Couplings & $pp\to Z\gamma\to\ell^{+}\ell^{-}\gamma$ & $pp\to ZZ\to4\ell$ & $\mu^{+}\mu^{-}\to Z\gamma\to\gamma\nu\bar{\nu}$\tabularnewline
$(\mathrm{TeV}^{-4})$ & (ATLAS) & (CMS) & Ref.~\cite{Senol:2022snc}\tabularnewline
\hline 
\hline 
$C_{\tilde{B}W}/\Lambda^{4}$ & $[-0.54,0.53]$ & $[-2.30,+2.50]$ & $[-0.0653,+0.0064]$\tabularnewline
\hline 
$C_{WW}/\Lambda^{4}$ & $[-1.90,1.78]$ & $[-1.40,+1.20]$ & $[-0.220,+0.220]$\tabularnewline
\hline 
$C_{BW}/\Lambda^{4}$ & $[-0.87,0.95]$ & $[-1.40,+1.30]$ & $[-0.0846,+0.0846]$\tabularnewline
\hline 
$C_{BB}/\Lambda^{4}$ & $[-0.37,0.37]$ & $[-1.20,+1.20]$ & $[-0.0247,+0.0247]$\tabularnewline
\hline 
$C_{G+}/\Lambda^{4}$ & $[-0.022,0.020]$ &  & \tabularnewline
\hline 
$C_{G-}/\Lambda^{4}$ & $[-1.41,1.08]$ &  & \tabularnewline
\hline 
\end{tabular}

\caption{\protect\label{tab:LHC-constraint}The 95\% confidence level constraints
on various dimension-8 nTGCs operators obtained from the ATLAS~\cite{ATLAS:2025ply}
(column 2) and CMS~\cite{CMS:2020gtj} (column 3) experiments at
the LHC. The expected constraints with 95\% confidence level from
Ref.~\cite{Senol:2022snc} (column 4) are also listed.}
\end{table}

\subsection{\label{subsec:Comparison-of-VBF}Comparison of VBF with annihilation
process in \texorpdfstring{$\mu^{+}\mu^{-}\to\gamma\nu\bar{\nu}$}{muon pair to (anti-)neutrinos and a photon}}

\begin{figure}[tp]
\centering
\includegraphics[scale=1.1]{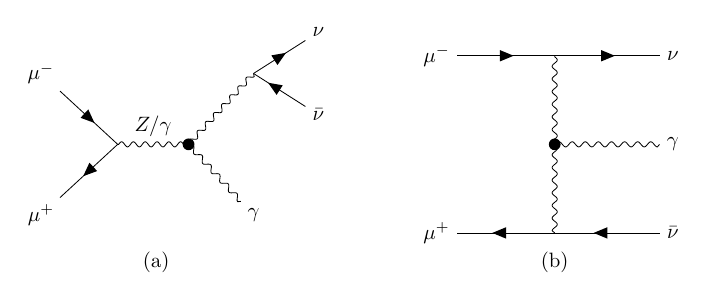}

\caption{\protect\label{fig:Feynman-diagrams-np}Feynman diagrams illustrating
the nTGCs contributions to the process $\mu^{+}\mu^{-}\to\gamma\nu\bar{\nu}$.
The solid black dot represents the NP vertex. Diagram (a) corresponds
to annihilation, and diagram (b) corresponds to VBF.}
\end{figure}

We choose the process $\mu^{+}\mu^{-}\to\gamma\nu\bar{\nu}$ to study
the sensitivity of nTGCs operators at muon colliders. For $\mu^{+}\mu^{-}\to\gamma\nu\bar{\nu}$
at tree level, the Feynman diagrams of nTGCs contribution are shown
in Figure~\ref{fig:Feynman-diagrams-np}. The nTGCs induced processes
can be classified into two categories: the VBF process and the annihilation
process. Figure~\ref{fig:Feynman-diagrams-np}(a) shows the annihilation
process $\mu^{+}\mu^{-}\to Z\gamma$, followed by $Z\to\nu\bar{\nu}$.
Figure~\ref{fig:Feynman-diagrams-np}(b) shows the VBF process, in
which the gauge bosons in the final states are associated with energetic
neutrinos in the forward region with respect to the beam. 

For the VBF in $\mu^{+}\mu^{-}\to\gamma\ell^{+}\ell^{-}$, the $\ell^{+}\ell^{-}$
in the final states are mainly in the forward region with low $p_{T}$
and will be suppressed by $p_{T,\ell}$ cut. On the contrary, the
neutrino final states of VBF in $\mu^{+}\mu^{-}\to\gamma\nu\bar{\nu}$
are not suppressed since the signal of $\nu\bar{\nu}$ are counted
as missing energy and there is no cut for them. For the annihilation,
the $\nu\bar{\nu}$ final states have advantages over the processes
where $Z$ decays into $\ell^{+}\ell^{-}$ or hadronic final states.
At multi-TeV muon colliders, highly boosted $Z$ bosons produce collinear
leptons in the $Z\to\ell^{+}\ell^{-}$ decay, making it challenging
to resolve the two leptons due to detector limitations. In contrast,
the invisible decay $Z\to\nu\bar{\nu}$ avoids collinear lepton merging
issue since neutrinos are invisible and don’t need to be spatially
resolved. Also, a larger $Z$ boson branching ratio into $\nu\bar{\nu}$
compared to that into $\ell^{+}\ell^{-}$ provides a better opportunity
to study the $Z\gamma$ production in high $p_{T}$ region, where
the sensitivity of the anomalous couplings will be higher. Therefore,
the $\mu^{+}\mu^{-}\to\gamma\nu\bar{\nu}$ process provides an environment
with better sensitivity than the $\mu^{+}\mu^{-}\to\gamma\ell^{+}\ell^{-}$
process.

There is a rich interplay between annihilation and VBF production
of both SM and BSM particles at high-energy muon colliders. When a
new particle $X$ is produced, there is a relative scale for the cross
sections of VBF and annihilation processes as a function of collider
energy $\sqrt{s}$~\cite{AlAli:2021let}
\begin{equation}
\frac{\sigma_{\mathrm{VBF}}^{\mathrm{BSM}}}{\sigma_{\mathrm{annih}}^{\mathrm{BSM}}}\propto\alpha_{W}^{2}\frac{s}{m_{X}^{2}}\log^{2}\frac{s}{m_{V}^{2}}\log\frac{s}{m_{X}^{2}},
\end{equation}
where $m_{V}$ is the mass scale of an intermediate state in the production
process (often an electroweak vector boson, which satisfy $m_{V}\ll\sqrt{s}$
at TeV colliders). Due to this logarithmic enhancement, it is generally
believed that starting from a few TeV energies, VBF will become the
dominant mode for many new physics models~\cite{Costantini:2020stv,AlAli:2021let,Han:2020uid}.
The collision energy at which the annihilation and VBF cross sections
for BSM final states intersect increases with the mass scale of the
final state. However, for any given mass scale, there always exists
a collision energy where VBF production becomes dominant. This supports
the idea of high-energy muon colliders functioning effectively as
gauge boson colliders.

In this work, we analytically calculate the cross sections of $\mu^{+}\mu^{-}\to\gamma\nu\bar{\nu}$
contributed from nTGCs operators. Among the 14 operators, $\mathcal{O}_{\tilde{B}W}$
and $\mathcal{O}_{\tilde{W}B}$ are equivalent, and $\mathcal{O}_{\tilde{B}B}$
gives no contribution to $\mu^{+}\mu^{-}\to\gamma\nu\bar{\nu}$. We
are left with 12 nonequivalent operators, including 7 CPC operators
and 5 CPV operators. Contrary to what is described above, we demonstrate
that annihilation process induced by nTGCs operators still dominate
at TeV scale. The expected intersect energy where VBF surpass is much
higher than the accessible energy range of current colliders. 

\begin{figure}[tp]
\centering
\includegraphics[scale=0.35]{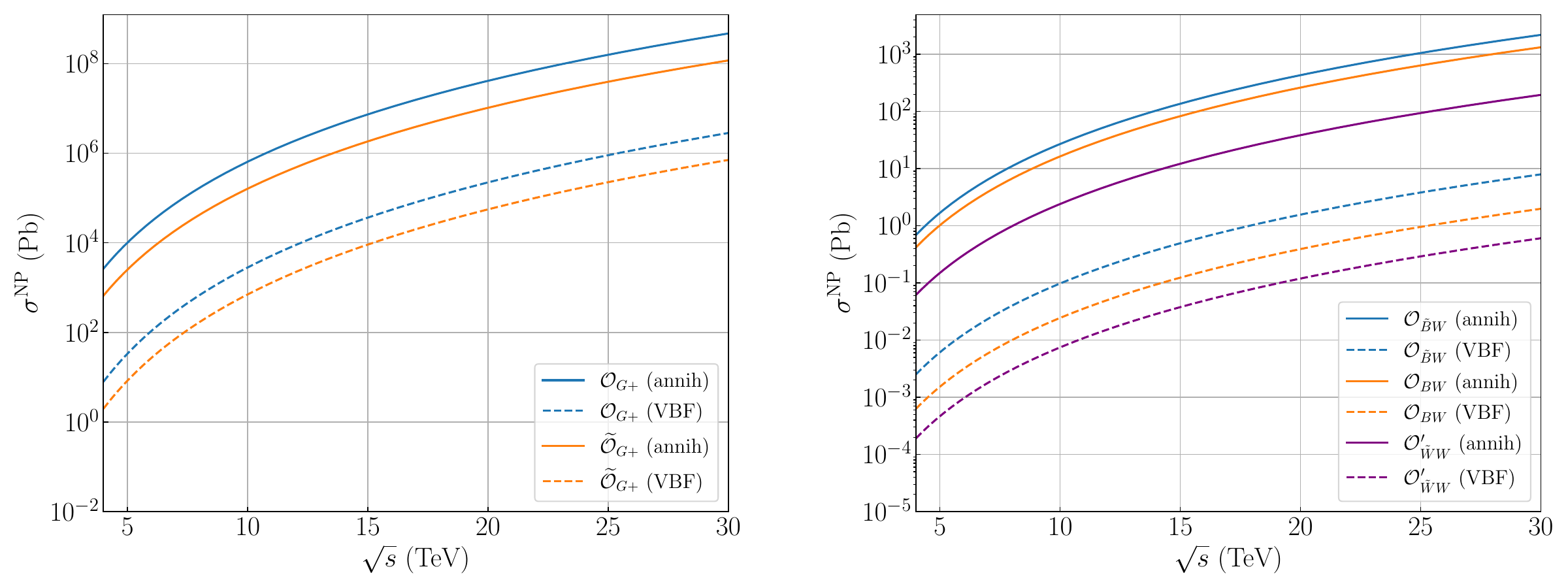}

\caption{\protect\label{fig:cs-comparison}The comparison of $\sigma_{\mathrm{annih}}^{\mathrm{NP}}$
with $\sigma_{\mathrm{VBF}}^{\mathrm{NP}}$ for $\mu^{+}\mu^{-}\to\gamma\nu\bar{\nu}$,
as a function $\sqrt{s}$ with $C_{i}/\Lambda^{4}=1.0\ \mathrm{TeV^{-4}}$
allowed by LHC constraints. Only those operators which have both annihilation
and VBF contributions are plotted.}
\end{figure}

The NP amplitudes are generated by \verb"FeynArts"~\cite{Hahn:2000kx}
and the matrix elements are simplified with the help of \verb"FeynCalc"~\cite{Shtabovenko:2016sxi}.
For the VBF process, the exact analytical expressions of cross sections
are difficult to obtain due to the complicated three-body phase space
integration. We first ignore the masses of muons, and then perform
a Taylor expansion of $M_{Z}/\sqrt{s}$. After these approximations,
we can obtain the analytical expressions of VBF cross sections. For
the annihilation process, we calculate the cross section of $\mu^{+}\mu^{-}\to Z\gamma$
and $\sigma(\mu^{+}\mu^{-}\to\gamma\nu\bar{\nu)}$ is estimated as
$\sigma(\mu^{+}\mu^{-}\to Z\gamma)\times\mathrm{Br}(Z\to\nu\bar{\nu})$,
and $\mathrm{Br}(Z\to\nu\bar{\nu})$ is taken to be 20\%~\cite{ParticleDataGroup:2024cfk}.
The NP cross sections from annihilation and VBF contributions of the
7 CPC nTGCs operators are given by:
\begin{align}
\mathcal{O}_{\tilde{W}W}:\quad & \left\{ \begin{aligned}\sigma_{\mathrm{annih}}^{\mathrm{NP}} & =0\\
\sigma_{\mathrm{VBF}}^{\mathrm{NP}} & =\frac{7C_{1}^{2}e^{2}s^{2}c_{W}^{4}M_{Z}^{4}}{18432\pi^{3}\Lambda^{8}M_{W}^{2}}
\end{aligned}
\right.\\
\mathcal{O}_{\tilde{W}W}^{\prime}:\quad & \left\{ \begin{aligned}\sigma_{\mathrm{annih}}^{\mathrm{NP}} & =\frac{C_{2}^{2}c_{W}^{2}M_{Z}^{2}s_{W}^{2}(s-M_{Z}^{2})^{3}(M_{Z}^{2}+s)}{768\pi\Lambda^{8}s^{2}}\times\mathrm{Br}(Z\to\nu\bar{\nu})\\
\sigma_{\mathrm{VBF}}^{\mathrm{NP}} & =\frac{7C_{2}^{2}e^{2}s^{2}c_{W}^{4}M_{Z}^{4}}{73728\pi^{3}\Lambda^{8}M_{W}^{2}}
\end{aligned}
\right.\\
\mathcal{O}_{\tilde{B}B}^{\prime}:\quad & \left\{ \begin{aligned}\sigma_{\mathrm{annih}}^{\mathrm{NP}} & =\frac{5C_{4}^{2}c_{W}^{2}M_{Z}^{2}s_{W}^{2}(s-M_{Z}^{2})^{3}(M_{Z}^{2}+s)}{48\pi\Lambda^{8}s^{2}}\mathrm{\times Br}(Z\to\nu\bar{\nu})\\
\sigma_{\mathrm{VBF}}^{\mathrm{NP}} & =0
\end{aligned}
\right.
\end{align}
\begin{align}
\mathcal{O}_{\tilde{B}W}:\quad & \left\{ \begin{aligned}\sigma_{\mathrm{annih}}^{\mathrm{NP}} & =\frac{C_{5}^{2}M_{Z}^{2}(-2c_{W}^{2}s_{W}^{2}+c_{W}^{4}+5s_{W}^{4})(s-M_{Z}^{2})^{3}(M_{Z}^{2}+s)}{192\pi\Lambda^{8}s^{2}}\mathrm{\times Br}(Z\to\nu\bar{\nu})\\
\sigma_{\mathrm{VBF}}^{\mathrm{NP}} & =\frac{7C_{5}^{2}e^{2}s^{2}c_{W}^{6}M_{Z}^{4}}{18432\pi^{3}\Lambda^{8}M_{W}^{2}s_{W}^{2}}
\end{aligned}
\right.\\
\mathcal{O}_{\tilde{B}W}^{\prime}:\quad & \left\{ \begin{aligned}\sigma_{\mathrm{annih}}^{\mathrm{NP}} & =\frac{C_{6}^{2}c_{W}^{4}M_{Z}^{2}(s-M_{Z}^{2})^{3}(M_{Z}^{2}+s)}{192\pi\Lambda^{8}s^{2}}\mathrm{\times Br}(Z\to\nu\bar{\nu})\\
\sigma_{\mathrm{VBF}}^{\mathrm{NP}} & =\frac{7C_{6}^{2}e^{2}s^{2}c_{W}^{6}M_{Z}^{4}}{18432\pi^{3}\Lambda^{8}M_{W}^{2}s_{W}^{2}}
\end{aligned}
\right.
\end{align}
\begin{align}
\mathcal{O}_{G+}:\quad & \left\{ \begin{aligned}\sigma_{\mathrm{annih}}^{\mathrm{NP}} & =\frac{C_{11}^{2}(s-M_{Z}^{2})^{3}(M_{Z}^{2}+s)}{192\pi\Lambda^{8}s}\mathrm{\times Br}(Z\to\nu\bar{\nu})\\
\sigma_{\mathrm{VBF}}^{\mathrm{NP}} & =\frac{C_{11}^{2}e^{2}s^{3}c_{W}^{2}\left[660\log(\frac{s}{M_{W}^{2}})+2527\right]}{5529600\pi^{3}\Lambda^{8}s_{W}^{2}}
\end{aligned}
\right.\\
\mathcal{O}_{G-}:\quad & \left\{ \begin{aligned}\sigma_{\mathrm{annih}}^{\mathrm{NP}} & =\frac{C_{12}^{2}M_{Z}^{2}s_{W}^{4}(s-M_{Z}^{2})^{3}(M_{Z}^{2}+s)}{24\pi\Lambda^{8}s^{2}}\mathrm{\times Br}(Z\to\nu\bar{\nu})\\
\sigma_{\mathrm{VBF}}^{\mathrm{NP}} & =0
\end{aligned}
\right.
\end{align}
where $s_{W}$ and $c_{W}$ are the sine and cosine of Weinberg angle
$\theta_{W}$, $M_{Z}$ and $M_{W}$ are the masses of $Z$ and $W$
bosons. The NP cross sections of the 5 CPV nTGCs operators are given
by:
\begin{align}
\mathcal{O}_{BW}:\quad & \left\{ \begin{aligned}\sigma_{\mathrm{annih}}^{\mathrm{NP}} & =\frac{C_{8}^{2}M_{Z}^{2}(4s_{W}^{4}+1)(s-M_{Z}^{2})^{3}(M_{Z}^{2}+s)}{768\pi\Lambda^{8}s^{2}}\mathrm{\times Br}(Z\to\nu\bar{\nu})\\
\sigma_{\mathrm{VBF}}^{\mathrm{NP}} & =\frac{7C_{8}^{2}e^{2}s^{2}c_{W}^{6}M_{Z}^{4}}{73728\pi^{3}\Lambda^{8}M_{W}^{2}s_{W}^{2}}
\end{aligned}
\right.\\
\mathcal{O}_{WW}:\quad & \left\{ \begin{aligned}\sigma_{\mathrm{annih}}^{\mathrm{NP}} & =\frac{C_{9}^{2}c_{W}^{2}M_{Z}^{2}s_{W}^{2}(s-M_{Z}^{2})^{3}(M_{Z}^{2}+s)}{768\pi\Lambda^{8}s^{2}}\mathrm{\times Br}(Z\to\nu\bar{\nu})\\
\sigma_{\mathrm{VBF}}^{\mathrm{NP}} & =0
\end{aligned}
\right.\\
\mathcal{O}_{BB}:\quad & \left\{ \begin{aligned}\sigma_{\mathrm{annih}}^{\mathrm{NP}} & =\frac{5C_{10}^{2}c_{W}^{2}M_{Z}^{2}s_{W}^{2}(s-M_{Z}^{2})^{3}(M_{Z}^{2}+s)}{48\pi\Lambda^{8}s^{2}}\mathrm{\times Br}(Z\to\nu\bar{\nu})\\
\sigma_{\mathrm{VBF}}^{\mathrm{NP}} & =0
\end{aligned}
\right.
\end{align}
\begin{align}
\widetilde{\mathcal{O}}_{G+}:\quad & \left\{ \begin{aligned}\sigma_{\mathrm{annih}}^{\mathrm{NP}} & =\frac{C_{13}^{2}(s-M_{Z}^{2})^{3}(M_{Z}^{2}+s)}{768\pi\Lambda^{8}s}\mathrm{\times Br}(Z\to\nu\bar{\nu})\\
\sigma_{\mathrm{VBF}}^{\mathrm{NP}} & =\frac{C_{13}^{2}e^{2}s^{3}c_{W}^{2}\left[660\log(\frac{s}{M_{W}^{2}})+2527\right]}{22118400\pi^{3}\Lambda^{8}s_{W}^{2}}
\end{aligned}
\right.\\
\widetilde{\mathcal{O}}_{G-}:\quad & \left\{ \begin{aligned}\sigma_{\mathrm{annih}}^{\mathrm{NP}} & =\frac{C_{14}^{2}M_{Z}^{2}s_{W}^{4}(s-M_{Z}^{2})^{3}(M_{Z}^{2}+s)}{96\pi\Lambda^{8}s^{2}}\mathrm{\times Br}(Z\to\nu\bar{\nu})\\
\sigma_{\mathrm{VBF}}^{\mathrm{NP}} & =0
\end{aligned}
\right.
\end{align}

From the above results, we can see that logarithmic enhancements only
exist for pure gauge operators $\mathcal{O}_{G+}$ and $\widetilde{\mathcal{O}}_{G+}$.
Phase space integration results large numbers in the denominators
of VBF than annihilation processes, which make annihilation to be
the dominant process at TeV scale. The comparison of the NP cross
sections of annihilation with VBF is shown in Figure~\ref{fig:cs-comparison}.
In general, the contribution from pure gauge operators is three orders
of magnitude higher than operators involving the Higgs doublet, and
the contribution from annihilation processes is three orders of magnitude
higher than VBF. 

We also analytically calculate the interference cross sections between
nTGCs operators and SM for the annihilation and VBF respectively.
For the CPV operators, the interference cross sections are zero and
we do not list them here. The interference cross sections for the
7 CPC operators are given by:
\begin{align}
\mathcal{O}_{\tilde{W}W}:\quad & \left\{ \begin{aligned}\sigma_{\mathrm{annih}}^{\mathrm{int}} & =0\\
\sigma_{\mathrm{VBF}}^{\mathrm{int}} & =\frac{C_{1}e^{4}sc_{W}^{2}M_{Z}^{2}}{384\pi^{3}\Lambda^{4}M_{W}^{2}s_{W}^{2}}
\end{aligned}
\right.\\
\mathcal{O}_{\tilde{W}W}^{\prime}:\quad & \left\{ \begin{aligned}\sigma_{\mathrm{annih}}^{\mathrm{int}} & =\frac{C_{2}e^{2}M_{Z}^{2}\left(c_{W}^{2}-s_{W}^{2}\right)\left(M_{Z}^{2}-s\right)\left(M_{Z}^{2}+s\right)}{32\pi\Lambda^{4}s^{2}}\times\mathrm{Br}(Z\to\nu\bar{\nu})\\
\sigma_{\mathrm{VBF}}^{\mathrm{int}} & =\frac{C_{2}e^{4}sc_{W}^{2}M_{Z}^{2}}{768\pi^{3}\Lambda^{4}M_{W}^{2}s_{W}^{2}}
\end{aligned}
\right.
\end{align}
\begin{align}
\mathcal{O}_{\tilde{B}B}^{\prime}:\quad & \left\{ \begin{aligned}\sigma_{\mathrm{annih}}^{\mathrm{int}} & =\frac{C_{4}e^{2}M_{Z}^{2}\left(c_{W}^{2}+3s_{W}^{2}\right)\left(M_{Z}^{2}-s\right)\left(M_{Z}^{2}+s\right)}{8\pi\Lambda^{4}s^{2}}\times\mathrm{Br}(Z\to\nu\bar{\nu})\\
\sigma_{\mathrm{VBF}}^{\mathrm{int}} & =0
\end{aligned}
\right.
\end{align}
\begin{align}
\mathcal{O}_{\tilde{B}W}:\quad & \left\{ \begin{aligned}\sigma_{\mathrm{annih}}^{\mathrm{int}} & =\frac{C_{5}e^{2}M_{Z}^{2}\left(-2c_{W}^{2}s_{W}^{2}+c_{W}^{4}-3s_{W}^{4}\right)\left(M_{Z}^{2}-s\right)\left(M_{Z}^{2}+s\right)}{16\pi\Lambda^{4}s^{2}c_{W}s_{W}}\mathrm{\times Br}(Z\to\nu\bar{\nu})\\
\sigma_{\mathrm{VBF}}^{\mathrm{int}} & =\frac{C_{5}e^{4}sc_{W}^{3}M_{Z}^{2}}{384\pi^{3}\Lambda^{4}M_{W}^{2}s_{W}^{3}}
\end{aligned}
\right.\\
\mathcal{O}_{\tilde{B}W}^{\prime}:\quad & \left\{ \begin{aligned}\sigma_{\mathrm{annih}}^{\mathrm{int}} & =\frac{C_{6}e^{2}c_{W}M_{Z}^{2}\left(c_{W}^{2}-s_{W}^{2}\right)\left(s-M_{Z}^{2}\right)\left(M_{Z}^{2}+s\right)}{16\pi\Lambda^{4}s^{2}s_{W}}\mathrm{\times Br}(Z\to\nu\bar{\nu})\\
\sigma_{\mathrm{VBF}}^{\mathrm{int}} & =\frac{C_{6}e^{4}sc_{W}^{3}M_{Z}^{2}}{384\pi^{3}\Lambda^{4}M_{W}^{2}s_{W}^{3}}
\end{aligned}
\right.
\end{align}
\begin{align}
\mathcal{O}_{G+}:\quad & \left\{ \begin{aligned}\sigma_{\mathrm{annih}}^{\mathrm{int}} & =\frac{C_{11}e^{2}M_{Z}^{2}\left(c_{W}^{2}-s_{W}^{2}\right)\left(M_{Z}^{2}-s\right)}{8\pi\Lambda^{4}sc_{W}s_{W}}\mathrm{\times Br}(Z\to\nu\bar{\nu})\\
\sigma_{\mathrm{VBF}}^{\mathrm{int}} & =-\frac{C_{11}e^{4}sc_{W}\left[12\log(\frac{M_{W}^{2}}{s})+49\right]}{13824\pi^{3}\Lambda^{4}s_{W}^{3}}
\end{aligned}
\right.\\
\mathcal{O}_{G-}:\quad & \left\{ \begin{aligned}\sigma_{\mathrm{annih}}^{\mathrm{int}} & =\frac{C_{12}e^{2}M_{Z}^{2}s_{W}\left(M_{Z}^{2}-s\right)\left(M_{Z}^{2}+s\right)}{8\pi\Lambda^{4}s^{2}c_{W}}\mathrm{\times Br}(Z\to\nu\bar{\nu})\\
\sigma_{\mathrm{VBF}}^{\mathrm{int}} & =0
\end{aligned}
\right.
\end{align}

\begin{figure}[tp]
\centering
\includegraphics[scale=0.4]{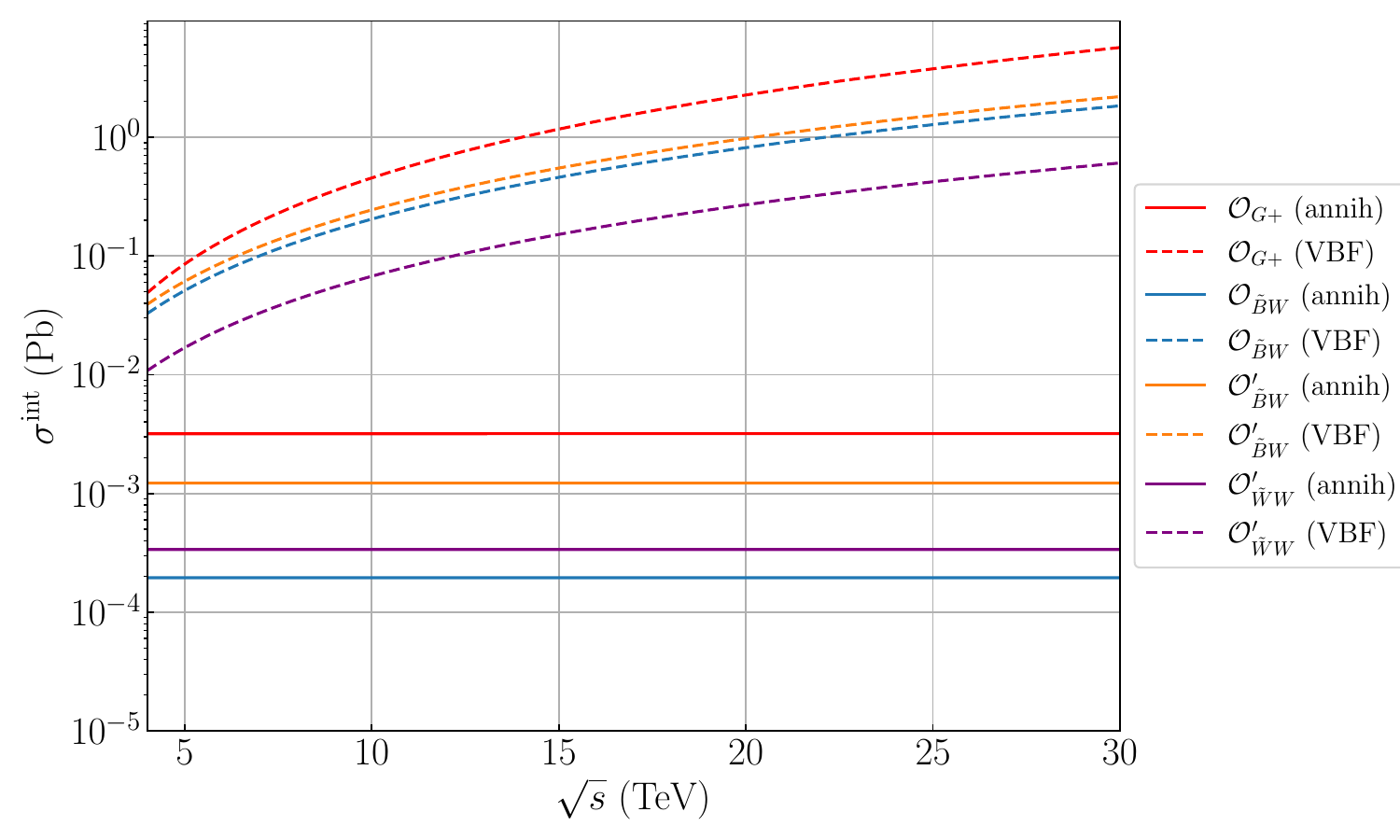}

\caption{\protect\label{fig:cs-comparison-1}The comparison of $\sigma_{\mathrm{annih}}^{\mathrm{int}}$
with $\sigma_{\mathrm{VBF}}^{\mathrm{int}}$ for $\mu^{+}\mu^{-}\to\gamma\nu\bar{\nu}$,
as a function $\sqrt{s}$ with $C_{i}/\Lambda^{4}=1.0\ \mathrm{TeV^{-4}}$
allowed by LHC constraints. Only those operators which have both annihilation
and VBF interference cross sections are plotted.}
\end{figure}

The interference cross sections of both annihilation and VBF with
SM are shown in Figure~\ref{fig:cs-comparison-1}. Taking into account
the impact of interference terms, the cross section can be expressed
as a bilinear function of the nTGCs operator coefficients $C_{i}$.
Specifically, the total cross section can be written as:
\begin{equation}
\sigma_{\mathrm{nTGC}}=\sigma_{\mathrm{SM}}+\frac{C_{i}}{\Lambda^{4}}\sigma_{\mathrm{int}}+\frac{C_{i}^{2}}{\Lambda^{8}}\sigma_{\mathrm{NP}},
\end{equation}
where $\sigma_{\mathrm{SM}}$, $\sigma_{\mathrm{int}}$, and $\sigma_{\mathrm{NP}}$
denote the SM, interference, and NP contributions, respectively. We
can perform a simple order-of-magnitude estimation of these contributions
under the approximation $\sqrt{s}\gg m$, where all the SM particle
masses $m$ can be neglected at TeV scale. Ignoring possible inferred
divergences and logarithmic enhancements, the only remaining scale
is $\sqrt{s}$. For the both sides of the equation to have the same
physical dimensions, we obtain the relation:
\begin{equation}
\frac{C_{i}^{2}}{\Lambda^{8}}\sigma_{\mathrm{NP}}\sim\frac{C_{i}}{\Lambda^{4}}\sigma_{\mathrm{int}}\sim\frac{1}{s},
\end{equation}
from which we can obtain the scaling behavior:
\begin{equation}
\sigma_{\mathrm{NP}}\sim s^{3},\quad\sigma_{\mathrm{int}}\sim s.
\end{equation}
From the above scaling behavior, we can roughly estimate, for a given
c.m. energy, when $C_{i}/\Lambda^{4}=1/s^{2}$, the NP contribution
becomes comparable to the interference term. In this work, for the
energy range we considered, the coefficient values $C_{i}/\Lambda^{4}$
are larger than $1/s^{2}$ (as will be shown in the following unitarity
bounds and expected constraints). Therefore, the NP contributions
from nTGCs operators to the total cross section are dominant compared
to the interference contributions, for the energy and coefficient
ranges in this work.

Based on the above comparisons, we reach the conclusion that the annihilation
process dominates over the VBF process in $\mu^{+}\mu^{-}\to\gamma\nu\bar{\nu}$
for nTGCs operators at TeV scale. This observation deviates from the
conventional expectation for many BSM models, where VBF processes
are typically expected to dominate due to their logarithmic enhancement
at high energies. In the present case, the annihilation process is
found to provide the dominant contribution for nTGCs operators at
TeV muon colliders.

\subsection{Beam polarization}

While technically beam polarization in muon colliders remains an important
and evolving research topic with challenges, polarized muons can enhance
the sensitivity of measurements, such as in precision tests of the
SM, Higgs boson properties and NP searches. Beam polarization enables
control over initial-state spin configurations, which helps disentangle
contributions in various scattering processes. Our study here aim
to explicitly check whether muon polarization provides a meaningful
advantage in the case of nTGCs, and to help quantify its impact on
the sensitivity reach for these couplings.

\begin{table}[tp]
\centering
\begin{tabular}{|c|c|c||c||c||c||c||c||c||c||c||c||c||c|}
\hline 
\multirow{2}{*}{Initial beam polarization} & \multicolumn{13}{c|}{Operators}\tabularnewline
\cline{2-14}
 & CPC & \multicolumn{12}{c|}{CPV}\tabularnewline
\hline 
$(+-)$ & $\mathcal{O}_{\tilde{W}W}^{\prime},\mathcal{O}_{\tilde{B}W}^{\prime},\mathcal{O}_{G+}$ & \multicolumn{12}{c|}{$\mathcal{O}_{WW},\widetilde{\mathcal{O}}_{G+}$}\tabularnewline
\hline 
$(+-)$ and $(-+)$  & $\mathcal{O}_{\tilde{B}B}^{\prime},\mathcal{O}_{\tilde{B}W},\mathcal{O}_{G-}$ & \multicolumn{12}{c|}{$\mathcal{O}_{BW},\mathcal{O}_{BB},\widetilde{\mathcal{O}}_{G-}$}\tabularnewline
\hline 
\end{tabular}

\caption{\protect\label{tab:Initial-state-polarization-of}Initial-state polarizations
of muons that contribute to $\mu^{+}\mu^{-}\to Z\gamma(Z\to\nu\bar{\nu)}$
from nTGCs operators. Here $(+-)$ denotes $\mu_{+1/2}^{+}\mu_{-1/2}^{-}$
polarization scheme, and $(-+)$ denotes $\mu_{-1/2}^{+}\mu_{+1/2}^{-}$
polarization scheme. The subscript $\pm1/2$ indicate muon helicities. }
\end{table}

For the VBF processes, we find that all the considered nTGCs operators
have only $\mu_{+1/2}^{+}\mu_{-1/2}^{-}$ (denoted by $(+-)$) polarization
contributions. So the polarized cross section expressions of VBF have
the same form as the unpolarized case, which are already shown in
Section~\ref{subsec:Comparison-of-VBF}.

For the annihilation processes, we obtain the analytical expressions
of their polarized cross sections. Among the 11 operators which have
annihilation contributions, 5 operators contribute to the nTGCs operators
exclusively through the $(+-)$ initial beam polarization channel,
while the other 6 operators contribute through both the $\mu_{+1/2}^{+}\mu_{-1/2}^{-}$
and $\mu_{-1/2}^{+}\mu_{+1/2}^{-}$ (denoted by $(-+)$) initial beam
polarization channels. The detailed classification is presented in
Table~\ref{tab:Initial-state-polarization-of}. For the 6 operators
with both the $(+-)$ and $(-+)$ polarization channels, the polarized
NP cross sections of annihilation are obtained:
\begin{align}
\mathcal{O}_{\tilde{B}B}^{\prime}:\quad & \left\{ \begin{aligned}\sigma_{\mathrm{+-}}^{\mathrm{NP}} & =\frac{C_{4}^{2}c_{W}^{2}M_{Z}^{2}s_{W}^{2}(s-M_{Z}^{2})^{3}(M_{Z}^{2}+s)}{12\pi\Lambda^{8}s^{2}}\mathrm{\times Br}(Z\to\nu\bar{\nu})\\
\sigma_{\mathrm{-+}}^{\mathrm{NP}} & =\frac{C_{4}^{2}c_{W}^{2}M_{Z}^{2}s_{W}^{2}(s-M_{Z}^{2})^{3}(M_{Z}^{2}+s)}{3\pi\Lambda^{8}s^{2}}\mathrm{\times Br}(Z\to\nu\bar{\nu})
\end{aligned}
\right.\\
\mathcal{O}_{\tilde{B}W}:\quad & \left\{ \begin{aligned}\sigma_{\mathrm{+-}}^{\mathrm{NP}} & =\frac{C_{5}^{2}M_{Z}^{2}\left(c_{W}-s_{W}\right){}^{2}\left(c_{W}+s_{W}\right){}^{2}\left(s-M_{Z}^{2}\right){}^{3}\left(M_{Z}^{2}+s\right)}{48\pi\Lambda^{8}s^{2}}\mathrm{\times Br}(Z\to\nu\bar{\nu})\\
\sigma_{\mathrm{-+}}^{\mathrm{NP}} & =\frac{C_{5}^{2}M_{Z}^{2}s_{W}^{4}\left(s-M_{Z}^{2}\right){}^{3}\left(M_{Z}^{2}+s\right)}{12\pi\Lambda^{8}s^{2}}\mathrm{\times Br}(Z\to\nu\bar{\nu})
\end{aligned}
\right.
\end{align}
\begin{align}
\mathcal{O}_{BW}:\quad & \left\{ \begin{aligned}\sigma_{\mathrm{+-}}^{\mathrm{NP}} & =\frac{C_{8}^{2}M_{Z}^{2}\left(s-M_{Z}^{2}\right){}^{3}\left(M_{Z}^{2}+s\right)}{192\pi\Lambda^{8}s^{2}}\mathrm{\times Br}(Z\to\nu\bar{\nu})\\
\sigma_{\mathrm{-+}}^{\mathrm{NP}} & =\frac{C_{8}^{2}M_{Z}^{2}s_{W}^{4}\left(s-M_{Z}^{2}\right){}^{3}\left(M_{Z}^{2}+s\right)}{48\pi\Lambda^{8}s^{2}}\mathrm{\times Br}(Z\to\nu\bar{\nu})
\end{aligned}
\right.\\
\mathcal{O}_{BB}:\quad & \left\{ \begin{aligned}\sigma_{\mathrm{+-}}^{\mathrm{NP}} & =\frac{C_{10}^{2}c_{W}^{2}M_{Z}^{2}s_{W}^{2}\left(s-M_{Z}^{2}\right){}^{3}\left(M_{Z}^{2}+s\right)}{12\pi\Lambda^{8}s^{2}}\mathrm{\times Br}(Z\to\nu\bar{\nu})\\
\sigma_{\mathrm{-+}}^{\mathrm{NP}} & =\frac{C_{10}^{2}c_{W}^{2}M_{Z}^{2}s_{W}^{2}\left(s-M_{Z}^{2}\right){}^{3}\left(M_{Z}^{2}+s\right)}{3\pi\Lambda^{8}s^{2}}\mathrm{\times Br}(Z\to\nu\bar{\nu})
\end{aligned}
\right.
\end{align}
\begin{align}
\mathcal{O}_{G-}:\quad & \left\{ \begin{aligned}\sigma_{\mathrm{+-}}^{\mathrm{NP}} & =\frac{C_{12}^{2}M_{Z}^{2}s_{W}^{4}(s-M_{Z}^{2})^{3}(M_{Z}^{2}+s)}{12\pi\Lambda^{8}s^{2}}\mathrm{\times Br}(Z\to\nu\bar{\nu})\\
\sigma_{\mathrm{-+}}^{\mathrm{NP}} & =\frac{C_{12}^{2}M_{Z}^{2}s_{W}^{4}(s-M_{Z}^{2})^{3}(M_{Z}^{2}+s)}{12\pi\Lambda^{8}s^{2}}\mathrm{\times Br}(Z\to\nu\bar{\nu})
\end{aligned}
\right.\\
\widetilde{\mathcal{O}}_{G-}:\quad & \left\{ \begin{aligned}\sigma_{\mathrm{+-}}^{\mathrm{NP}} & =\frac{C_{14}^{2}M_{Z}^{2}s_{W}^{4}(s-M_{Z}^{2})^{3}(M_{Z}^{2}+s)}{48\pi\Lambda^{8}s^{2}}\mathrm{\times Br}(Z\to\nu\bar{\nu})\\
\sigma_{\mathrm{-+}}^{\mathrm{NP}} & =\frac{C_{14}^{2}M_{Z}^{2}s_{W}^{4}(s-M_{Z}^{2})^{3}(M_{Z}^{2}+s)}{48\pi\Lambda^{8}s^{2}}\mathrm{\times Br}(Z\to\nu\bar{\nu})
\end{aligned}
\right.
\end{align}
 The polarized interference cross sections of annihilation with SM
are also obtained:
\begin{align}
\mathcal{O}_{\tilde{B}B}^{\prime}:\quad & \left\{ \begin{aligned}\sigma_{\mathrm{+-}}^{\mathrm{int}} & =\frac{C_{4}e^{2}M_{Z}^{2}\left(c_{W}^{2}-s_{W}^{2}\right)\left(M_{Z}^{2}-s\right)\left(M_{Z}^{2}+s\right)}{2\pi\Lambda^{4}s^{2}}\mathrm{\times Br}(Z\to\nu\bar{\nu})\\
\sigma_{\mathrm{-+}}^{\mathrm{int}} & =\frac{2C_{4}e^{2}M_{Z}^{2}s_{W}^{2}\left(M_{Z}^{2}-s\right)\left(M_{Z}^{2}+s\right)}{\pi\Lambda^{4}s^{2}}\mathrm{\times Br}(Z\to\nu\bar{\nu})
\end{aligned}
\right.\\
\mathcal{O}_{\tilde{B}W}:\quad & \left\{ \begin{aligned}\sigma_{\mathrm{+-}}^{\mathrm{int}} & =\frac{C_{5}e^{2}M_{Z}^{2}\left(c_{W}-s_{W}\right){}^{2}\left(c_{W}+s_{W}\right){}^{2}\left(M_{Z}^{2}-s\right)\left(M_{Z}^{2}+s\right)}{4\pi\Lambda^{4}s^{2}c_{W}s_{W}}\mathrm{\times Br}(Z\to\nu\bar{\nu})\\
\sigma_{\mathrm{-+}}^{\mathrm{int}} & =\frac{C_{5}e^{2}M_{Z}^{2}s_{W}^{3}\left(s-M_{Z}^{2}\right)\left(M_{Z}^{2}+s\right)}{\pi\Lambda^{4}s^{2}c_{W}}\mathrm{\times Br}(Z\to\nu\bar{\nu})
\end{aligned}
\right.
\end{align}
\begin{align}
\mathcal{O}_{G-}:\quad & \left\{ \begin{aligned}\sigma_{\mathrm{+-}}^{\mathrm{int}} & =\frac{C_{12}e^{2}M_{Z}^{2}s_{W}\left(c_{W}^{2}-s_{W}^{2}\right)\left(M_{Z}^{2}-s\right)\left(M_{Z}^{2}+s\right)}{2\pi\Lambda^{4}s^{2}c_{W}}\mathrm{\times Br}(Z\to\nu\bar{\nu})\\
\sigma_{\mathrm{-+}}^{\mathrm{int}} & =\frac{C_{12}e^{2}M_{Z}^{2}s_{W}^{3}\left(M_{Z}^{2}-s\right)\left(M_{Z}^{2}+s\right)}{\pi\Lambda^{4}s^{2}c_{W}}\mathrm{\times Br}(Z\to\nu\bar{\nu})
\end{aligned}
\right.
\end{align}

\begin{align}
\mathcal{O}_{BW}:\quad & \left\{ \begin{aligned}\sigma_{\mathrm{+-}}^{\mathrm{int}} & =0\\
\sigma_{\mathrm{-+}}^{\mathrm{int}} & =0
\end{aligned}
\right.\\
\mathcal{O}_{BB}:\quad & \left\{ \begin{aligned}\sigma_{\mathrm{+-}}^{\mathrm{int}} & =0\\
\sigma_{\mathrm{-+}}^{\mathrm{int}} & =0
\end{aligned}
\right.\\
\widetilde{\mathcal{O}}_{G-}:\quad & \left\{ \begin{aligned}\sigma_{\mathrm{+-}}^{\mathrm{int}} & =0\\
\sigma_{\mathrm{-+}}^{\mathrm{int}} & =0
\end{aligned}
\right.
\end{align}
Here the interference cross sections of CPV operators are zero. For
completeness, we also list them here. 

\subsection{Partial wave unitarity bounds}

As an effective field theory, SMEFT is valid only at a certain energy
scale. A signal that the effective theory is no longer valid is the
violation of unitarity. When considering the contributions from nTGCs,
the cross sections of the processes $\mu^{+}\mu^{-}\to\gamma\nu\bar{\nu}$
increase with the c.m. energy. At sufficiently high energies $\sqrt{s}$,
this leads to a violation of unitarity. The violation of unitarity
indicates that SMEFT is no longer applicable in a perturbative framework.
Partial wave unitarity is often used as a criterion for determining
the validity of SMEFT~\cite{Corbett:2014ora,Chen:2023bhu,Remmen:2020uze,Garcia-Garcia:2019oig}.
In this subsection we present the calculation of the partial wave
unitarity bounds for nTGCs and provide the upper bounds on $C_{i}$
from different helicity amplitudes. 

The helicity amplitudes of $\mu^{+}\mu^{-}\to Z\gamma$ from the nTGCs
operators are obtained. For the 5 operators with $(+-)$ polarization
channel, the helicity amplitudes are given by:
\begin{align}
\mathcal{O}_{\tilde{W}W}^{\prime}:\quad & \left\{ \begin{aligned}\mathcal{M}\left(\mu_{+\frac{1}{2}}^{+}\mu_{-\frac{1}{2}}^{-}\to Z_{0}\gamma_{\pm}\right) & =\frac{C_{2}e^{2}\sqrt{s}v^{2}e^{-i\phi}(\cos(\theta)\pm1)\left(s-M_{Z}^{2}\right)}{16\sqrt{2}c_{W}M_{Z}s_{W}\Lambda^{4}}\\
\mathcal{M}\left(\mu_{+\frac{1}{2}}^{+}\mu_{-\frac{1}{2}}^{-}\to Z_{\pm}\gamma_{\pm}\right) & =\pm\frac{C_{2}e^{2}v^{2}e^{-i\phi}\sin(\theta)\left(s-M_{Z}^{2}\right)}{16c_{W}s_{W}\Lambda^{4}}
\end{aligned}
\right.\label{eq:2-52}
\end{align}
\begin{align}
\mathcal{O}_{\tilde{B}W}^{\prime}:\quad & \left\{ \begin{aligned}\mathcal{M}\left(\mu_{+\frac{1}{2}}^{+}\mu_{-\frac{1}{2}}^{-}\to Z_{0}\gamma_{\pm}\right) & =\frac{C_{6}e^{2}\sqrt{s}v^{2}e^{-i\phi}(\cos(\theta)\pm1)\left(M_{Z}^{2}-s\right)}{8\sqrt{2}M_{Z}s_{W}^{2}\Lambda^{4}}\\
\mathcal{M}\left(\mu_{+\frac{1}{2}}^{+}\mu_{-\frac{1}{2}}^{-}\to Z_{\pm}\gamma_{\pm}\right) & =\pm\frac{\text{\ensuremath{C_{6}}}e^{2}v^{2}e^{-i\phi}\sin(\theta)\left(M_{Z}^{2}-s\right)}{8s_{W}^{2}\Lambda^{4}}
\end{aligned}
\right.
\end{align}
\begin{align}
\mathcal{O}_{WW}:\quad & \left\{ \begin{aligned}\mathcal{M}\left(\mu_{+\frac{1}{2}}^{+}\mu_{-\frac{1}{2}}^{-}\to Z_{0}\gamma_{\pm}\right) & =\frac{iC_{9}e^{2}\sqrt{s}v^{2}e^{-i\phi}(1\pm\cos(\theta))\left(M_{Z}^{2}-s\right)}{16\sqrt{2}\Lambda^{4}c_{W}M_{Z}s_{W}}\\
\mathcal{M}\left(\mu_{+\frac{1}{2}}^{+}\mu_{-\frac{1}{2}}^{-}\to Z_{\pm}\gamma_{\pm}\right) & =\frac{iC_{9}e^{2}v^{2}e^{-i\phi}\sin(\theta)\left(M_{Z}^{2}-s\right)}{16\Lambda^{4}c_{W}s_{W}}
\end{aligned}
\right.
\end{align}
\begin{align}
\mathcal{O}_{G+}:\quad & \left\{ \begin{aligned}\mathcal{M}\left(\mu_{+\frac{1}{2}}^{+}\mu_{-\frac{1}{2}}^{-}\to Z_{0}\gamma_{\pm}\right) & =\frac{C_{11}\sqrt{s}e^{-i\phi}(\cos(\theta)\pm1)M_{Z}\left(s-M_{Z}^{2}\right)}{2\sqrt{2}\Lambda^{4}}\\
\mathcal{M}\left(\mu_{+\frac{1}{2}}^{+}\mu_{-\frac{1}{2}}^{-}\to Z_{\pm}\gamma_{\pm}\right) & =\pm\frac{C_{11}se^{-i\phi}\sin(\theta)\left(s-M_{Z}^{2}\right)}{2\Lambda^{4}}
\end{aligned}
\right.
\end{align}
\begin{align}
\widetilde{\mathcal{O}}_{G+}:\quad & \left\{ \begin{aligned}\mathcal{M}\left(\mu_{+\frac{1}{2}}^{+}\mu_{-\frac{1}{2}}^{-}\to Z_{0}\gamma_{\pm}\right) & =\frac{iC_{13}\sqrt{s}e^{-i\phi}(1\pm\cos(\theta))M_{Z}\left(M_{Z}^{2}-s\right)}{4\sqrt{2}\Lambda^{4}}\\
\mathcal{M}\left(\mu_{+\frac{1}{2}}^{+}\mu_{-\frac{1}{2}}^{-}\to Z_{\pm}\gamma_{\pm}\right) & =\frac{iC_{13}se^{-i\phi}\sin(\theta)\left(M_{Z}^{2}-s\right)}{4\Lambda^{4}}
\end{aligned}
\right.
\end{align}

In the above equations $\theta$ and $\phi$ are the polar and azimuthal
angles of the $Z$ boson. For the 6 operators with $(+-)$ and $(-+)$
polarization channels, the helicity amplitudes are given by:
\begin{align}
\mathcal{O}_{\tilde{B}B}^{\prime}:\quad & \left\{ \begin{aligned}\mathcal{M}\left(\mu_{+\frac{1}{2}}^{+}\mu_{-\frac{1}{2}}^{-}\to Z_{0}\gamma_{\pm}\right) & =\frac{C_{4}e^{2}\sqrt{s}v^{2}e^{-i\phi}(\cos(\theta)\pm1)\left(s-M_{Z}^{2}\right)}{4\sqrt{2}c_{W}M_{Z}s_{W}\Lambda^{4}}\\
\mathcal{M}\left(\mu_{+\frac{1}{2}}^{+}\mu_{-\frac{1}{2}}^{-}\to Z_{\pm}\gamma_{\pm}\right) & =\pm\frac{\text{\ensuremath{C_{4}}}e^{2}v^{2}e^{-i\phi}\sin(\theta)\left(s-M_{Z}^{2}\right)}{4c_{W}s_{W}\Lambda^{4}}\\
\mathcal{M}\left(\mu_{-\frac{1}{2}}^{+}\mu_{+\frac{1}{2}}^{-}\to Z_{0}\gamma_{\pm}\right) & =\frac{\text{\ensuremath{C_{4}}}e^{2}\sqrt{s}v^{2}e^{i\phi}(\cos(\theta)\mp1)\left(M_{Z}^{2}-s\right)}{2\sqrt{2}c_{W}M_{Z}s_{W}\Lambda^{4}}\\
\mathcal{M}\left(\mu_{-\frac{1}{2}}^{+}\mu_{+\frac{1}{2}}^{-}\to Z_{\pm}\gamma_{\pm}\right) & =\pm\frac{\text{\text{\ensuremath{C_{4}}}}e^{2}v^{2}e^{i\phi}\sin(\theta)\left(M_{Z}^{2}-s\right)}{2c_{W}s_{W}\Lambda^{4}}
\end{aligned}
\right.
\end{align}
\begin{align}
\mathcal{O}_{\tilde{B}W}:\quad & \left\{ \begin{aligned}\mathcal{M}\left(\mu_{+\frac{1}{2}}^{+}\mu_{-\frac{1}{2}}^{-}\to Z_{0}\gamma_{\pm}\right) & =\frac{C_{5}e^{2}\sqrt{s}v^{2}e^{-i\phi}(\cos(\theta)\pm1)\left(1-2s_{W}^{2}\right)\left(s-M_{Z}^{2}\right)}{8\sqrt{2}c_{W}^{2}M_{Z}s_{W}^{2}\Lambda^{4}}\\
\mathcal{M}\left(\mu_{+\frac{1}{2}}^{+}\mu_{-\frac{1}{2}}^{-}\to Z_{\pm}\gamma_{\pm}\right) & =\pm\frac{C_{5}e^{2}v^{2}e^{-i\phi}\sin(\theta)\left(1-2s_{W}^{2}\right)\left(s-M_{Z}^{2}\right)}{8c_{W}^{2}s_{W}^{2}\Lambda^{4}}\\
\mathcal{M}\left(\mu_{-\frac{1}{2}}^{+}\mu_{+\frac{1}{2}}^{-}\to Z_{0}\gamma_{\pm}\right) & =\frac{C_{5}e^{2}\sqrt{s}v^{2}e^{i\phi}(\cos(\theta)\mp1)\left(s-M_{Z}^{2}\right)}{4\sqrt{2}c_{W}^{2}M_{Z}\Lambda^{4}}\\
\mathcal{M}\left(\mu_{-\frac{1}{2}}^{+}\mu_{+\frac{1}{2}}^{-}\to Z_{\pm}\gamma_{\pm}\right) & =\pm\frac{C_{5}e^{2}v^{2}e^{i\phi}\sin(\theta)\left(s-M_{Z}^{2}\right)}{4c_{W}^{2}\Lambda^{4}}
\end{aligned}
\right.
\end{align}
\begin{align}
\mathcal{O}_{BW}:\quad & \left\{ \begin{aligned}\mathcal{M}\left(\mu_{+\frac{1}{2}}^{+}\mu_{-\frac{1}{2}}^{-}\to Z_{0}\gamma_{\pm}\right) & =\frac{iC_{8}e^{2}\sqrt{s}v^{2}e^{-i\phi}(1\pm\cos(\theta))\left(s-M_{Z}^{2}\right)}{16\sqrt{2}c_{W}^{2}M_{Z}s_{W}^{2}\Lambda^{4}}\\
\mathcal{M}\left(\mu_{+\frac{1}{2}}^{+}\mu_{-\frac{1}{2}}^{-}\to Z_{\pm}\gamma_{\pm}\right) & =\frac{i\text{\ensuremath{C_{8}}}e^{2}v^{2}e^{-i\phi}\sin(\theta)\left(s-M_{Z}^{2}\right)}{16c_{W}^{2}s_{W}^{2}\Lambda^{4}}\\
\mathcal{M}\left(\mu_{-\frac{1}{2}}^{+}\mu_{+\frac{1}{2}}^{-}\to Z_{0}\gamma_{\pm}\right) & =\frac{iC_{8}e^{2}\sqrt{s}v^{2}e^{i\phi}(1\mp\cos(\theta))\left(s-M_{Z}^{2}\right)}{8\sqrt{2}\Lambda^{4}c_{W}^{2}M_{Z}}\\
\mathcal{M}\left(\mu_{-\frac{1}{2}}^{+}\mu_{+\frac{1}{2}}^{-}\to Z_{\pm}\gamma_{\pm}\right) & =\frac{i\text{\ensuremath{C_{8}}}e^{2}v^{2}e^{i\phi}\sin(\theta)\left(M_{Z}^{2}-s\right)}{8c_{W}^{2}\Lambda^{4}}
\end{aligned}
\right.
\end{align}
\begin{align}
\mathcal{O}_{BB}:\quad & \left\{ \begin{aligned}\mathcal{M}\left(\mu_{+\frac{1}{2}}^{+}\mu_{-\frac{1}{2}}^{-}\to Z_{0}\gamma_{\pm}\right) & =\frac{iC_{10}e^{2}\sqrt{s}v^{2}e^{-i\phi}(1\mp\cos(\theta))\left(M_{Z}^{2}-s\right)}{4\sqrt{2}c_{W}M_{Z}s_{W}\Lambda^{4}}\\
\mathcal{M}\left(\mu_{+\frac{1}{2}}^{+}\mu_{-\frac{1}{2}}^{-}\to Z_{\pm}\gamma_{\pm}\right) & =\frac{\text{i\ensuremath{C_{10}}}e^{2}v^{2}e^{-i\phi}\sin(\theta)\left(M_{Z}^{2}-s\right)}{4c_{W}s_{W}\Lambda^{4}}\\
\mathcal{M}\left(\mu_{-\frac{1}{2}}^{+}\mu_{+\frac{1}{2}}^{-}\to Z_{0}\gamma_{\pm}\right) & =\frac{iC_{10}e^{2}\sqrt{s}v^{2}e^{i\phi}(1\mp\cos(\theta))\left(M_{Z}^{2}-s\right)}{2\sqrt{2}\Lambda^{4}c_{W}M_{Z}s_{W}}\\
\mathcal{M}\left(\mu_{-\frac{1}{2}}^{+}\mu_{+\frac{1}{2}}^{-}\to Z_{\pm}\gamma_{\pm}\right) & =\frac{iC_{10}e^{2}v^{2}e^{i\phi}\sin(\theta)\left(s-M_{Z}^{2}\right)}{2\Lambda^{4}c_{W}s_{W}}
\end{aligned}
\right.
\end{align}
\begin{align}
\mathcal{O}_{G-}:\quad & \left\{ \begin{aligned}\mathcal{M}\left(\mu_{+\frac{1}{2}}^{+}\mu_{-\frac{1}{2}}^{-}\to Z_{0}\gamma_{\pm}\right) & =\frac{C_{12}\sqrt{s}e^{-i\phi}(\cos(\theta)\pm1)M_{Z}s_{W}^{2}\left(s-M_{Z}^{2}\right)}{\sqrt{2}\Lambda^{4}}\\
\mathcal{M}\left(\mu_{+\frac{1}{2}}^{+}\mu_{-\frac{1}{2}}^{-}\to Z_{\pm}\gamma_{\pm}\right) & =\frac{\text{\ensuremath{\pm}\ensuremath{C_{12}}}e^{-i\phi}\sin(\theta)M_{Z}^{2}s_{W}^{2}\left(s-M_{Z}^{2}\right)}{\Lambda^{4}}\\
\mathcal{M}\left(\mu_{-\frac{1}{2}}^{+}\mu_{+\frac{1}{2}}^{-}\to Z_{0}\gamma_{\pm}\right) & =\frac{\text{\ensuremath{C_{12}}}\sqrt{s}e^{i\phi}(\cos(\theta)\mp1)s_{W}^{2}\left(M_{Z}^{3}-sM_{Z}\right)}{\sqrt{2}\Lambda^{4}}\\
\mathcal{M}\left(\mu_{-\frac{1}{2}}^{+}\mu_{+\frac{1}{2}}^{-}\to Z_{\pm}\gamma_{\pm}\right) & =\frac{\pm\text{\ensuremath{C_{12}}}e^{i\phi}\sin(\theta)M_{Z}^{2}s_{W}^{2}\left(M_{Z}^{2}-s\right)}{\Lambda^{4}}
\end{aligned}
\right.
\end{align}
\begin{align}
\widetilde{\mathcal{O}}_{G-}:\quad & \left\{ \begin{aligned}\mathcal{M}\left(\mu_{+\frac{1}{2}}^{+}\mu_{-\frac{1}{2}}^{-}\to Z_{0}\gamma_{\pm}\right) & =\frac{iC_{14}\sqrt{s}e^{-i\phi}(1\pm\cos(\theta))M_{Z}s_{W}^{2}\left(s-M_{Z}^{2}\right)}{2\sqrt{2}\Lambda^{4}}\\
\mathcal{M}\left(\mu_{+\frac{1}{2}}^{+}\mu_{-\frac{1}{2}}^{-}\to Z_{\pm}\gamma_{\pm}\right) & =\frac{i\text{\ensuremath{C_{14}}}e^{-i\phi}\sin(\theta)M_{Z}^{2}s_{W}^{2}\left(s-M_{Z}^{2}\right)}{2\Lambda^{4}}\\
\mathcal{M}\left(\mu_{-\frac{1}{2}}^{+}\mu_{+\frac{1}{2}}^{-}\to Z_{0}\gamma_{\pm}\right) & =\frac{iC_{14}\sqrt{s}e^{i\phi}(1\mp\cos(\theta))M_{Z}s_{W}^{2}\left(s-M_{Z}^{2}\right)}{2\sqrt{2}\Lambda^{4}}\\
\mathcal{M}\left(\mu_{-\frac{1}{2}}^{+}\mu_{+\frac{1}{2}}^{-}\to Z_{\pm}\gamma_{\pm}\right) & =\frac{i\text{\ensuremath{C_{14}}}e^{i\phi}\sin(\theta)M_{Z}^{2}s_{W}^{2}\left(M_{Z}^{2}-s\right)}{2\Lambda^{4}}
\end{aligned}
\right.\label{eq:2-62}
\end{align}

In the case of $f\bar{f}\to V_{1}V_{2}$, where $f$ is a fermion,
$\bar{f}$ is an antifermion, and $V_{1,2}$ are bosons, the amplitude
can be expanded as~\cite{Jacob:1959at,Baur:1987mt}:
\begin{equation}
\mathcal{M}(f_{\sigma_{1}}\bar{f}_{\sigma_{2}}\to V_{1,\lambda_{3}}V_{2,\lambda_{4}})=16\pi\sum_{J}\left(J+\frac{1}{2}\right)\delta_{\sigma_{1},-\sigma_{2}}e^{i(m_{1}-m_{2})\phi}d_{m_{1},m_{2}}^{J}(\theta,\phi)T_{J},
\end{equation}
where $\sigma_{1,2}$ are the helicities of the fermion and antifermion,
$\lambda_{1,2}$ are the helicities of the bosons. $m_{1}=\sigma_{1}-\sigma_{2}$,
$m_{2}=\lambda_{3}-\lambda_{4}$, and $d_{m_{1},m_{2}}^{J}$ are the
Wigner D functions. $\phi$ and $\theta$ are azimuth and zenith angles
of $V_{1}$, and $T_{J}$ are coefficients of the partial wave expansion.
Helicity amplitudes need to be expanded to extract the expansion coefficients:
\begin{equation}
\mathcal{M}(f_{\sigma_{1}}\bar{f}_{\sigma_{2}}\to V_{1,\lambda_{3}}V_{2,\lambda_{4}})=\sum_{J=0}^{\infty}T_{J}f_{J}\left(x\right),
\end{equation}
in which:
\begin{equation}
f_{J}\left(x\right)=16\pi\left(J+\frac{1}{2}\right)e^{i\left(\sigma_{1}-\sigma_{2}-\lambda_{3}+\lambda_{4}\right)\phi}d_{\sigma_{1}-\sigma_{2},\lambda_{3}-\lambda_{4}}^{J}\left(\theta,\phi\right).
\end{equation}
Unitarity bound imposes the constraint that~\cite{Corbett:2014ora}:
\begin{equation}
\left|T_{J}\right|\leq1.
\end{equation}

\begin{table}[tp]
\centering
\begin{tabular}{c|c|c|c|c}
\hline 
 & $\sqrt{s}=3\mathrm{~TeV}$ & $\sqrt{s}=10\mathrm{~TeV}$ & $\sqrt{s}=14\mathrm{~TeV}$ & $\sqrt{s}=3\mathrm{0~TeV}$\tabularnewline
\hline 
\hline 
$\mathcal{O}_{\tilde{W}W}^{\prime}$ & $205.6$ & $5.54$ & $2.02$ & $0.20$5\tabularnewline
\hline 
$\mathcal{O}_{\tilde{B}B}^{\prime}$ & $25.7$ & $0.69$ & $0.25$ & $0.026$\tabularnewline
\hline 
$\mathcal{O}_{\tilde{B}W}$ & $80.6$ & $2.17$ & $0.79$ & $0.08$1\tabularnewline
\hline 
$\mathcal{O}_{\tilde{B}W}^{\prime}$ & $56.3$ & $1.52$ & $0.55$ & $0.056$\tabularnewline
\hline 
$\mathcal{O}_{BW}$ & $86.6$ & $2.33$ & $0.85$ & $0.086$\tabularnewline
\hline 
$\mathcal{O}_{WW}$ & $205.6$ & $5.54$ & $2.02$ & $0.205$\tabularnewline
\hline 
$\mathcal{O}_{BB}$ & $25.7$ & $0.69$ & $0.25$ & $0.026$\tabularnewline
\hline 
$\mathcal{O}_{G+}$ & $1.31$ & $0.011$ & $0.0027$ & $0.00013$\tabularnewline
\hline 
$\mathcal{O}_{G-}$ & $93.7$ & $2.52$ & $0.92$ & $0.093$\tabularnewline
\hline 
$\widetilde{\mathcal{O}}_{G+}$ & $2.63$ & $0.021$ & $0.0055$ & $0.00026$\tabularnewline
\hline 
$\widetilde{\mathcal{O}}_{G-}$ & $187.4$ & $5.05$ & $1.84$ & $0.18$\tabularnewline
\hline 
\end{tabular}

\caption{\protect\label{tab:Unitarity-bounds}The upper bounds on $C_{i}/\Lambda^{4}~(\mathrm{TeV}^{-4})$
of each operators from unitarity constraints at different c.m.energies.
Two of the pure gauge operators $\mathcal{O}_{G+}$ and $\widetilde{\mathcal{O}}_{G+}$
have the strongest constraints.}
\end{table}

We expand the helicity amplitudes of the nTGCs and obtain the coefficients
$T_{J}$ corresponding to different angular momenta. The
explicit expressions of Wigner function $d_{m_{1}m_{2}}^{J}(\phi,\theta)$
and partial wave expansion coefficients $T_{J}$ of the helicity amplitudes
are given in Appendix~\ref{sec:Wigner-D-functions}. From Eq.~(\ref{eq:2-52})
to Eq.~(\ref{eq:2-62}) it can be seen that for CPC operators $T_{J}$
are real, and for CPV operators $T_{J}$ are imaginary. By imposing
the condition $\left|T_{J}\right|\leq1$ and choosing the strongest
one among different angular momenta configurations, we obtain the
unitarity bounds on the $C_{i}/\Lambda^{4}$ for each nTGCs operators
as follows:
\begin{align*}
\mathcal{O}_{\tilde{W}W}^{\prime}:\quad & \left|\frac{C_{2}}{\Lambda^{4}}\right|\leq\frac{192\sqrt{2}\pi c_{W}M_{Z}s_{W}}{e^{2}\sqrt{s}v^{2}\left(s-M_{Z}^{2}\right)} & \mathcal{O}_{\tilde{B}B}^{\prime}:\quad & \left|\frac{C_{4}}{\Lambda^{4}}\right|\leq\frac{24\sqrt{2}\pi c_{W}M_{Z}s_{W}}{e^{2}\sqrt{s}v^{2}\left(s-M_{Z}^{2}\right)}
\end{align*}
\begin{align*}
\mathcal{O}_{\tilde{B}W}:\quad & \left|\frac{C_{5}}{\Lambda^{4}}\right|\leq\frac{96\sqrt{2}\pi c_{W}^{2}M_{Z}s_{W}^{2}}{e^{2}\sqrt{s}v^{2}\left(c_{W}^{2}-s_{W}^{2}\right)\left(s-M_{Z}^{2}\right)} & \mathcal{O}_{\tilde{B}W}^{\prime}:\quad & \left|\frac{C_{6}}{\Lambda^{4}}\right|\leq\frac{96\sqrt{2}\pi M_{Z}s_{W}^{2}}{e^{2}\sqrt{s}v^{2}\left(s-M_{Z}^{2}\right)}
\end{align*}
\begin{align*}
\mathcal{O}_{BW}:\quad & \left|\frac{C_{8}}{\Lambda^{4}}\right|\leq\frac{192\sqrt{2}\pi c_{W}^{2}M_{Z}s_{W}^{2}}{e^{2}\sqrt{s}v^{2}\left(s-M_{Z}^{2}\right)}
\end{align*}
\begin{align*}
\mathcal{O}_{WW}:\quad & \left|\frac{C_{9}}{\Lambda^{4}}\right|\leq\frac{192\sqrt{2}\pi c_{W}M_{Z}s_{W}}{e^{2}\sqrt{s}v^{2}\left(s-M_{Z}^{2}\right)} & \mathcal{O}_{BB}:\quad & \left|\frac{C_{10}}{\Lambda^{4}}\right|\leq\frac{24\sqrt{2}\pi c_{W}M_{Z}s_{W}}{e^{2}\sqrt{s}v^{2}\left(s-M_{Z}^{2}\right)}
\end{align*}
\begin{align*}
\mathcal{O}_{G+}:\quad & \left|\frac{C_{11}}{\Lambda^{4}}\right|\leq\frac{24\sqrt{2}\pi}{s\left(s-M_{Z}^{2}\right)} & \mathcal{O}_{G-}:\quad & \left|\frac{C_{12}}{\Lambda^{4}}\right|\leq\frac{12\sqrt{2}\pi}{M_{Z}^{2}s_{W}^{2}\left(M_{Z}^{2}-s\right)}
\end{align*}
\begin{align*}
\widetilde{\mathcal{O}}_{G+}:\quad & \left|\frac{C_{13}}{\Lambda^{4}}\right|\leq\frac{48\sqrt{2}\pi}{s\left(s-M_{Z}^{2}\right)} & \widetilde{\mathcal{O}}_{G-}:\quad & \left|\frac{C_{14}}{\Lambda^{4}}\right|\leq\frac{24\sqrt{2}\pi}{\sqrt{s}M_{Z}s_{W}^{2}\left(s-M_{Z}^{2}\right)}
\end{align*}

Experiments can constrain the coefficients of nTGCs. However, nTGCs
are only effective, and the experimental constraints are only meaningful,
when the experimental bounds on the coefficients are tighter than
the unitarity bounds. In Table~\ref{tab:Unitarity-bounds} we list
the numerical results of the unitarity bounds on the $C_{i}/\Lambda^{4}$
for each nTGCs operators, at c.m. energies $\sqrt{s}=3,10,14,30\ \mathrm{TeV}$.
The operator coefficients $C_{i}/\Lambda^{4}$ used in the Monte Carlo
(MC) simulation will be compared with the unitarity bounds obtained
here. Also the projected sensitivities on the nTGCs coefficients will
be compared with the unitarity bound to ensures the validity of the
SMEFT approach in this work.

\begin{table}[tp]
\centering
\begin{tabular}{c|c|c|c|c}
\hline 
 & $\sqrt{s}=3\mathrm{~TeV}$ & $\sqrt{s}=10\mathrm{~TeV}$ & $\sqrt{s}=14\mathrm{~TeV}$ & $\sqrt{s}=3\mathrm{0~TeV}$\tabularnewline
\hline 
\hline 
$\mathcal{O}_{\tilde{W}W}^{\prime}$ & $1.2$ & $0.61$ & $0.031$ & $0.007$\tabularnewline
\hline 
$\mathcal{O}_{\tilde{B}B}^{\prime}$ & $0.14$ & $0.007$ & $0.0035$ & $0.0007$\tabularnewline
\hline 
$\mathcal{O}_{\tilde{B}W}$ & $0.35$ & $0.019$ & $0.0093$ & $0.0021$\tabularnewline
\hline 
$\mathcal{O}_{\tilde{B}W}^{\prime}$ & $0.33$ & $0.017$ & $0.0086$ & $0.0019$\tabularnewline
\hline 
$\mathcal{O}_{BW}$ & $0.45$ & $0.023$ & $0.012$ & $0.0026$\tabularnewline
\hline 
$\mathcal{O}_{WW}$ & $1.18$ & $0.061$ & $0.031$ & $0.0068$\tabularnewline
\hline 
$\mathcal{O}_{BB}$ & $0.13$ & $0.0068$ & $0.0035$ & $0.00076$\tabularnewline
\hline 
$\mathcal{O}_{G+}$ & $0.0073$ & $0.00012$ & $0.00004$ & $4.2\times10^{-6}$\tabularnewline
\hline 
$\mathcal{O}_{G-}$ & $0.2$ & $0.0034$ & $0.0013$ & $0.00013$\tabularnewline
\hline 
$\widetilde{\mathcal{O}}_{G+}$ & $0.015$ & $0.00022$ & $0.00008$ & $8.2\times10^{-6}$\tabularnewline
\hline 
$\widetilde{\mathcal{O}}_{G-}$ & $0.38$ & $0.0068$ & $0.0025$ & $0.00026$\tabularnewline
\hline 
\end{tabular}

\caption{\protect\label{tab:The-ranges-of-no-pol}The values of coefficients
$C_{i}/\Lambda^{4}~(\mathrm{TeV}^{-4})$ used for each operator in
the MC simulation for the unpolarized case.}
\end{table}

\section{\label{sec:Numerical}Numerical Results}

To effectively probe nTGCs at muon colliders, it is essential to generate
large numbers of signal events. Moreover, the signal must be sufficiently
distinguishable from the non-interfering SM backgrounds to enable
precise measurements of the anomalous couplings. With this objective,
we carry out detailed MC simulation and collider analysis aimed at
identifying optimal conditions for the most accurate determination
of the nTGCs parameters. We first consider the unpolarized case for
the 11 operators. Then, we consider the polarized case for the 6 operators
which have two polarization channels. 

In the analytical calculation, we separately obtained
the cross sections for both the VBF and annihilation channels, with
the annihilation channel being larger by $2-3$ orders of magnitude.
Their comparison is shown in Figure~\ref{fig:cs-comparison} and
Figure~\ref{fig:cs-comparison-1}. In the Monte Carlo simulation,
we only specify the final states, so both VBF and annihilation contributions
are implicitly included. Therefore, the expected constraints obtained
also include the contribution from the VBF, although this contribution
is very small compared with annihilation and can be safely neglected.

\subsection{The unpolarized case}

The MC simulations are carried out using the \verb"MadGraph5_aMC@NLO"~\cite{Alwall:2014hca,Christensen:2008py}
framework. A fast detector simulation is subsequently performed with
\verb"Delphes"~\cite{deFavereau:2013fsa}, utilizing the muon collider
card. Signal and background analyses are conducted using \verb"MLAnalysis"~\cite{Guo:2023nfu}
package. In the numerical simulation of $\mu^{+}\mu^{-}\to\gamma\nu\bar{\nu}$,
the basic cuts are set as same as the default settings of \verb"MadGraph5_aMC@NLO",
by demanding $p_{T,\gamma}>10\ \mathrm{GeV}$ and $|\eta_{\gamma}|<2.5$,
where $p_{T,\gamma}$ is the transverse momentum of the photon, and
$\eta_{\gamma}$ is the pseudo-rapidity of the photon. 

To investigate the kinematic characteristics of both signal and background
processes, signal events are generated by activating one operator
at a time in the MC simulation, with the operator coefficients set
to be the values listed in Table~\ref{tab:The-ranges-of-no-pol}.
These values are obtained by a simple signal significance
estimation $\mathcal{S}_{stat}=N_{\mathrm{NP}}/\sqrt{N_{\mathrm{NP}}+N_{\mathrm{SM}}}=3$,
where $N_{\mathrm{SM}}$ is the number of events of the SM backgrounds,
and $N_{\mathrm{NP}}$ is defined $N_{\mathrm{NP}}=\mathcal{L\times}(\sigma_{\mathrm{SM+NP}}-\sigma_{\mathrm{SM}})$, which represents a deviation from the SM background with the interference between the SM and NP contributions considered.
$\mathcal{L}$ is the integrated luminosity and we use the values
of the ``conservative'' case~\cite{AlAli:2021let}. It can be seen
that the coefficient values used in the MC simulation (Table~\ref{tab:The-ranges-of-no-pol})
are at least one order of magnitude smaller than the unitarity bounds
(Table~\ref{tab:Unitarity-bounds}).

\begin{figure}[tp]
\centering
\includegraphics[scale=0.8]{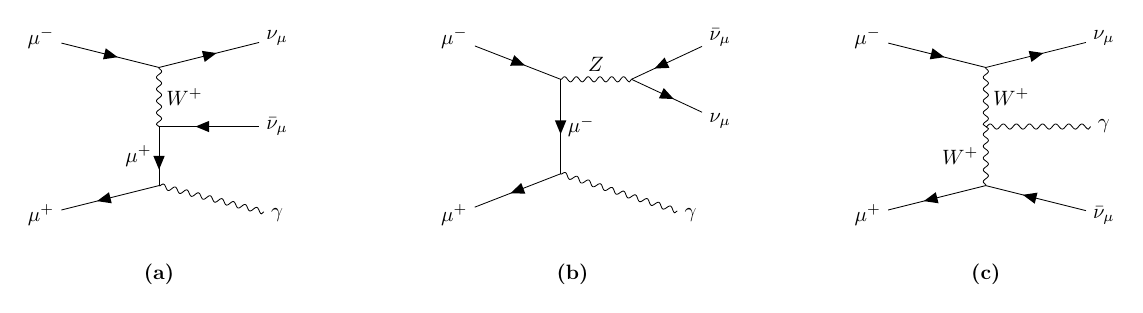}

\caption{\protect\label{fig:Typical-Feynman-diagrams}Typical Feynman diagrams
of the SM backgrounds for the process $\mu^{+}\mu^{-}\to\gamma\nu\bar{\nu}$.}
\end{figure}

The signal of this work is $\mu^{+}\mu^{-}\to\gamma\nu\bar{\nu}$
contributed from nTGCs operators as shown in Figure~\ref{fig:Feynman-diagrams-np}.
The Feynman diagrams of SM backgrounds of $\mu^{+}\mu^{-}\to\gamma\nu\bar{\nu}$
are shown in Figure~\ref{fig:Typical-Feynman-diagrams}. 
In addition to SM background, the dimension-6 operator induced vertices
$\mu\mu\gamma$, $Z\mu\mu$, $Z\nu\nu$ and $WW\gamma$ also contribute
to the background. However, our focus in this work is on the projected
sensitivity to individual operators and we adopt the ``one operator
at a time'' approach. So the background contributions from dimension-6
operators are not included in the simulation. On the other hand, for the contribution from the
dimension-6 vertices $\mu\mu\gamma$, $Z\mu\mu$, and $Z\nu\nu$,
the typical transverse momentum ($p_{T}$) is relatively small. In
our analysis, we applied a $p_{T}$ cut, suppressing events from those
vertices. For the contribution arising from the dimension-6 $WW\gamma$
vertex, the cut on the recoil mass $M_{\text{recoil}}$ will suppress
the $WW\gamma$ contribution. As a result, the effects from those
dimension-6 operators are neglected in our work. 

In principle, reducible backgrounds involving soft/collinear muons/photons, such as $\mu^{+}\mu^{-}\gamma$, $\gamma\gamma$,
and $\gamma\gamma\gamma$ should also be included. 
In our analysis we require exactly one photon in the final state.
Therefore, $p_{T,\gamma}$ is exactly $p_{T,missing}$, and a large $p_{T,\gamma}$ cut will suppress such backgrounds, because the soft/collinear muons/photons will either be observed or lead to a small $p_{T,missing}$.
Besides, there are contributions from the channels whose final states contain both (anti-)neutrinos and soft/collinear muons/photons, which are expected at higher orders of $\alpha _{EM}$.
Therefore, they are not considered in this work.

We require the photon number in the final states to be $N_{\gamma}=1$.
This requirements is denoted as $N_{\gamma}$ cut. Signal and background
analyses are conducted after $N_{\gamma}$ cut. For both the unpolarized
and polarized cases, we analyze the kinematic feature of the signal
and background, and subsequently determine the event selection strategy.

\begin{figure}[tp]
\centering
\includegraphics[scale=0.5]{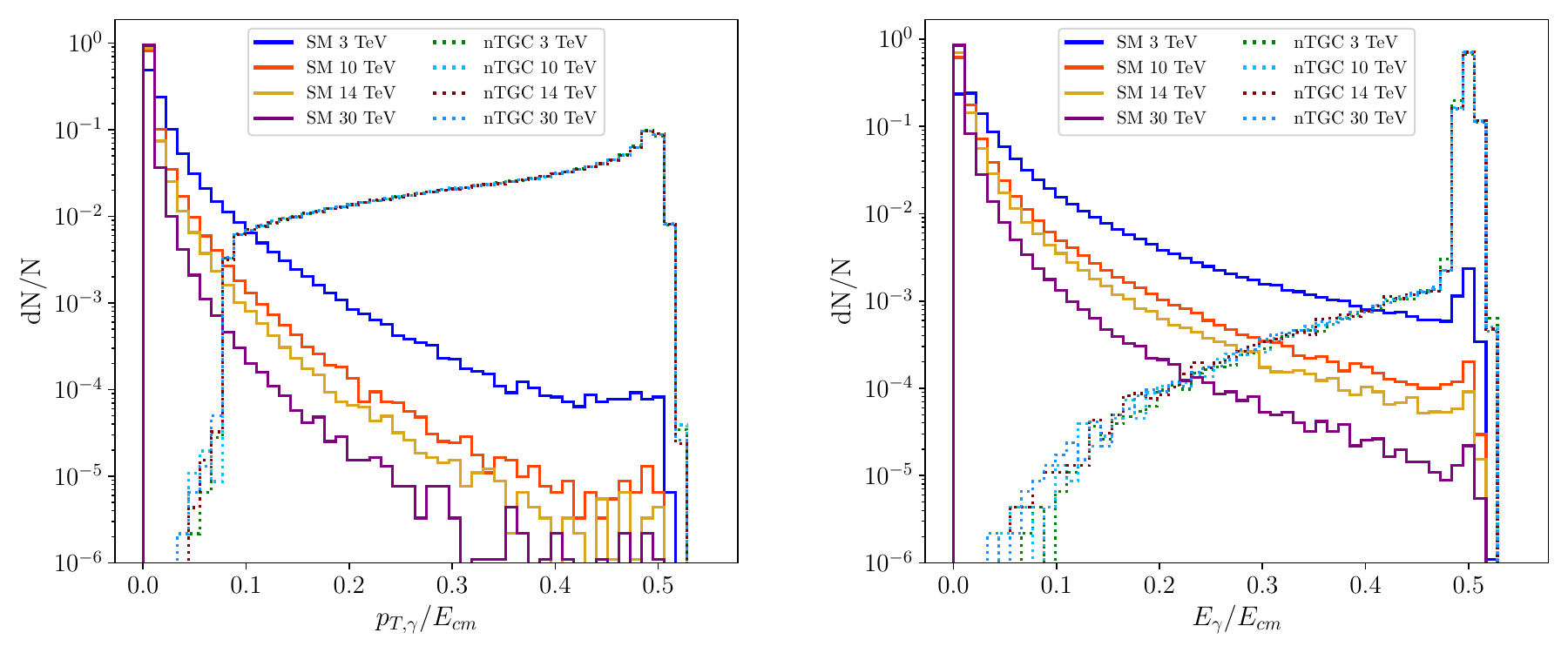}

\caption{\protect\label{fig:gamma-distributions-no-pol}The normalized distributions
of $p_{T,\gamma}/E_{cm}$ and $E_{\gamma}/E_{cm}$ for the signal
and the background of operator $\mathcal{O}_{\tilde{B}W}$, for the
unpolarized case. The values of coefficients $C_{5}/\Lambda^{4}~(\mathrm{TeV}^{-4})$
used for $\mathcal{O}_{\tilde{B}W}$ in the MC simulation are: 0.35
at 3~TeV, 0.019 at 10~TeV, 0.0093 at 14~TeV, and 0.0021 at 30~TeV.
The distributions of other nTGCs operators are similar.}
\end{figure}

From the SM Feynman diagrams in Figure~\ref{fig:Typical-Feynman-diagrams},
we can see that the background photons are mainly emitted from $\mu^{\pm}$
beams or from a $t$-channel $W$ boson, and at high energies they
will be mostly collinear with the incoming $\mu^{\pm}$ beams. It
is expected that the distribution of SM background photons is dominantly
around small $t$ region and the transverse momenta of the photons
are small. Meanwhile, signal photons come from $s$-channel annihilation
diagram or from $W^{+}W^{-}$ fusion, and their distribution will
be around large transverse momenta region. Denoting $p_{T,\gamma}$
as the transverse momentum of the photon and $E_{\mathrm{cm}}$
as the energy, the normalized distributions of $p_{T,\gamma}/E_{\mathrm{cm}}$
for the signal of $\mathcal{O}_{\tilde{B}W}$ and the SM backgrounds
are displayed in the left panel of Figure~\ref{fig:gamma-distributions-no-pol}.
We normalize the photon observable $p_{T,\gamma}$
to $E_{\mathrm{cm}}$ for plotting convenience, so that curves corresponding
to different energy configurations approximately cluster together.
This avoids the need to produce separate plots for each energy point.
The distributions for other nTGCs operators are similar. The $p_{T,\gamma}$
distribution shown in the figure is consistent with our expectations,
and we can use $p_{T,\gamma}$ cuts to distinguish the signal from
the background. We implement the event selection strategy by obtaining
the optimal sensitivity of nTGCs signal. The $p_{T,\gamma}$ cuts
are determined to be $p_{T,\gamma}>850~\mathrm{GeV}$ for $\sqrt{s}=3\mathrm{~TeV}$,
$p_{T,\gamma}>3200~\mathrm{GeV}$ for $\sqrt{s}=10\mathrm{~TeV}$,
$p_{T,\gamma}>4000~\mathrm{GeV}$ for $\sqrt{s}=14\mathrm{~TeV}$,
and $p_{T,\gamma}>8200~\mathrm{GeV}$ for $\sqrt{s}=30\mathrm{~TeV}$.

Another observable is the energy of the final state photon. In general,
cross sections from high-dimensional NP operators tend to grow with
c.m. energy, while the SM cross sections generally decrease or saturate
at high energies and unitarity is protected by gauge structure and
the Higgs boson. It is expected that the photon in the final state
will have relatively low energy in the SM, while NP contributions
tend to produce photons with higher energy. Denoting $E_{\gamma}$
as the energy of the photon and $E_{\mathrm{cm}}$ as the
c.m. energy, the normalized distributions of $E_{\gamma}/E_{cm}$
for the signal of $\mathcal{O}_{\tilde{B}W}$ and the SM backgrounds
are displayed in the right panel of Figure~\ref{fig:gamma-distributions-no-pol}.
We can use $E_{\gamma}$ cuts to further distinguish the signal from
the background. By maximizing sensitivity to the nTGCs signals, the
$E_{\gamma}$ cuts are set to be $E_{\gamma}>1460~\mathrm{GeV}$ for
$\sqrt{s}=3\mathrm{~TeV}$, $E_{\gamma}>4920~\mathrm{GeV}$ for $\sqrt{s}=10\mathrm{~TeV}$,
$E_{\gamma}>6800~\mathrm{GeV}$ for $\sqrt{s}=14\mathrm{~TeV}$, and
$E_{\gamma}>13500~\mathrm{GeV}$ for $\sqrt{s}=30\mathrm{~TeV}$.

\begin{figure}[tp]
\centering
\includegraphics[scale=0.5]{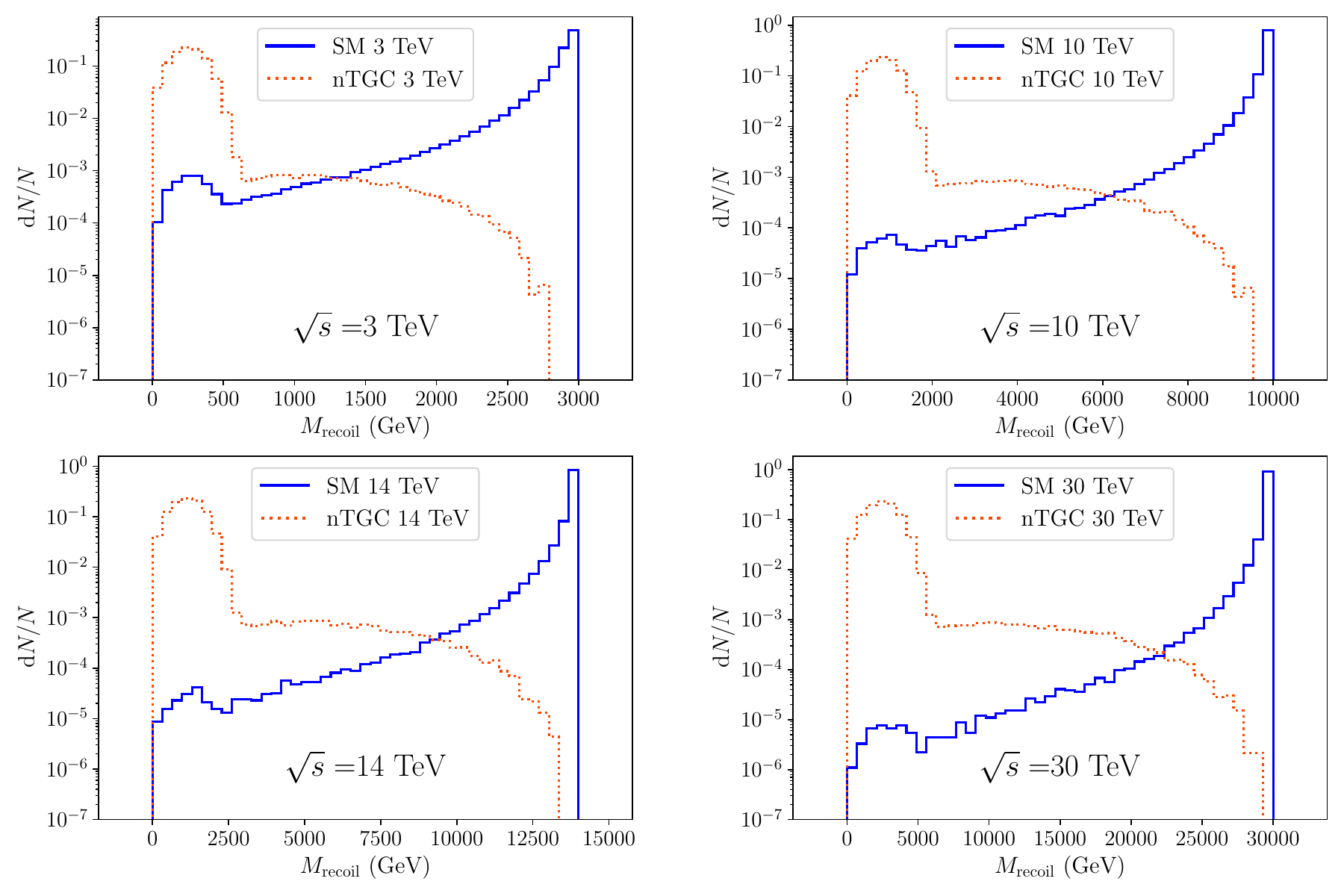}

\caption{\protect\label{fig:miss-distribution-no-pol}The $M_{\mathrm{recoil}}$
distribution for the signal and the background of operator $\mathcal{O}_{\tilde{B}W}$,
for the unpolarized case. The values of coefficients
$C_{5}/\Lambda^{4}~(\mathrm{TeV}^{-4})$ used for $\mathcal{O}_{\tilde{B}W}$
in the MC simulation are: 0.35 at 3~TeV, 0.019 at 10~TeV, 0.0093
at 14~TeV, and 0.0021 at 30~TeV. The distributions of other nTGCs
operators are similar.}
\end{figure}

As indicated by the Feynman diagrams, some neutrinos in the SM background
does not originate from $Z$ boson decays. If the neutrinos are produced
via $Z$ boson decays, their invariant mass is expected to lie near
the $Z$ boson mass. In NP the neutrino pairs predominantly arise
from such decays, resulting in relatively small invariant masses.
In contrast, the invariant mass of neutrinos of the SM background
can be significantly larger. This distinction allows us using the
recoil mass as an effective observable to further separate signal
from background. The initial 4-momentum $p_{i}$ before collision
is:
\begin{equation}
p_{i}=(\sqrt{s},0,0,0).
\end{equation}
We can obtain the recoil 4-momentum $p_{\mathrm{recoil}}$ by subtracting
the sum of all the visible final states 4-momenta:
\begin{equation}
p_{\mathrm{recoil}}=p_{i}-\sum_{\mathrm{visible}}p_{f},
\end{equation}
 and recoil invariant mass $M_{\mathrm{recoil}}$ is given by:
\begin{equation}
M_{\mathrm{recoil}}=\sqrt{p_{\mathrm{recoil}}^{2}}=\sqrt{E_{\mathrm{recoil}}^{2}-\mathbf{p}_{\mathrm{recoil}}^{2}}.
\end{equation}
The $M_{\mathrm{recoil}}$ distribution of the SM background and nTGCs
signal is displayed in Figure~\ref{fig:miss-distribution-no-pol}.
The calculation of $M_{\mathrm{recoil}}$ involves subtracting two
large values to obtain a small value, $E_{\mathrm{recoil}}^{2}-\mathbf{p}_{\mathrm{recoil}}^{2}$,
where the errors in $E_{\mathrm{recoil}}^{2}$ and $\mathbf{p}_{\mathrm{recoil}}^{2}$
from detector simulations are almost as large as $M_{Z}^{2}$. The
resulting peak near $M_{Z}$ is relatively broad and not sufficiently
sharp. Therefore, we have chosen to use the upper limit of $M_{\mathrm{recoil}}$
much larger than $M_{Z}$ for event selection. A cut close to $M_{Z}$
will cut off too many signal events. The $M_{\mathrm{recoil}}$ cuts
are determined to be $M_{\mathrm{recoil}}<500~\mathrm{GeV}$ for $\sqrt{s}=3\mathrm{~TeV}$,
$M_{\mathrm{recoil}}<1600~\mathrm{GeV}$ for $\sqrt{s}=10\mathrm{~TeV}$,
$M_{\mathrm{recoil}}<2000~\mathrm{GeV}$ for $\sqrt{s}=14\mathrm{~TeV}$,
and $M_{\mathrm{recoil}}<5500~\mathrm{GeV}$ for $\sqrt{s}=30\mathrm{~TeV}$.
The proposed event selection strategy and the cross-sections after
cuts are summarized in Table~\ref{tab:The-event-selection}. The
SM backgrounds can be effectively reduced by our selection strategy.

\begin{table}[tp]
\centering
\begin{tabular}{c|c|c|c|c|c|c|c|c}
\hline 
 & \multicolumn{2}{c|}{3 TeV} & \multicolumn{2}{c|}{10 TeV} & \multicolumn{2}{c|}{14 TeV} & \multicolumn{2}{c}{30 TeV}\tabularnewline
\hline 
 & SM & NP & SM & NP & SM & NP & SM & NP\tabularnewline
\hline 
Basic Cuts & 3.038 & 0.00525 & 3.299 & 0.00191 & 3.328 & 0.00176 & 3.336 & 0.00189\tabularnewline
\hline 
$N_{\gamma}$ cuts & 2.770 & 0.00483 & 3.007 & 0.00175 & 3.034 & 0.00161 & 3.039 & 0.00174\tabularnewline
\hline 
$p_{T,\gamma}$ cuts & 1.093 & 0.00462 & 0.593 & 0.00168 & 0.507 & 0.00154 & 0.469 & 0.00166\tabularnewline
\hline 
$E_{\gamma}$ cuts & 0.796 & 0.00441 & 0.202 & 0.00160 & 0.197 & 0.00151 & 0.0892 & 0.00145\tabularnewline
\hline 
$M_{\mathrm{recoil}}$ cuts & 0.102 & 0.00335 & 0.0764 & 0.00106 & 0.0615 & 0.00103 & 0.0339 & 0.000937\tabularnewline
\hline 
Efficiency $\epsilon$ & 3.8\% & 63.8\% & 2.3\% & 55.4\% & 1.8\% & 58.5\% & 1.1\% & 49.5\%\tabularnewline
\hline 
\end{tabular}

\caption{\protect\label{tab:The-event-selection}The event selection strategy
and cross-sections (pb) after cuts. The results of NP are obtained
using the coefficients in Table~\ref{tab:The-ranges-of-no-pol}.}

\end{table}

\begin{figure}[tp]
\centering
\includegraphics[scale=0.5]{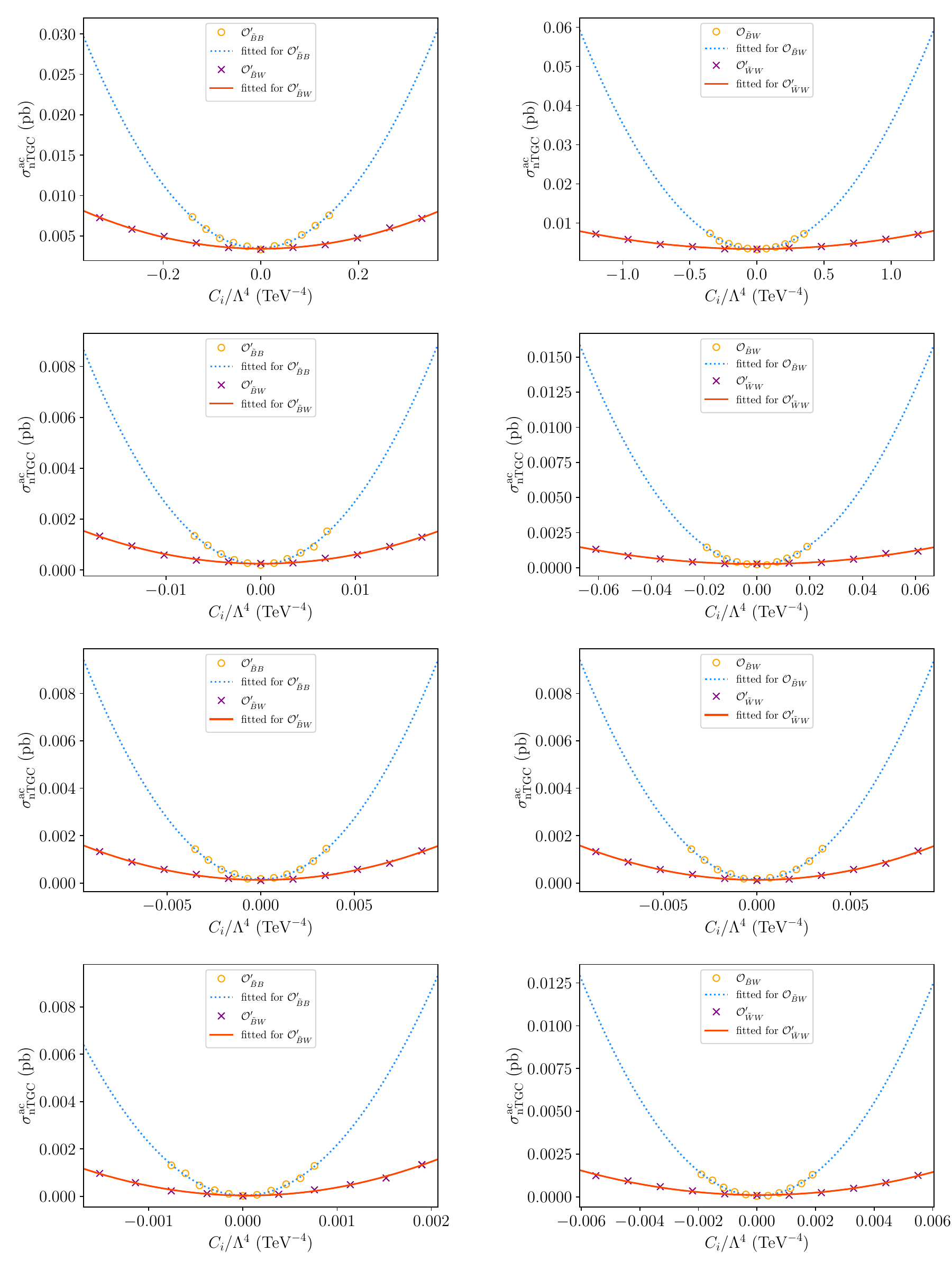}

\caption{\protect\label{fig:parabola}The calculated and fitted $\sigma_{\mathrm{nTGC}}^{\mathrm{ac}}$
as a function of $C_{i}/\Lambda^{4}$ at $\sqrt{s}=3\ \mathrm{TeV}$
(row 1), $\sqrt{s}=10\ \mathrm{TeV}$ (row 2), $\sqrt{s}=14\ \mathrm{TeV}$
(row 3) and $\sqrt{s}=30\ \mathrm{TeV}$ (row 4).}
\end{figure}

The total cross sections are obtained by scanning the values of coefficient
of the operators as shown in Table \ref{tab:The-ranges-of-no-pol}.
The scanned ranges of coefficients are one order of magnitude smaller
than the coefficient ranges allowed by the unitarity bounds, indicating
that our study does not violate the requirement of unitarity. Then
we implement event selection strategy to get the total cross sections
after cut (denoted by $\sigma_{\mathrm{nTGC}}^{\mathrm{ac}}$). Taking
into account the impact of interference terms, the cross sections
$\sigma_{\mathrm{nTGC}}^{\mathrm{ac}}$ can be fitted as bilinear
functions of $C_{i}$:
\begin{equation}
\sigma_{\mathrm{nTGC}}^{\mathrm{ac}}=\sigma_{\mathrm{SM}}^{\mathrm{ac}}+\frac{C_{i}}{\Lambda^{4}}\sigma_{\mathrm{int}}^{\mathrm{ac}}+\frac{C_{i}^{2}}{\Lambda^{8}}\sigma_{\mathrm{NP}}^{\mathrm{ac}},\label{eq:Total-cs}
\end{equation}
where $\sigma_{\mathrm{SM}}^{\mathrm{ac}}$, $\sigma_{\mathrm{int}}^{\mathrm{ac}}$
and $\sigma_{\mathrm{NP}}^{\mathrm{ac}}$ are the fitted SM, interference
and NP contributions respectively. The fitted curves are shown in
Figure~\ref{fig:parabola}, and the fitted results of $\sigma_{\mathrm{int}}^{\mathrm{ac}}$
and $\sigma_{\mathrm{NP}}^{\mathrm{ac}}$ are given in Table~\ref{tab:fitted-values-nopol}.

\begin{table}[tp]
\centering
\begin{tabular}{c|c|c|c|c|c}
\hline 
 &  & $\sqrt{s}=3\mathrm{~TeV}$ & $\sqrt{s}=10\mathrm{~TeV}$ & $\sqrt{s}=14\mathrm{~TeV}$ & $\sqrt{s}=30\mathrm{~TeV}$\tabularnewline
\hline 
\hline 
\multirow{2}{*}{$\mathcal{O}_{\tilde{W}W}^{\prime}$} & $\sigma_{\mathrm{NP}}^{\mathrm{ac}}$ & $0.0026$ & $0.266$ & $1.267$ & $27.24$\tabularnewline
 & $\sigma_{\mathrm{int}}^{\mathrm{ac}}$ & $-0.000044$ & $-0.000086$ & $-0.000049$ & $-0.000029$\tabularnewline
\hline 
\multirow{2}{*}{$\mathcal{O}_{\tilde{B}B}^{\prime}$} & $\sigma_{\mathrm{NP}}^{\mathrm{ac}}$ & $0.202$ & $24.4$ & $103.4$ & $2191$\tabularnewline
 & $\sigma_{\mathrm{int}}^{\mathrm{ac}}$ & $-0.0013$ & $-0.0058$ & $-0.0057$ & $-0.051$\tabularnewline
\hline 
\multirow{2}{*}{$\mathcal{O}_{\tilde{B}W}$} & $\sigma_{\mathrm{NP}}^{\mathrm{ac}}$ & $0.032$ & $3.48$ & $14.6$ & $305.5$\tabularnewline
 & $\sigma_{\mathrm{int}}^{\mathrm{ac}}$ & $-0.00011$ & $-0.00027$ & $-0.0028$ & $-0.013$\tabularnewline
\hline 
\multirow{2}{*}{$\mathcal{O}_{\tilde{B}W}^{\prime}$} & $\sigma_{\mathrm{NP}}^{\mathrm{ac}}$ & $0.035$ & $3.65$ & $16.1$ & $358.3$\tabularnewline
 & $\sigma_{\mathrm{int}}^{\mathrm{ac}}$ & $0.00012$ & $0.00042$ & $0.00094$ & $0.021$\tabularnewline
\hline 
\multirow{2}{*}{$\mathcal{O}_{G+}$} & $\sigma_{\mathrm{NP}}^{\mathrm{ac}}$ & $84.5$ & $115407$ & $873696$ & $8.6609\times10^{7}$\tabularnewline
 & $\sigma_{\mathrm{int}}^{\mathrm{ac}}$ & $-0.0078$ & $-0.103$ & $-0.139$ & $-4.34$\tabularnewline
\hline 
\multirow{2}{*}{$\mathcal{O}_{G-}$} & $\sigma_{\mathrm{NP}}^{\mathrm{ac}}$ & $0.036$ & $7.05$ & $47.8$ & $8823$\tabularnewline
 & $\sigma_{\mathrm{int}}^{\mathrm{ac}}$ & $-0.00025$ & $-0.0023$ & $-0.0048$ & $-0.098$\tabularnewline
\hline 
\end{tabular}

\caption{\protect\label{tab:fitted-values-nopol}The fitted values of $\sigma_{\mathrm{NP}}^{\mathrm{ac}}$
($\mathrm{pb}\times\mathrm{TeV^{8}}$) and $\sigma_{\mathrm{int}}^{\mathrm{ac}}$
($\mathrm{pb}\times\mathrm{TeV^{4}}$ ) for each energy point, where
$\sigma_{\mathrm{NP}}^{\mathrm{ac}}$ and $\sigma_{\mathrm{int}}^{\mathrm{ac}}$
are defined in Eq.~(\ref{eq:Total-cs}).}
\end{table}

The sensitivity of the future muon colliders to the nTGCs operators
and the expected constraints on the coefficients can be obtained with
respect to the significance defined as~\cite{ParticleDataGroup:2024cfk,Cowan:2010js}
\begin{equation}
\mathcal{S}_{stat}=\sqrt{2\left[(N_{\mathrm{bg}}+N_{s})\ln(1+N_{s}/N_{\mathrm{bg}})-N_{s}\right]},\label{eq:SS-difinition}
\end{equation}
where $N_{s}=N_{\mathrm{nTGC}}-N_{\mathrm{SM}}$ and $N_{\mathrm{bg}}=N_{\mathrm{SM}}$.
Here $N_{\mathrm{SM}}$ is the number of events of the SM backgrounds,
and $N_{\mathrm{nTGC}}$ is the number of total events. $N_{\mathrm{SM}}$
are obtained by $N_{\mathrm{SM}}=\mathcal{L}\times\sigma_{\mathrm{SM}}^{\mathrm{ac}}$,
and $\sigma_{\mathrm{SM}}^{\mathrm{ac}}$ are fitted by Eq.~(\ref{eq:Total-cs}).
$N_{\mathrm{nTGC}}$ are obtained by $N_{\mathrm{nTGC}}=\mathcal{L}\times\sigma_{\mathrm{nTGC}}^{\mathrm{ac}}$.
The integrated luminosities $\mathcal{L}$ for both the ``conservative''
and ``optimistic'' cases are considered~\cite{AlAli:2021let}.
By taking $\mathcal{S}_{stat}$ to be 2, 3 and 5, we can obtain the
projected sensitivity on $C_{i}$ with $2\sigma$, $3\sigma$ or $5\sigma$
significance. The numerical results of expected constraints of $C_{i}$
are shown in Table~\ref{tab:sensitivities-nopol} and Table~\ref{tab:sensitivities-nopol-1}.
These expected constraints are significantly tighter than the unitarity
bound. For the muon collider with $\sqrt{s}=3\mathrm{~TeV}$, we can
see that the optimal expected constraints on dimension-8 nTGCs are
better than the LHC experimental constraints listed in Table~\ref{tab:LHC-constraint}.
In particular, the two pure gauge operators $\mathcal{O}_{G+}$ and
$\widetilde{\mathcal{O}}_{G+}$ yield the most sensitive limits on
the dimension-8 operators, improving the expected constraints by two
orders of magnitude. For collider processes, the EFT
remains valid only when $\Lambda\gg\sqrt{s}$. In the coefficient
$C_{i}/\Lambda^{4}$, assuming $C_{i}\sim1$, the condition $\Lambda\gg\sqrt{s}$
does not meet for several sensitivities in Table~\ref{tab:sensitivities-nopol}.
However, the values listed in Table~\ref{tab:sensitivities-nopol} correspond to the projected sensitivities. It indicates that, if $C_{i}\sim1$,
the sensitivities are not high enough for the EFT to be discovered.
From the perspective of partial wave unitarity, no signs of the EFT
being invalid have been observed. Meanwhile, extensive phenomenological
studies on the nTGCs indicate that when $C_{i}=1$, the achievable
$\Lambda$ is generally at the same order of $\sqrt{s}$~\cite{Senol:2022snc,Ellis:2019zex,Jahedi:2023myu}.
Therefore, the $Z\gamma$ production process in muon colliders remains
one of the most competitive channels for observing the nTGCs.

\begin{table}[tp]
\centering
\begin{tabular}{c|c|c|c|c|c}
\hline 
\multirow{3}{*}{} & \multirow{3}{*}{{\footnotesize$\mathcal{S}_{stat}$}} & {\footnotesize$\sqrt{s}=3\mathrm{~TeV}$} & {\footnotesize$\sqrt{s}=10\mathrm{~TeV}$} & {\footnotesize$\sqrt{s}=14\mathrm{~TeV}$} & {\footnotesize$\sqrt{s}=30\mathrm{~TeV}$}\tabularnewline
 &  & {\footnotesize 1 ab$^{-1}$} & {\footnotesize 10 ab$^{-1}$} & {\footnotesize 10 ab$^{-1}$} & {\footnotesize 10 ab$^{-1}$}\tabularnewline
 &  & {\footnotesize (0.1 TeV$^{-4}$)} & {\footnotesize (0.01TeV$^{-4}$)} & {\footnotesize (0.001TeV$^{-4}$)} & {\footnotesize (0.0001TeV$^{-4}$)}\tabularnewline
\hline 
\hline 
\multirow{3}{*}{{\footnotesize$\mathcal{O}_{\tilde{W}W}^{\prime}$}} & {\footnotesize 2} & {\footnotesize$\left[-2.02,2.19\right]$} & {\footnotesize$\left[-0.603,0.636\right]$} & {\footnotesize$\left[-2.23,2.62\right]$} & {\footnotesize$\left[-3.87,3.89\right]$}\tabularnewline
 & {\footnotesize 3} & {\footnotesize$\left[-2.49,2.66\right]$} & {\footnotesize$\left[-0.744,0.776\right]$} & {\footnotesize$\left[-2.78,3.17\right]$} & {\footnotesize$\left[-4.77,4.78\right]$}\tabularnewline
 & {\footnotesize 5} & {\footnotesize$\left[-3.25,3.42\right]$} & {\footnotesize$\left[-0.968,1.00\right]$} & {\footnotesize$\left[-3.66,4.05\right]$} & {\footnotesize$\left[-6.20,6.21\right]$}\tabularnewline
\hline 
\multirow{3}{*}{{\footnotesize$\mathcal{O}_{\tilde{B}B}^{\prime}$}} & {\footnotesize 2} & {\footnotesize$\left[-0.212,0.277\right]$} & {\footnotesize$\left[-0.0511,0.0750\right]$} & {\footnotesize$\left[-0.273,0.279\right]$} & {\footnotesize$\left[-0.316,0.552\right]$}\tabularnewline
 & {\footnotesize 3} & {\footnotesize$\left[-0.266,0.332\right]$} & {\footnotesize$\left[-0.0650,0.0888\right]$} & {\footnotesize$\left[-0.336,0.342\right]$} & {\footnotesize$\left[-0.409,0.646\right]$}\tabularnewline
 & {\footnotesize 5} & {\footnotesize$\left[-0.354,0.419\right]$} & {\footnotesize$\left[-0.0873,0.111\right]$} & {\footnotesize$\left[-0.437,0.442\right]$} & {\footnotesize$\left[-0.562,0.798\right]$}\tabularnewline
\hline 
\multirow{3}{*}{{\footnotesize$\mathcal{O}_{\tilde{B}W}$}} & {\footnotesize 2} & {\footnotesize$\left[-0.581,0.616\right]$} & {\footnotesize$\left[-0.154,0.162\right]$} & {\footnotesize$\left[-0.617,0.807\right]$} & {\footnotesize$\left[-0.942,1.38\right]$}\tabularnewline
 & {\footnotesize 3} & {\footnotesize$\left[-0.716,0.752\right]$} & {\footnotesize$\left[-0.190,0.198\right]$} & {\footnotesize$\left[-0.777,0.966\right]$} & {\footnotesize$\left[-1.20,1.64\right]$}\tabularnewline
 & {\footnotesize 5} & {\footnotesize$\left[-0.932,0.968\right]$} & {\footnotesize$\left[-0.248,0.256\right]$} & {\footnotesize$\left[-1.03,1.22\right]$} & {\footnotesize$\left[-1.62,2.06\right]$}\tabularnewline
\hline 
\multirow{3}{*}{{\footnotesize$\mathcal{O}_{\tilde{B}W}^{\prime}$}} & {\footnotesize 2} & {\footnotesize$\left[-0.594,0.562\right]$} & {\footnotesize$\left[-0.171,0.160\right]$} & {\footnotesize$\left[-0.706,0.648\right]$} & {\footnotesize$\left[-1.43,0.834\right]$}\tabularnewline
 & {\footnotesize 3} & {\footnotesize$\left[-0.725,0.693\right]$} & {\footnotesize$\left[-0.209,0.197\right]$} & {\footnotesize$\left[-0.860,0.802\right]$} & {\footnotesize$\left[-1.67,1.07\right]$}\tabularnewline
 & {\footnotesize 5} & {\footnotesize$\left[-0.933,0.901\right]$} & {\footnotesize$\left[-0.269,0.257\right]$} & {\footnotesize$\left[-1.10,1.04\right]$} & {\footnotesize$\left[-2.07,1.47\right]$}\tabularnewline
\hline 
\multirow{3}{*}{{\footnotesize$\mathcal{O}_{BW}$}} & {\footnotesize 2} & {\footnotesize$0.803$} & {\footnotesize$0.227$} & {\footnotesize$0.937$} & {\footnotesize$1.33$}\tabularnewline
 & {\footnotesize 3} & {\footnotesize$0.986$} & {\footnotesize$0.279$} & {\footnotesize$1.15$} & {\footnotesize$1.64$}\tabularnewline
 & {\footnotesize 5} & {\footnotesize$1.27$} & {\footnotesize$0.361$} & {\footnotesize$1.49$} & {\footnotesize$2.13$}\tabularnewline
\hline 
\multirow{3}{*}{{\footnotesize$\mathcal{O}_{WW}$}} & {\footnotesize 2} & {\footnotesize$2.10$} & {\footnotesize$0.593$} & {\footnotesize$2.44$} & {\footnotesize$3.47$}\tabularnewline
 & {\footnotesize 3} & {\footnotesize$2.57$} & {\footnotesize$0.727$} & {\footnotesize$3.00$} & {\footnotesize$4.27$}\tabularnewline
 & {\footnotesize 5} & {\footnotesize$3.33$} & {\footnotesize$0.942$} & {\footnotesize$3.89$} & {\footnotesize$5.57$}\tabularnewline
\hline 
\multirow{3}{*}{{\footnotesize$\mathcal{O}_{BB}$}} & {\footnotesize 2} & {\footnotesize$0.235$} & {\footnotesize$0.066$} & {\footnotesize$0.273$} & {\footnotesize$0.388$}\tabularnewline
 & {\footnotesize 3} & {\footnotesize$0.288$} & {\footnotesize$0.081$} & {\footnotesize$0.335$} & {\footnotesize$0.478$}\tabularnewline
 & {\footnotesize 5} & {\footnotesize$0.373$} & {\footnotesize$0.105$} & {\footnotesize$0.434$} & {\footnotesize$0.623$}\tabularnewline
\hline 
\multirow{3}{*}{{\footnotesize$\mathcal{O}_{G+}$}} & {\footnotesize 2} & {\footnotesize$\left[-0.0113,0.0122\right]$} & {\footnotesize$\left[-0.00084,0.00093\right]$} & {\footnotesize$\left[-0.00283,0.00290\right]$} & {\footnotesize$\left[-0.00195,0.00245\right]$}\tabularnewline
 & {\footnotesize 3} & {\footnotesize$\left[-0.0139,0.0147\right]$} & {\footnotesize$\left[-0.00104,0.00113\right]$} & {\footnotesize$\left[-0.00349,0.00365\right]$} & {\footnotesize$\left[-0.00245,0.00295\right]$}\tabularnewline
 & {\footnotesize 5} & {\footnotesize$\left[-0.0182,0.0191\right]$} & {\footnotesize$\left[-0.00136,0.00145\right]$} & {\footnotesize$\left[-0.00455,0.00471\right]$} & {\footnotesize$\left[-0.00326,0.00376\right]$}\tabularnewline
\hline 
\multirow{3}{*}{{\footnotesize$\mathcal{O}_{G-}$}} & {\footnotesize 2} & {\footnotesize$\left[-0.535,0.605\right]$} & {\footnotesize$\left[-0.101,0.135\right]$} & {\footnotesize$\left[-0.356,0.459\right]$} & {\footnotesize$\left[-0.154,0.266\right]$}\tabularnewline
 & {\footnotesize 3} & {\footnotesize$\left[-0.663,0.734\right]$} & {\footnotesize$\left[-0.127,0.161\right]$} & {\footnotesize$\left[-0.448,0.550\right]$} & {\footnotesize$\left[-0.199,0.311\right]$}\tabularnewline
 & {\footnotesize 5} & {\footnotesize$\left[-0.869,0.939\right]$} & {\footnotesize$\left[-0.169,0.203\right]$} & {\footnotesize$\left[-0.594,0.697\right]$} & {\footnotesize$\left[-0.273,0.385\right]$}\tabularnewline
\hline 
\multirow{3}{*}{{\footnotesize$\widetilde{\mathcal{O}}_{G+}$}} & {\footnotesize 2} & {\footnotesize$0.0235$} & {\footnotesize$0.00192$} & {\footnotesize$0.00579$} & {\footnotesize$0.00384$}\tabularnewline
 & {\footnotesize 3} & {\footnotesize$0.0289$} & {\footnotesize$0.00235$} & {\footnotesize$0.00710$} & {\footnotesize$0.00473$}\tabularnewline
 & {\footnotesize 5} & {\footnotesize$0.0374$} & {\footnotesize$0.00305$} & {\footnotesize$0.00921$} & {\footnotesize$0.00616$}\tabularnewline
\hline 
\multirow{3}{*}{{\footnotesize$\widetilde{\mathcal{O}}_{G-}$}} & {\footnotesize 2} & {\footnotesize$1.17$} & {\footnotesize$0.223$} & {\footnotesize$0.673$} & {\footnotesize$0.394$}\tabularnewline
 & {\footnotesize 3} & {\footnotesize$1.44$} & {\footnotesize$0.274$} & {\footnotesize$0.826$} & {\footnotesize$0.485$}\tabularnewline
 & {\footnotesize 5} & {\footnotesize$1.86$} & {\footnotesize$0.355$} & {\footnotesize$1.07$} & {\footnotesize$0.632$}\tabularnewline
\hline 
\end{tabular}

\caption{\protect\label{tab:sensitivities-nopol}The projected sensitivities
on the nTGCs coefficients $C_{i}/\Lambda^{4}~(\mathrm{TeV}^{-4})$
at the muon colliders with selected c.m. energies and ``conservative''
integrated luminosities, for the unpolarized case. For CPV operators,
the coefficient ranges are symmetric and we only list their upper
bound. It should be noted that the numerical results in different
columns are scaled by different factors, as indicated in the first
row of the table.}
\end{table}

\begin{table}[tp]
\centering
\begin{tabular}{c|c|c|c}
\hline 
\multirow{3}{*}{} & \multirow{3}{*}{{\footnotesize$\mathcal{S}_{stat}$}} & {\footnotesize$\sqrt{s}=14\mathrm{~TeV}$} & {\footnotesize$\sqrt{s}=30\mathrm{~TeV}$}\tabularnewline
 &  & {\footnotesize 20 ab$^{-1}$} & {\footnotesize 90 ab$^{-1}$}\tabularnewline
 &  & {\footnotesize (0.001TeV$^{-4}$)} & {\footnotesize (0.0001TeV$^{-4}$)}\tabularnewline
\hline 
\hline 
\multirow{3}{*}{{\footnotesize$\mathcal{O}_{\tilde{W}W}^{\prime}$}} & {\footnotesize 2} & {\footnotesize$\left[-1.84,2.24\right]$} & {\footnotesize$\left[-2.22,2.24\right]$}\tabularnewline
 & {\footnotesize 3} & {\footnotesize$\left[-2.30,2.70\right]$} & {\footnotesize$\left[-2.73,2.74\right]$}\tabularnewline
 & {\footnotesize 5} & {\footnotesize$\left[-3.04,3.43\right]$} & {\footnotesize$\left[-3.53,3.55\right]$}\tabularnewline
\hline 
\multirow{3}{*}{{\footnotesize$\mathcal{O}_{\tilde{B}B}^{\prime}$}} & {\footnotesize 2} & {\footnotesize$\left[-0.229,0.235\right]$} & {\footnotesize$\left[-0.249,0.385\right]$}\tabularnewline
 & {\footnotesize 3} & {\footnotesize$\left[-0.282,0.287\right]$} & {\footnotesize$\left[-0.299,0.435\right]$}\tabularnewline
 & {\footnotesize 5} & {\footnotesize$\left[-0.366,0.371\right]$} & {\footnotesize$\left[-0.381,0.517\right]$}\tabularnewline
\hline 
\multirow{3}{*}{{\footnotesize$\mathcal{O}_{\tilde{B}W}$}} & {\footnotesize 2} & {\footnotesize$\left[-0.506,0.695\right]$} & {\footnotesize$\left[-0.571,0.812\right]$}\tabularnewline
 & {\footnotesize 3} & {\footnotesize$\left[-0.639,0.828\right]$} & {\footnotesize$\left[-0.713,0.905\right]$}\tabularnewline
 & {\footnotesize 5} & {\footnotesize$\left[-0.852,1.04\right]$} & {\footnotesize$\left[-0.943,1.184\right]$}\tabularnewline
\hline 
\multirow{3}{*}{{\footnotesize$\mathcal{O}_{\tilde{B}W}^{\prime}$}} & {\footnotesize 2} & {\footnotesize$\left[-0.598,0.539\right]$} & {\footnotesize$\left[-0.894,0.596\right]$}\tabularnewline
 & {\footnotesize 3} & {\footnotesize$\left[-0.727,0.668\right]$} & {\footnotesize$\left[-1..24,0.727\right]$}\tabularnewline
 & {\footnotesize 5} & {\footnotesize$\left[-0.932,0.874\right]$} & {\footnotesize$\left[-1.239,0.841\right]$}\tabularnewline
\hline 
\multirow{3}{*}{{\footnotesize$\mathcal{O}_{BW}$}} & {\footnotesize 2} & {\footnotesize$0.787$} & {\footnotesize$0.764$}\tabularnewline
 & {\footnotesize 3} & {\footnotesize$0.965$} & {\footnotesize$0.937$}\tabularnewline
 & {\footnotesize 5} & {\footnotesize$1.25$} & {\footnotesize$1.21$}\tabularnewline
\hline 
\multirow{3}{*}{{\footnotesize$\mathcal{O}_{WW}$}} & {\footnotesize 2} & {\footnotesize$2.05$} & {\footnotesize$1.99$}\tabularnewline
 & {\footnotesize 3} & {\footnotesize$2.51$} & {\footnotesize$2.44$}\tabularnewline
 & {\footnotesize 5} & {\footnotesize$3.26$} & {\footnotesize$3.16$}\tabularnewline
\hline 
\multirow{3}{*}{{\footnotesize$\mathcal{O}_{BB}$}} & {\footnotesize 2} & {\footnotesize$0.229$} & {\footnotesize$0.222$}\tabularnewline
 & {\footnotesize 3} & {\footnotesize$0.281$} & {\footnotesize$0.273$}\tabularnewline
 & {\footnotesize 5} & {\footnotesize$0.364$} & {\footnotesize$0.354$}\tabularnewline
\hline 
\multirow{3}{*}{{\footnotesize$\mathcal{O}_{G+}$}} & {\footnotesize 2} & {\footnotesize$\left[-0.00236,0.00252\right]$} & {\footnotesize$\left[-0.00103,0.00153\right]$}\tabularnewline
 & {\footnotesize 3} & {\footnotesize$\left[-0.00291,0.00307\right]$} & {\footnotesize$\left[-0.00131,0.00181\right]$}\tabularnewline
 & {\footnotesize 5} & {\footnotesize$\left[-0.00380,0.00396\right]$} & {\footnotesize$\left[-0.00176,0.00226\right]$}\tabularnewline
\hline 
\multirow{3}{*}{{\footnotesize$\mathcal{O}_{G-}$}} & {\footnotesize 2} & {\footnotesize$\left[-0.292,0.394\right]$} & {\footnotesize$\left[-0.073,0.125\right]$}\tabularnewline
 & {\footnotesize 3} & {\footnotesize$\left[-0.368,0.471\right]$} & {\footnotesize$\left[-0.097,0.179\right]$}\tabularnewline
 & {\footnotesize 5} & {\footnotesize$\left[-0.491,0.593\right]$} & {\footnotesize$\left[-0.137,0.219\right]$}\tabularnewline
\hline 
\multirow{3}{*}{{\footnotesize$\widetilde{\mathcal{O}}_{G+}$}} & {\footnotesize 2} & {\footnotesize$0.00486$} & {\footnotesize$0.00220$}\tabularnewline
 & {\footnotesize 3} & {\footnotesize$0.00596$} & {\footnotesize$0.00271$}\tabularnewline
 & {\footnotesize 5} & {\footnotesize$0.00772$} & {\footnotesize$0.00350$}\tabularnewline
\hline 
\multirow{3}{*}{{\footnotesize$\widetilde{\mathcal{O}}_{G-}$}} & {\footnotesize 2} & {\footnotesize$0.565$} & {\footnotesize$0.226$}\tabularnewline
 & {\footnotesize 3} & {\footnotesize$0.693$} & {\footnotesize$0.277$}\tabularnewline
 & {\footnotesize 5} & {\footnotesize$0.898$} & {\footnotesize$0.359$}\tabularnewline
\hline 
\end{tabular}

\caption{\protect\label{tab:sensitivities-nopol-1}The projected sensitivities
on the nTGCs coefficients $C_{i}/\Lambda^{4}~(\mathrm{TeV}^{-4})$
at the muon colliders with selected c.m. energies and ``optimistic''
integrated luminosities, for the unpolarized case. For CPV operators,
the coefficient ranges are symmetric and we only list their upper
bound. It should be noted that the numerical results in different
columns are scaled by different factors, as indicated in the first
row of the table.}
\end{table}

\subsection{The polarized case }

\begin{table}[tp]
\centering
\begin{tabular}{c|c|c}
\hline 
 & $\sqrt{s}=3\mathrm{~TeV}$ & $\sqrt{s}=30\mathrm{~TeV}$\tabularnewline
\hline 
\hline 
$\mathcal{O}_{\tilde{B}B}^{\prime}$ & $0.022$ & $0.00006$\tabularnewline
\hline 
$\mathcal{O}_{\tilde{B}W}$ & $0.081$ & $0.00023$\tabularnewline
\hline 
$\mathcal{O}_{BW}$ & $0.14$ & $0.00047$\tabularnewline
\hline 
$\mathcal{O}_{BB}$ & $0.02$ & $0.00006$\tabularnewline
\hline 
$\mathcal{O}_{G-}$ & $0.056$ & $0.00002$\tabularnewline
\hline 
$\widetilde{\mathcal{O}}_{G-}$ & $0.074$ & $0.00003$\tabularnewline
\hline 
\end{tabular}

\caption{\protect\label{tab:ranges-pol}The ranges of coefficients $C_{i}/\Lambda^{4}~(\mathrm{TeV}^{-4})$
used for each energy point in the MC simulation for the polarized
case $(-+)$ .}
\end{table}

We also study the impact and advantage of beam polarization in this
work. By using the analytical expressions we derived in Section~\ref{sec:formalism-nTGC}
and a simple signal significance estimation $\mathcal{S}_{stat}=N_{\mathrm{NP}}/\sqrt{N_{\mathrm{NP}}+N_{\mathrm{SM}}}=3$,
we find that for the 6 nTGCs operators with both the $(+-)$ and $(-+)$
polarization channels, the $(-+)$ channel provides tighter expected
constraints on the operator coefficients. Hence, we perform MC simulations
for the $(-+)$ initial beam polarization of these 6 operators. Signal
events are generated by activating one operator at a time, with the
operator coefficients set to be the values listed in Table~\ref{tab:ranges-pol}.
In the MC simulation, we adopt a simplified polarization configuration
with the helicity fraction of $\mu^{+}$ ($P_{\mu^{+}}$) and $\mu^{-}$
($P_{\mu^{-}}$) to be $(P_{\mu^{+}},P_{\mu^{-}})=(-100\%,+100\%)$.
This choice is not intended to represent a realistic experimental
setup, but rather to serve as an idealized scenario for testing the
potential impact of muon beam polarization on the sensitivity to nTGCs.
Our goal is to assess whether polarization can significantly improve
the sensitivity and lead to more stringent expected constraints on
the nTGCs coefficients.

\begin{figure}[tp]
\centering
\includegraphics[scale=0.5]{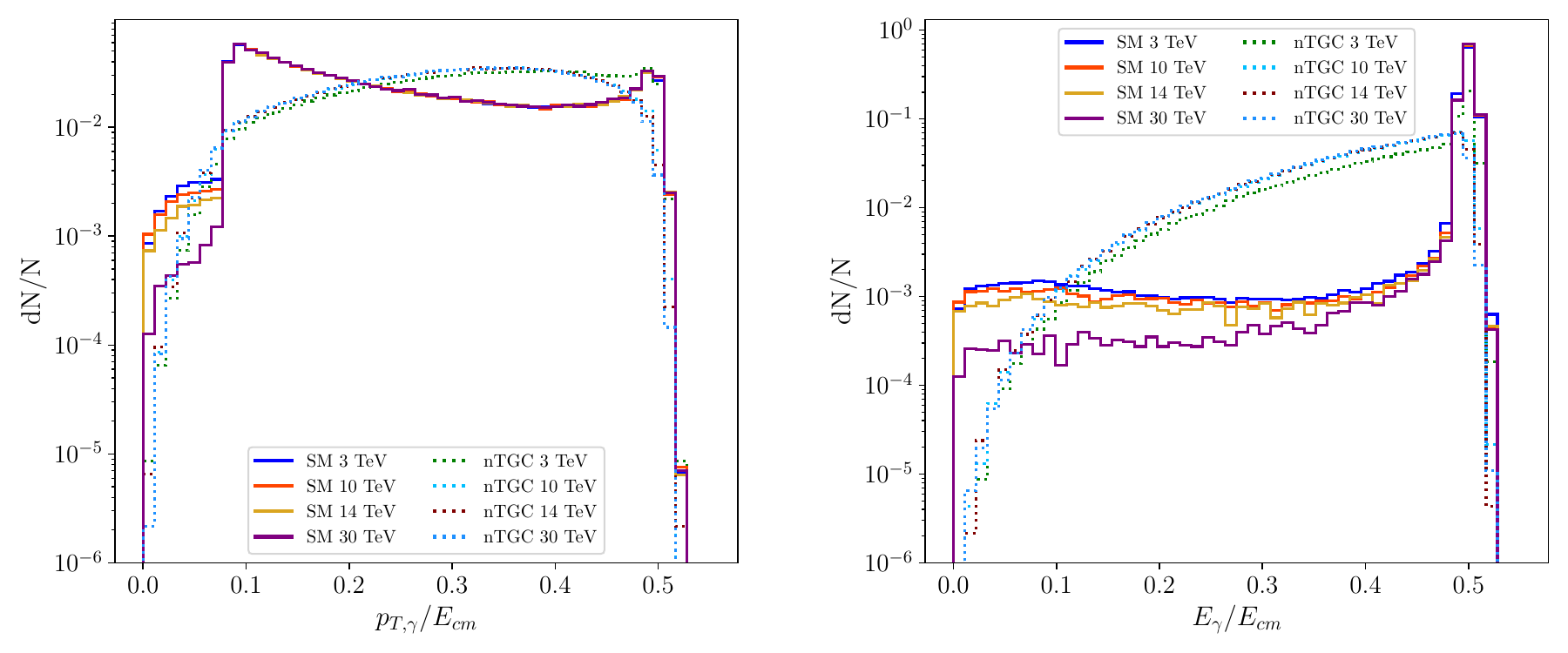}

\caption{\protect\label{fig:gamma-feature-pol}The normalized distributions
of $p_{T,\gamma}/E_{cm}$ and $E_{\gamma}/E_{cm}$ for the signal
and the background of operator $\mathcal{O}_{\tilde{B}W}$, for the
polarized case $(-+)$. The values of coefficients
$C_{5}/\Lambda^{4}~(\mathrm{TeV}^{-4})$ used for $\mathcal{O}_{\tilde{B}W}$
in the MC simulation are: 0.081 at 3~TeV, 0.0036 at 10~TeV, 0.0016
at 14~TeV, and 0.00023 at 30~TeV. The distributions of other nTGCs
operators are similar.}
\end{figure}

\begin{figure}[tp]
\centering
\includegraphics[scale=0.5]{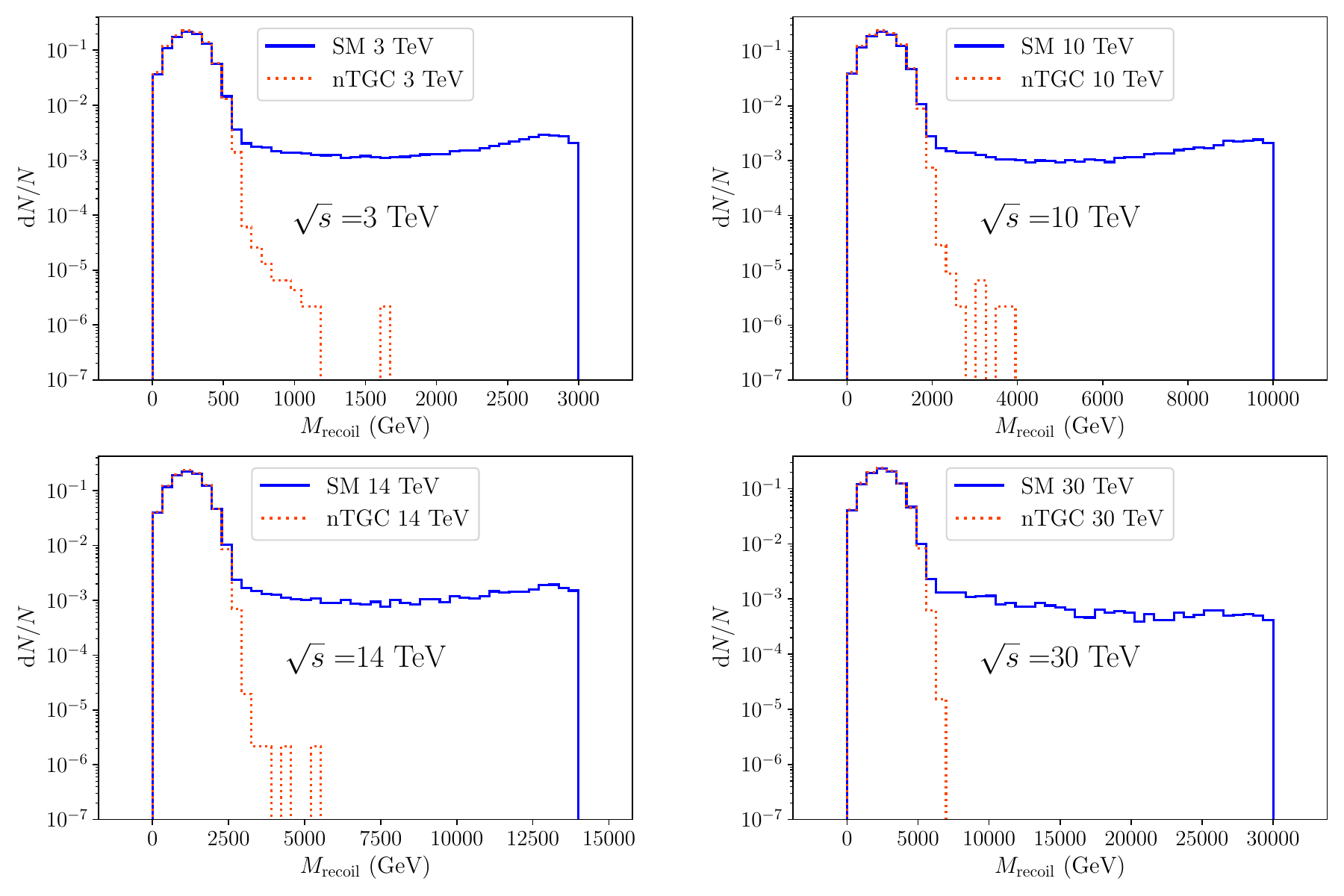}

\caption{\protect\label{fig:missing-feature-pol}The $M_{\mathrm{recoil}}$
distribution for the signal and the background of operator $\mathcal{O}_{\tilde{B}W}$,
for the polarized case $(-+)$. The values of coefficients
$C_{5}/\Lambda^{4}~(\mathrm{TeV}^{-4})$ used for $\mathcal{O}_{\tilde{B}W}$
in the MC simulation are: 0.081 at 3~TeV, 0.0036 at 10~TeV, 0.0016
at 14~TeV, and 0.00023 at 30~TeV. The distributions of other nTGCs
operators are similar.}
\end{figure}

In the polarized case, in order to make a comparison with the unpolarized
scenario, we performed MC simulations at two energy points, $\sqrt{s}=3\mathrm{~TeV}$
and $\sqrt{s}=30\mathrm{~TeV}$. We still use $p_{T,\gamma}$ cut,
$E_{\gamma}$ cut missing invariant mass $M_{\mathrm{recoil}}$ cut
to distinguish the signal from the background. The normalized distributions
of $p_{T,\gamma}/E_{cm}$ for the signal the background are displayed
in Figure~\ref{fig:gamma-feature-pol}(a). The normalized distributions
of $E_{\gamma}/E_{cm}$ for the signal and the background are displayed
in Figure~\ref{fig:gamma-feature-pol}(b). The $M_{\mathrm{recoil}}$
distribution for the signal and the background are displayed in Figure~\ref{fig:missing-feature-pol}.
We can see there are some gaps in the nTGCs distributions in Figure~\ref{fig:missing-feature-pol}.
These gaps can be attributed to statistical effects stemming from
the limited statistics of the simulation. In generating the SM background,
we used $10^{6}$ events. In generating the NP background, we used
$0.5\times10^{6}$ events. Consequently, in certain energy region
of the distributions, the actual number of events is only at the order
of a few. The appearance of a gap is therefore purely a statistical
fluctuation, since the event counts in this region are already at
the single-digit level. From theses figures we can see that the kinematic
features in the polarized case differ from those in the unpolarized
case. We adopt cut values accordingly to obtain the best expected
constraints on the operator coefficients. For $\sqrt{s}=3\mathrm{~TeV}$,
the cut values are $p_{T,\gamma}>900$ GeV, $E_{\gamma}>1450$ GeV
and $M_{\mathrm{recoil}}<500$ GeV. For $\sqrt{s}=3\mathrm{0~TeV}$,
the cut values are $p_{T,\gamma}>9000$ GeV, $E_{\gamma}>14500$ GeV
and $M_{\mathrm{recoil}}<5500$ GeV. The basic cuts are set as same
as the default settings of \verb"MadGraph5_aMC@NLO". 

For the polarized case, we performed a scan over the coefficient ranges
shown in Table~\ref{tab:ranges-pol}, which are two order of magnitude
smaller than the ranges required by unitarity bounds. Then we perform
a parabolic fit to obtain the NP cross sections and interference cross
sections under the polarized scenario. Substituting these into the
significance Eq.~(\ref{eq:SS-difinition}), we obtain the expected
constraints of the nTGCs operator coefficients at the $2\sigma$,
$3\sigma$, and $5\sigma$ levels. The numerical results are shown
in Table~\ref{tab:sensitivities-pol}. The results are consistent
with the analytical estimates, showing that the $(-+)$ polarization
scheme provides stronger expected constraints on the dimension-8 operator
coefficients than the unpolarized case, and these expected constraints
are much tighter than the unitarity bound. Our discussion
on polarization does not assume the existence of a fully realized
muon beamline or a collider plan. Such a setup remains technically
very challenging~\cite{Norum:1996mi}. Precisely because of these
challenges, our motivation is to assess whether implementing polarization
would be necessary or beneficial for future muon colliders to detect
the dimension-8 operators contributing to the nTGCs.

\begin{table}[tp]
\centering
\begin{tabular}{c|c|c|c}
\hline 
 & \multirow{3}{*}{{\footnotesize$\mathcal{S}_{stat}$}} & {\footnotesize$\sqrt{s}=3\mathrm{~TeV}$} & {\footnotesize$\sqrt{s}=30\mathrm{~TeV}$}\tabularnewline
 &  & {\footnotesize 1 ab$^{-1}$} & {\footnotesize 10 ab$^{-1}$}\tabularnewline
 &  & {\footnotesize (0.1 TeV$^{-4}$)} & {\footnotesize (0.0001TeV$^{-4}$)}\tabularnewline
\hline 
\hline 
\multirow{3}{*}{{\footnotesize$\mathcal{O}_{\tilde{B}B}^{\prime}$}} & {\footnotesize 2} & {\footnotesize$\left[-0.122,0.180\right]$} & {\footnotesize$\left[-0.261,0.267\right]$}\tabularnewline
 & {\footnotesize 3} & {\footnotesize$\left[-0.155,0.213\right]$} & {\footnotesize$\left[-0.321,0.328\right]$}\tabularnewline
 & {\footnotesize 5} & {\footnotesize$\left[-0.208,0.266\right]$} & {\footnotesize$\left[-0.419,0.425\right]$}\tabularnewline
\hline 
\multirow{3}{*}{{\footnotesize$\mathcal{O}_{\tilde{B}W}$}} & {\footnotesize 2} & {\footnotesize$\left[-0.624,0.443\right]$} & {\footnotesize$\left[-0.962,0.945\right]$}\tabularnewline
 & {\footnotesize 3} & {\footnotesize$\left[-0.742,0.560\right]$} & {\footnotesize$\left[-1.18,1.16\right]$}\tabularnewline
 & {\footnotesize 5} & {\footnotesize$\left[-0.930,0.748\right]$} & {\footnotesize$\left[-1.53,1.51\right]$}\tabularnewline
\hline 
\multirow{3}{*}{{\footnotesize$\mathcal{O}_{BW}$}} & {\footnotesize 2} & {\footnotesize$0.547$} & {\footnotesize$1.03$}\tabularnewline
 & {\footnotesize 3} & {\footnotesize$0.671$} & {\footnotesize$1.26$}\tabularnewline
 & {\footnotesize 5} & {\footnotesize$0.868$} & {\footnotesize$1.64$}\tabularnewline
\hline 
\multirow{3}{*}{{\footnotesize$\mathcal{O}_{BB}$}} & {\footnotesize 2} & {\footnotesize$0.149$} & {\footnotesize$0.268$}\tabularnewline
 & {\footnotesize 3} & {\footnotesize$0.183$} & {\footnotesize$0.330$}\tabularnewline
 & {\footnotesize 5} & {\footnotesize$0.236$} & {\footnotesize$0.429$}\tabularnewline
\hline 
\multirow{3}{*}{{\footnotesize$\mathcal{O}_{G-}$}} & {\footnotesize 2} & {\footnotesize$\left[-0.373,0.545\right]$} & {\footnotesize$\left[-0.174,0.191\right]$}\tabularnewline
 & {\footnotesize 3} & {\footnotesize$\left[-0.474,0.646\right]$} & {\footnotesize$\left[-0.216,0.233\right]$}\tabularnewline
 & {\footnotesize 5} & {\footnotesize$\left[-0.635,0.807\right]$} & {\footnotesize$\left[-0.284,0.301\right]$}\tabularnewline
\hline 
\multirow{3}{*}{{\footnotesize$\widetilde{\mathcal{O}}_{G-}$}} & {\footnotesize 2} & {\footnotesize$0.938$} & {\footnotesize$0.346$}\tabularnewline
 & {\footnotesize 3} & {\footnotesize$1.15$} & {\footnotesize$0.425$}\tabularnewline
 & {\footnotesize 5} & {\footnotesize$1.48$} & {\footnotesize$0.553$}\tabularnewline
\hline 
\end{tabular}

\caption{\protect\label{tab:sensitivities-pol}The projected sensitivities
on the nTGCs coefficients at the muon colliders with selected c.m.
energies and integrated luminosities, for the polarized case $(-+)$.
For CPV operators, the coefficient ranges are symmetric and we only
list their upper bound. It should be noted that the numerical results
in different columns are scaled by different factors, as indicated
in the first row of the table.}
\end{table}

\section{\label{sec:EDM}Constraints on CPV nTGCs operators from the
electron EDM}

While searches at future muon colliders provide a direct way to probe
CPV nTGCs operators, these operators can also be constrained by precision
measurements of the electron EDM. The existence of an EDM requires
CP violation, and the SM prediction for the electron EDM lies many
orders of magnitude below current experimental sensitivity. The ACME
Collaboration has reported the current upper limit on the electron
EDM~\cite{ACME:2018yjb}:
\begin{equation}
|d_{e}|<1.1\times10^{-29}~e\cdot\textrm{cm}.\label{eq:deconstraint}
\end{equation}
This measurement sets strict limits on possible CPV operators. The
dimension-8 CPV nTGCs operators can induce EDM through loop diagrams
involving gauge and Higgs bosons, meaning that EDM bounds can probe
NP effects complementary to those directly accessible at colliders.
Therefore, it is instructive to compare the expected collider constraints
with the EDM constraints. In this section, we perform a preliminary
analysis by deriving the EDM constraints on the two pure gauge CPV
operators $\widetilde{\mathcal{O}}_{G+}$ and $\widetilde{\mathcal{O}}_{G-}$.
For the CPV operators involving the Higgs doublet, a large number
of Feynman diagrams with more complicated structures will be generated,
and we leave their detailed study to future work.

The Lorentz decomposition of the amplitude can be written as:
\begin{align}
 & \langle\alpha_{f}|M_{\mu}|\alpha_{i}\rangle=\bar{u}_{f}(k_{2})\left[F_{1}(t)\gamma_{\mu}-\frac{i}{2m}F_{2}(t)\sigma_{\mu\nu}k_{1}^{\nu}+\frac{1}{m}F_{3}(t)k_{1\mu}\right.\nonumber \\
 & +\left.\gamma_{5}\left(G_{1}(t)\gamma_{\mu}-\frac{i}{2m}G_{2}(t)\sigma_{\mu\nu}k_{1}^{\nu}+\frac{1}{m}G_{3}(t)k_{1\mu}\right)u_{i}(p_{1})\right],
\end{align}
with $k_{1}=k_{2}-p_{1}$, $t=k_{1}^{2}$ and $p_{1}^{2}=k_{2}^{2}=m^{2}$.
Conservation of the electromagnetic current requires $F_{3}(t)=0$.
$F_{1}(t)$ is the charge form factor, $F_{2}(t)$ is the magnetic
moment form factor, and $G_{2}(t)$ is the EDM form factor. The EDM
$d_{e}$ of the fermion is given by: 
\begin{equation}
d_{e}=\frac{G_{2}(0)}{2m}.
\end{equation}
To extract $G_{2}(t)$ from $M_{\mu}$, a projection operator has
been introduced~\cite{Czarnecki:1996rx}:
\begin{equation}
G_{2}(t)=\frac{2mp^{\mu}}{t(t-4m^{2})}\mathrm{Tr}\!\left[(\slashed{p}_{1}+m)\gamma_{5}(\slashed{p}_{2}+m)M_{\mu}\right].\label{eq:EDM-projector}
\end{equation}

\begin{figure}[tp]
\centering
\includegraphics{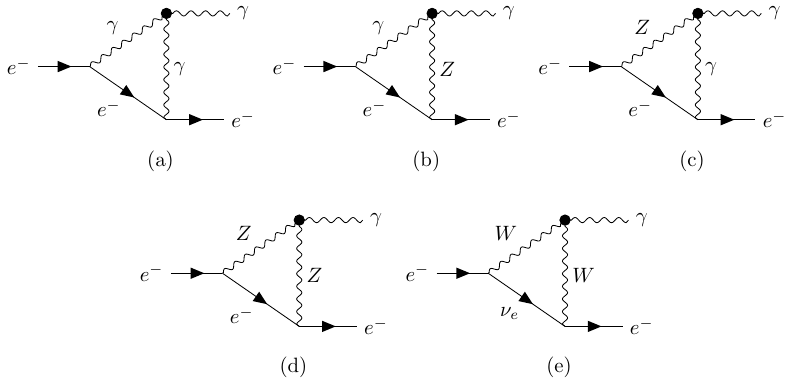}

\caption{\protect\label{fig:EDM-diagrams} Feynman diagrams contributing to
the electron EDM at the one-loop level involving nTGCs. These diagrams
are induced by CPV operators $\widetilde{\mathcal{O}}_{G+}$ and $\widetilde{\mathcal{O}}_{G-}$.
The solid black dot represents the nTGCs vertex.}
\end{figure}

The corrections to the QED vertex can arise through triangle diagrams
at the one-loop level, which accurately corresponds to the Barr–Zee
diagrams. When CPV nTGC operators are present, they provide the vertices
involving neutral triple gauge boson such as $Z\gamma\gamma$ and
$ZZ\gamma$. These nTGCs vertices can appear inside the loop, altering
the effective photon-electron vertex. For the pure gauge CPV operators
$\widetilde{\mathcal{O}}_{G+}$ and $\widetilde{\mathcal{O}}_{G-}$,
there are five diagrams at the one-loop level contributing to the
electron EDM, as shown in Figure~\ref{fig:EDM-diagrams}. The amplitudes
of these diagrams are listed in Appendix~\ref{sec:Bar-ZeeAmplitudes}.
The diagrams are calculated with electrons on-shell and with the photon
off-shell. Then the EDM form factor $G_{2}(t)$ is extracted and the
photon is set to be on-shell. The results are:
\begin{equation}
\begin{split}G_{2}^{\widetilde{\mathcal{O}}_{G+}}(0)= & \frac{iC_{13}e\left(c_{W}^{2}-1\right)}{96\pi^{2}\text{sw}c_{W}}\biggl[m_{Z}^{2}\left(5m_{e}^{2}B_{0}\left(m_{e}^{2},m_{e}^{2},m_{Z}^{2}\right)-2m_{Z}^{2}B_{0}\left(m_{e}^{2},m_{e}^{2},m_{Z}^{2}\right)+\right.\\
 & \left.3m_{e}^{2}B_{0}\left(m_{e}^{2},0,m_{e}^{2}\right)+3m_{e}^{2}m_{Z}^{2}C_{0}\left(0,m_{e}^{2},m_{e}^{2},0,m_{Z}^{2},m_{e}^{2}\right)+2A_{0}\left(m_{Z}^{2}\right)\right)+\\
 & 2\left(2m_{e}^{2}-m_{Z}^{2}\right)A_{0}\left(m_{e}^{2}\right)\biggr],
\end{split}
\label{eq:EDM-G3-AfterTrace}
\end{equation}
\begin{equation}
\begin{split}G_{2}^{\widetilde{\mathcal{O}}_{G-}}(0)= & \frac{iC_{14}e\left(c_{W}^{2}-1\right)}{384\pi^{2}c_{W}s_{W}}\biggl[m_{Z}^{2}\left(2m_{Z}^{2}B_{0}\left(m_{e}^{2},m_{e}^{2},m_{Z}^{2}\right)-20m_{e}^{2}B_{0}\left(m_{e}^{2},m_{e}^{2},m_{Z}^{2}\right)+\right.\\
 & 3\left(2m_{Z}^{2}-4m_{e}^{2}\right)B_{0}\left(m_{e}^{2},0,m_{e}^{2}\right)-12m_{Z}^{2}B_{0}\left(0,0,m_{Z}^{2}\right)+\\
 & 6m_{Z}^{4}C_{0}\left(0,m_{e}^{2},m_{e}^{2},0,m_{Z}^{2},m_{e}^{2}\right)-12m_{e}^{2}m_{Z}^{2}C_{0}\left(0,m_{e}^{2},m_{e}^{2},0,m_{Z}^{2},m_{e}^{2}\right)+\\
 & \left.4A_{0}\left(m_{Z}^{2}\right)\right)-8\left(2m_{e}^{2}-m_{Z}^{2}\right)A_{0}\left(m_{e}^{2}\right)\biggr],
\end{split}
\label{eq:EDM-G4-AfterTrace}
\end{equation}
where the superscript denotes the source of the contribution, and
$A_{0}$, $B_{0}$, $C_{0}$ are Passarino-Veltman functions.

The loop integrals in the above EDM form factor expressions are divergent.
In order to render these integrals finite, we introduce a momentum-space
cutoff, $\Lambda_{c}$, effectively restricting the loop momenta to
$|l|<\Lambda_{c}$. This procedure is widely adopted in effective
field theory, where higher-dimensional operators cannot be renormalized
in the conventional sense. Physically, the cutoff corresponds to the
energy scale beyond which the effective theory ceases to be valid,
reflecting the presence of unknown ultraviolet physics. By employing
this momentum cutoff, we can consistently compute the contributions
of dimension-8 operators while remaining within the regime of validity
of the effective theory.

Using the results of the $A_{0}$, $B_{0}$, $C_{0}$ functions in
Appendix~\ref{sec:PV-reduction}, the contributions to the electron
EDM from the operators $\widetilde{\mathcal{O}}_{G+}$ and $\widetilde{\mathcal{O}}_{G-}$
are obtained:
\begin{equation}
\begin{split}d_{e}^{\widetilde{\mathcal{O}}_{G+}}(0)= & \frac{C_{13}em_{e}m_{Z}^{2}s_{W}\left[12\log\left(\frac{\Lambda_{c}}{m_{Z}}\right)-1\right]}{192\pi^{2}c_{W}}+\mathcal{O}(\frac{m_{e}^{2}}{m_{Z}^{2}}),\\
d_{e}^{\widetilde{\mathcal{O}}_{G-}}(0)= & \frac{C_{14}em_{e}m_{Z}^{2}s_{W}\left[12\log\left(\frac{\Lambda_{c}}{m_{Z}}\right)-1\right]}{192\pi^{2}c_{W}}+\mathcal{O}(\frac{m_{e}^{2}}{m_{Z}^{2}}).
\end{split}
\label{eq:de-result}
\end{equation}
By applying different momentum-space cutoffs, and using the bound
in Eq.~(\ref{eq:deconstraint}), we obtain numerical constraints
on $C_{13}$ and $C_{14}$, listed in Table~\ref{tab:EDM}.

We find that the operators $\widetilde{\mathcal{O}}_{G+}$ and $\widetilde{\mathcal{O}}_{G-}$
contribute equally to the EDM form factors at the leading order in
$m_{e}^{2}/m_{Z}^{2}$, so the constraints on $C_{13}$ and $C_{14}$
are identical. In comparison with the numerical results at muon colliders
(see Table~\ref{tab:sensitivities-nopol} and Table~\ref{tab:sensitivities-pol}),
we observe that, except for the 3 TeV scenario for $\widetilde{\mathcal{O}}_{G-}$,
all the constraints at muon colliders are more stringent than those
inferred from the EDM measurement. This highlights that the EDM measurement
alone cannot provide comprehensive constraints on the nTGCs operators,
underscoring the necessity of collider investigations.

\begin{table}[tp]
\centering
\begin{tabular}{|c|c|c|c|c|}
\hline 
 & $\Lambda_{c}=3\mathrm{~TeV}$ & $\Lambda_{c}=10\mathrm{~TeV}$ & $\Lambda_{c}=14\mathrm{~TeV}$ & $\Lambda_{c}=30\mathrm{~TeV}$\tabularnewline
\hline 
\hline 
\multirow{1}{*}{$\widetilde{\mathcal{O}}_{G+}$} & $0.0113$ & $0.00839$ & $0.00782$ & $0.00678$\tabularnewline
\hline 
\end{tabular}

\caption{\protect\label{tab:EDM}Upper bounds on the coefficient $C_{13}/\Lambda^{4}~(\mathrm{TeV}^{-4})$
of $\widetilde{\mathcal{O}}_{G+}$, derived from the electron EDM
limit $|d_{e}|<1.1\times10^{-29}~e\cdot\textrm{cm}$. Upper bounds
on $\widetilde{\mathcal{O}}_{G-}$ are the same. The results with
different momentum-space cutoffs $\Lambda_{c}=3\mathrm{~TeV}$, $\Lambda_{c}=10\mathrm{~TeV}$,
$\Lambda_{c}=14\mathrm{~TeV}$ and $\Lambda_{c}=30\mathrm{~TeV}$
are listed.}
\end{table}

\section{\label{sec:Summary}Summary}

We conduct a detailed investigation into the sensitivity of high-energy
muon colliders to nTGCs within the SMEFT framework, specifically through
the process $\mu^{+}\mu^{-}\to\gamma\nu\bar{\nu}$. Our analysis incorporates
14 dimension-8 operators, encompassing both Higgs-related and pure
gauge structures, to explore their contributions in high-energy collisions.
By calculating the cross sections and performing MC simulations at
c.m. energies ranging from 3 to 30 TeV, we show that the annihilation
process dominates over VBF at TeV scales. The comparison of these
two processes reveals that the anticipated transition where VBF overtakes
annihilation occurs at much higher energies than the current collider
reach. We also present analytical results of the interference terms
between nTGCs operators and the SM, emphasizing the non-interfering
nature of CPV operators and the asymmetrical dependence of CPC interactions
on the effective coupling strengths.

Beam polarization effects play a crucial role in refining sensitivity,
particularly in distinguishing operator contributions. We examine
both unpolarized and polarized beam configurations and find that the
$(-+)$ polarization channel significantly enhances sensitivity for
certain operators, particularly $\mathcal{O}_{\tilde{B}B}^{\prime}$
and $\mathcal{O}_{BW}$, and the pure gauge operators $\widetilde{\mathcal{O}}_{G+}$
and $\widetilde{\mathcal{O}}_{G-}$. To extract precise expected constraints,
we analyze the kinematic distributions of signal and background events
and apply optimized event selection strategies at different c.m. energies.
We perform coefficient scans followed by parabolic fitting to extract
NP and interference cross sections, and obtain the expected constraints
on the operator coefficients at $2\sigma$, $3\sigma$, and $5\sigma$
levels. Our unpolarized results on the expected constraints
of $\mathcal{O}_{\tilde{B}W}$, $\mathcal{O}_{BB}$, $\mathcal{O}_{BW}$
and $\mathcal{O}_{WW}$ operators are consistent with previous studies~\cite{Senol:2022snc}.
Our work goes beyond by investigating a set of 14 nTGCs operators
and also the polarized beams.

Numerical results show that future muon colliders can set substantially
stronger expected constraints on nTGCs operators compared to the current
bounds from the LHC, particularly for the two purely gauge operators,
$\mathcal{O}_{G+}$ and $\widetilde{\mathcal{O}}_{G+}$, which yield
the most stringent expected constraints. Additionally, the partial
wave unitarity constraints are derived for each operator, and the
obtained expected constraints in this work are much tighter than the
unitarity bounds, ensuring the validity of SMEFT approach within the
studied energy range. We also analyze the constraints
on CPV operators $\widetilde{\mathcal{O}}_{G+}$ and $\widetilde{\mathcal{O}}_{G-}$
from the electron EDM, showing that the expected constraints at future
muon colliders are stronger than the EDM constraints. The study underscores
the critical role of muon colliders in probing nTGCs interactions,
and further reinforces the growing importance of dimension-8 operators
in SMEFT analyses. In light of recent proposals to construct future
muon colliders, this work shows the important role of muon colliders
in exploring NP beyond the SM.

\section*{Acknowledgments}

WX was supported in part by the National Natural Science Foundation
of China under Grants No. 12375137 and 12005114. J.-C. Y. was supported
in part by the National Natural Science Foundation of China under
Grants Nos. 12147214 and 12575106.

\appendix

\section{\label{sec:Wigner-D-functions}Helicity amplitudes, Wigner
functions and partial wave expansion coefficients}

The helicity amplitudes of nTGCs operators, and their corresponding
Wigner functions $d_{m_{1}m_{2}}^{J}(\phi,\theta)$ and partial wave
expansion coefficients $T_{J}$ are explicitly listed from Table~\ref{tab:A1-start}
to Table~\ref{tab:A1-end}. The expressions of the $d_{m_{1}m_{2}}^{J}(\phi,\theta)$
functions in the following tables are:
\begin{align*}
 & d_{-1,\mp1}^{1}(\phi,\theta)=\frac{1}{2}e^{\mp i\phi}(1\pm\cos\theta),\\
 & d_{-1,0}^{1}(\phi,\theta)=-\frac{\sqrt{2}}{2}\sin\theta,\\
 & d_{1,\mp1}^{1}(\phi,\theta)=\frac{1}{2}e^{\mp i\phi}(1\mp\cos\theta),\\
 & d_{1,0}^{1}=\frac{\sqrt{2}}{2}\sin\theta.
\end{align*}

\begin{table}[H]
\centering
\begin{tabular}{|c|c|c|c|}
\hline 
 & $\mathcal{M}$ & $d_{m_{1}m_{2}}^{J}(\phi,\theta)$ & $T_{J}$\tabularnewline
\hline 
\hline 
$\mu_{-\frac{1}{2}}^{-}\mu_{+\frac{1}{2}}^{+}\to Z_{0}\gamma_{\pm}$ & $\frac{C_{2}e^{2}\sqrt{s}v^{2}e^{-i\phi}(\cos(\theta)\pm1)\left(s-M_{Z}^{2}\right)}{16\sqrt{2}c_{W}M_{Z}s_{W}\Lambda^{4}}$ & $d_{-1,\mp1}^{1}(\phi,\theta)$ & $\pm\frac{C_{2}e^{2}\sqrt{s}v^{2}\left(s-M_{Z}^{2}\right)}{192\sqrt{2}\pi\Lambda^{4}c_{W}M_{Z}s_{W}}$\tabularnewline
\hline 
$\mu_{-\frac{1}{2}}^{-}\mu_{+\frac{1}{2}}^{+}\to Z_{\pm}\gamma_{\pm}$ & $\pm\frac{C_{2}e^{2}v^{2}e^{-i\phi}\sin(\theta)\left(s-M_{Z}^{2}\right)}{16c_{W}s_{W}\Lambda^{4}}$ & $d_{-1,0}^{1}(\phi,\theta)$ & $\mp\frac{C_{2}e^{2}v^{2}\left(s-M_{Z}^{2}\right)}{192\sqrt{2}\pi\Lambda^{4}c_{W}s_{W}}$\tabularnewline
\hline 
\end{tabular}

\caption{\protect\label{tab:A1-start}Helicity amplitudes and partial wave
expansion coefficients for $\mathcal{O}_{\tilde{W}W}^{\prime}$ }
\end{table}

\begin{table}[H]
\centering
\begin{tabular}{|c|c|c|c|}
\hline 
 & $\mathcal{M}$ & $d_{m_{1}m_{2}}^{J}(\phi,\theta)$ & $T_{J}$\tabularnewline
\hline 
\hline 
$\mu_{-\frac{1}{2}}^{-}\mu_{+\frac{1}{2}}^{+}\to Z_{0}\gamma_{\pm}$ & $\frac{C_{6}e^{2}\sqrt{s}v^{2}e^{-i\phi}(\cos(\theta)\pm1)\left(M_{Z}^{2}-s\right)}{8\sqrt{2}M_{Z}s_{W}^{2}\Lambda^{4}}$ & $d_{-1,\mp1}^{1}(\phi,\theta)$ & $\pm\frac{C_{6}e^{2}\sqrt{s}v^{2}\left(M_{Z}^{2}-s\right)}{96\sqrt{2}\pi\Lambda^{4}M_{Z}s_{W}^{2}}$\tabularnewline
\hline 
$\mu_{-\frac{1}{2}}^{-}\mu_{+\frac{1}{2}}^{+}\to Z_{\pm}\gamma_{\pm}$ & $\pm\frac{C_{6}e^{2}v^{2}e^{-i\phi}\sin(\theta)\left(M_{Z}^{2}-s\right)}{8\Lambda^{4}s_{W}^{2}}$ & $d_{-1,0}^{1}(\phi,\theta)$ & $\mp\frac{C_{6}e^{2}v^{2}\left(M_{Z}^{2}-s\right)}{96\sqrt{2}\pi\Lambda^{4}s_{W}^{2}}$\tabularnewline
\hline 
\end{tabular}

\caption{Helicity amplitudes and partial wave expansion coefficients for $\mathcal{O}_{\tilde{B}W}^{\prime}$}
\end{table}

\begin{table}[H]
\centering
\begin{tabular}{|c|c|c|c|}
\hline 
 & $\mathcal{M}$ & $d_{m_{1}m_{2}}^{J}(\phi,\theta)$ & $T_{J}$\tabularnewline
\hline 
\hline 
$\mu_{-\frac{1}{2}}^{-}\mu_{+\frac{1}{2}}^{+}\to Z_{0}\gamma_{\pm}$ & $\frac{iC_{9}e^{2}\sqrt{s}v^{2}e^{-i\phi}(1\pm\cos(\theta))\left(M_{Z}^{2}-s\right)}{16\sqrt{2}\Lambda^{4}c_{W}M_{Z}s_{W}}$ & $d_{-1,\mp1}^{1}(\phi,\theta)$ & $\frac{iC_{9}e^{2}\sqrt{s}v^{2}\left(M_{Z}^{2}-s\right)}{192\sqrt{2}\pi\Lambda^{4}c_{W}M_{Z}s_{W}}$\tabularnewline
\hline 
$\mu_{-\frac{1}{2}}^{-}\mu_{+\frac{1}{2}}^{+}\to Z_{\pm}\gamma_{\pm}$ & $\frac{iC_{9}e^{2}v^{2}e^{-i\phi}\sin(\theta)\left(M_{Z}^{2}-s\right)}{16\Lambda^{4}c_{W}s_{W}}$ & $d_{-1,0}^{1}(\phi,\theta)$ & $\frac{iC_{9}e^{2}v^{2}\left(s-M_{Z}^{2}\right)}{192\sqrt{2}\pi\Lambda^{4}c_{W}s_{W}}$\tabularnewline
\hline 
\end{tabular}

\caption{Helicity amplitudes and partial wave expansion coefficients for $\mathcal{O}_{WW}$}
\end{table}

\begin{table}[H]
\centering
\begin{tabular}{|c|c|c|c|}
\hline 
 & $\mathcal{M}$ & $d_{m_{1}m_{2}}^{J}(\phi,\theta)$ & $T_{J}$\tabularnewline
\hline 
\hline 
$\mu_{-\frac{1}{2}}^{-}\mu_{+\frac{1}{2}}^{+}\to Z_{0}\gamma_{\pm}$ & $\frac{C_{11}\sqrt{s}e^{-i\phi}(\cos(\theta)\pm1)M_{Z}\left(s-M_{Z}^{2}\right)}{2\sqrt{2}\Lambda^{4}}$ & $d_{-1,\mp1}^{1}(\phi,\theta)$ & $\pm\frac{C_{11}\sqrt{s}M_{Z}\left(s-M_{Z}^{2}\right)}{24\sqrt{2}\pi\Lambda^{4}}$\tabularnewline
\hline 
$\mu_{-\frac{1}{2}}^{-}\mu_{+\frac{1}{2}}^{+}\to Z_{\pm}\gamma_{\pm}$ & $\pm\frac{C_{11}se^{-i\phi}\sin(\theta)\left(s-M_{Z}^{2}\right)}{2\Lambda^{4}}$ & $d_{-1,0}^{1}(\phi,\theta)$ & $\mp\frac{C_{11}s\left(s-M_{Z}^{2}\right)}{24\sqrt{2}\pi\Lambda^{4}}$\tabularnewline
\hline 
\end{tabular}

\caption{Helicity amplitudes and partial wave expansion coefficients for $\mathcal{O}_{G+}$}
\end{table}

\begin{table}[H]
\centering
\begin{tabular}{|c|c|c|c|}
\hline 
 & $\mathcal{M}$ & $d_{m_{1}m_{2}}^{J}(\phi,\theta)$ & $T_{J}$\tabularnewline
\hline 
\hline 
$\mu_{-\frac{1}{2}}^{-}\mu_{+\frac{1}{2}}^{+}\to Z_{0}\gamma_{\pm}$ & $\frac{iC_{13}\sqrt{s}e^{-i\phi}(1\pm\cos(\theta))M_{Z}\left(M_{Z}^{2}-s\right)}{4\sqrt{2}\Lambda^{4}}$ & $d_{-1,\mp1}^{1}(\phi,\theta)$ & $-\frac{iC_{13}\sqrt{s}M_{Z}\left(s-M_{Z}^{2}\right)}{48\sqrt{2}\pi\Lambda^{4}}$\tabularnewline
\hline 
$\mu_{-\frac{1}{2}}^{-}\mu_{+\frac{1}{2}}^{+}\to Z_{\pm}\gamma_{\pm}$ & $\frac{iC_{13}se^{-i\phi}\sin(\theta)\left(M_{Z}^{2}-s\right)}{4\Lambda^{4}}$ & $d_{-1,0}^{1}(\phi,\theta)$ & $\frac{iC_{13}s\left(s-M_{Z}^{2}\right)}{48\sqrt{2}\pi\Lambda^{4}}$\tabularnewline
\hline 
\end{tabular}

\caption{Helicity amplitudes and partial wave expansion coefficients for $\widetilde{\mathcal{O}}_{G+}$}
\end{table}

\begin{table}[H]
\centering
\begin{tabular}{|c|c|c|c|}
\hline 
 & $\mathcal{M}$ & $d_{m_{1}m_{2}}^{J}(\phi,\theta)$ & $T_{J}$\tabularnewline
\hline 
\hline 
$\mu_{+\frac{1}{2}}^{-}\mu_{-\frac{1}{2}}^{+}\to Z_{0}\gamma_{\pm}$ & $\frac{\text{\ensuremath{C_{4}}}e^{2}\sqrt{s}v^{2}e^{i\phi}(\cos(\theta)\mp1)\left(M_{Z}^{2}-s\right)}{2\sqrt{2}c_{W}M_{Z}s_{W}\Lambda^{4}}$ & $d_{1,\mp1}^{1}(\phi,\theta)$ & $\pm\frac{C_{4}e^{2}\sqrt{s}v^{2}\left(s-M_{Z}^{2}\right)}{24\sqrt{2}\pi\Lambda^{4}c_{W}M_{Z}s_{W}}$\tabularnewline
\hline 
$\mu_{+\frac{1}{2}}^{-}\mu_{-\frac{1}{2}}^{+}\to Z_{\pm}\gamma_{\pm}$ & $\pm\frac{\text{\text{\ensuremath{C_{4}}}}e^{2}v^{2}e^{i\phi}\sin(\theta)\left(M_{Z}^{2}-s\right)}{2c_{W}s_{W}\Lambda^{4}}$ & $d_{1,0}^{1}(\phi,\theta)$ & $\pm\frac{C_{4}e^{2}v^{2}\left(M_{Z}^{2}-s\right)}{24\sqrt{2}\pi\Lambda^{4}c_{W}s_{W}}$\tabularnewline
\hline 
$\mu_{-\frac{1}{2}}^{-}\mu_{+\frac{1}{2}}^{+}\to Z_{0}\gamma_{\pm}$ & $\frac{C_{4}e^{2}\sqrt{s}v^{2}e^{-i\phi}(\cos(\theta)\pm1)\left(s-M_{Z}^{2}\right)}{4\sqrt{2}c_{W}M_{Z}s_{W}\Lambda^{4}}$ & $d_{-1,\mp1}^{1}(\phi,\theta)$ & $\pm\frac{C_{4}e^{2}\sqrt{s}v^{2}\left(s-M_{Z}^{2}\right)}{48\sqrt{2}\pi\Lambda^{4}c_{W}M_{Z}s_{W}}$\tabularnewline
\hline 
$\mu_{-\frac{1}{2}}^{-}\mu_{+\frac{1}{2}}^{+}\to Z_{\pm}\gamma_{\pm}$ & $\pm\frac{\text{\ensuremath{C_{4}}}e^{2}v^{2}e^{-i\phi}\sin(\theta)\left(s-M_{Z}^{2}\right)}{4c_{W}s_{W}\Lambda^{4}}$ & $d_{-1,0}^{1}(\phi,\theta)$ & $\pm\frac{C_{4}e^{2}v^{2}\left(M_{Z}^{2}-s\right)}{48\sqrt{2}\pi\Lambda^{4}c_{W}s_{W}}$\tabularnewline
\hline 
\end{tabular}

\caption{Helicity amplitudes and partial wave expansion coefficients for $\mathcal{O}_{\tilde{B}B}^{\prime}$}
\end{table}

\begin{table}[H]
\centering
\begin{tabular}{|c|c|c|c|}
\hline 
 & $\mathcal{M}$ & $d_{m_{1}m_{2}}^{J}(\phi,\theta)$ & $T_{J}$\tabularnewline
\hline 
\hline 
$\mu_{+\frac{1}{2}}^{-}\mu_{-\frac{1}{2}}^{+}\to Z_{0}\gamma_{\pm}$ & $\frac{C_{5}e^{2}\sqrt{s}v^{2}e^{i\phi}(\cos(\theta)\mp1)\left(s-M_{Z}^{2}\right)}{4\sqrt{2}c_{W}^{2}M_{Z}\Lambda^{4}}$ & $d_{1,\mp1}^{1}(\phi,\theta)$ & $\pm\frac{C_{5}e^{2}\sqrt{s}v^{2}\left(M_{Z}^{2}-s\right)}{48\sqrt{2}\pi\Lambda^{4}c_{W}^{2}M_{Z}}$\tabularnewline
\hline 
$\mu_{+\frac{1}{2}}^{-}\mu_{-\frac{1}{2}}^{+}\to Z_{\pm}\gamma_{\pm}$ & $\pm\frac{C_{5}e^{2}v^{2}e^{i\phi}\sin(\theta)\left(s-M_{Z}^{2}\right)}{4c_{W}^{2}\Lambda^{4}}$ & $d_{1,0}^{1}(\phi,\theta)$ & $\pm\frac{C_{5}e^{2}v^{2}\left(s-M_{Z}^{2}\right)}{48\sqrt{2}\pi\Lambda^{4}c_{W}^{2}}$\tabularnewline
\hline 
$\mu_{-\frac{1}{2}}^{-}\mu_{+\frac{1}{2}}^{+}\to Z_{0}\gamma_{\pm}$ & $\frac{C_{5}e^{2}\sqrt{s}v^{2}e^{-i\phi}(\cos(\theta)\pm1)\left(1-2s_{W}^{2}\right)\left(s-M_{Z}^{2}\right)}{8\sqrt{2}c_{W}^{2}M_{Z}s_{W}^{2}\Lambda^{4}}$ & $d_{-1,\mp1}^{1}(\phi,\theta)$ & $\pm\frac{C_{5}e^{2}\sqrt{s}v^{2}\left(c^{2}{}_{w}-s^{2}{}_{w}\right)\left(s-M_{Z}^{2}\right)}{96\sqrt{2}\pi\Lambda^{4}c_{W}^{2}M_{Z}s_{W}^{2}}$\tabularnewline
\hline 
$\mu_{-\frac{1}{2}}^{-}\mu_{+\frac{1}{2}}^{+}\to Z_{\pm}\gamma_{\pm}$ & $\pm\frac{C_{5}e^{2}v^{2}e^{-i\phi}\sin(\theta)\left(1-2s_{W}^{2}\right)\left(s-M_{Z}^{2}\right)}{8c_{W}^{2}s_{W}^{2}\Lambda^{4}}$ & $d_{-1,0}^{1}(\phi,\theta)$ & $\pm\frac{C_{5}e^{2}v^{2}\left(c^{2}{}_{w}-s^{2}{}_{w}\right)\left(M_{Z}^{2}-s\right)}{96\sqrt{2}\pi\Lambda^{4}c_{W}^{2}s_{W}^{2}}$\tabularnewline
\hline 
\end{tabular}

\caption{Helicity amplitudes and partial wave expansion coefficients for $\mathcal{O}_{\tilde{B}W}$}
\end{table}

\begin{table}[H]
\centering
\begin{tabular}{|c|c|c|c|}
\hline 
 & $\mathcal{M}$ & $d_{m_{1}m_{2}}^{J}(\phi,\theta)$ & $T_{J}$\tabularnewline
\hline 
\hline 
$\mu_{+\frac{1}{2}}^{-}\mu_{-\frac{1}{2}}^{+}\to Z_{0}\gamma_{\pm}$ & $\frac{iC_{8}e^{2}\sqrt{s}v^{2}e^{i\phi}(1\mp\cos(\theta))\left(s-M_{Z}^{2}\right)}{8\sqrt{2}\Lambda^{4}c_{W}^{2}M_{Z}}$ & $d_{1,\mp1}^{1}(\phi,\theta)$ & $\frac{iC_{8}e^{2}\sqrt{s}v^{2}\left(s-M_{Z}^{2}\right)}{96\sqrt{2}\pi\Lambda^{4}c_{W}^{2}M_{Z}}$\tabularnewline
\hline 
$\mu_{+\frac{1}{2}}^{-}\mu_{-\frac{1}{2}}^{+}\to Z_{\pm}\gamma_{\pm}$ & $\frac{i\text{\ensuremath{C_{8}}}e^{2}v^{2}e^{i\phi}\sin(\theta)\left(M_{Z}^{2}-s\right)}{8c_{W}^{2}\Lambda^{4}}$ & $d_{1,0}^{1}(\phi,\theta)$ & $-\frac{iC_{8}e^{2}v^{2}\left(s-M_{Z}^{2}\right)}{96\sqrt{2}\pi\Lambda^{4}c_{W}^{2}}$\tabularnewline
\hline 
$\mu_{-\frac{1}{2}}^{-}\mu_{+\frac{1}{2}}^{+}\to Z_{0}\gamma_{\pm}$ & $\frac{iC_{8}e^{2}\sqrt{s}v^{2}e^{-i\phi}(1\pm\cos(\theta))\left(s-M_{Z}^{2}\right)}{16\sqrt{2}c_{W}^{2}M_{Z}s_{W}^{2}\Lambda^{4}}$ & $d_{-1,\mp1}^{1}(\phi,\theta)$ & $\frac{iC_{8}e^{2}\sqrt{s}v^{2}\left(s-M_{Z}^{2}\right)}{192\sqrt{2}\pi\Lambda^{4}c_{W}^{2}M_{Z}s_{W}^{2}}$\tabularnewline
\hline 
$\mu_{-\frac{1}{2}}^{-}\mu_{+\frac{1}{2}}^{+}\to Z_{\pm}\gamma_{\pm}$ & $\frac{i\text{\ensuremath{C_{8}}}e^{2}v^{2}e^{-i\phi}\sin(\theta)\left(s-M_{Z}^{2}\right)}{16c_{W}^{2}s_{W}^{2}\Lambda^{4}}$ & $d_{-1,0}^{1}(\phi,\theta)$ & $-\frac{iC_{8}e^{2}v^{2}\left(s-M_{Z}^{2}\right)}{192\sqrt{2}\pi\Lambda^{4}c_{W}^{2}s_{W}^{2}}$\tabularnewline
\hline 
\end{tabular}

\caption{Helicity amplitudes and partial wave expansion coefficients for $\mathcal{O}_{BW}$}
\end{table}

\begin{table}[H]
\centering
\begin{tabular}{|c|c|c|c|}
\hline 
 & $\mathcal{M}$ & $d_{m_{1}m_{2}}^{J}(\phi,\theta)$ & $T_{J}$\tabularnewline
\hline 
\hline 
$\mu_{+\frac{1}{2}}^{-}\mu_{-\frac{1}{2}}^{+}\to Z_{0}\gamma_{\pm}$ & $\frac{iC_{10}e^{2}\sqrt{s}v^{2}e^{i\phi}(1\mp\cos(\theta))\left(M_{Z}^{2}-s\right)}{2\sqrt{2}\Lambda^{4}c_{W}M_{Z}s_{W}}$ & $d_{1,\mp1}^{1}(\phi,\theta)$ & $\frac{iC_{10}e^{2}\sqrt{s}v^{2}\left(M_{Z}^{2}-s\right)}{24\sqrt{2}\pi\Lambda^{4}c_{W}M_{Z}s_{W}}$\tabularnewline
\hline 
$\mu_{+\frac{1}{2}}^{-}\mu_{-\frac{1}{2}}^{+}\to Z_{\pm}\gamma_{\pm}$ & $\frac{iC_{10}e^{2}v^{2}e^{i\phi}\sin(\theta)\left(s-M_{Z}^{2}\right)}{2\Lambda^{4}c_{W}s_{W}}$ & $d_{1,0}^{1}(\phi,\theta)$ & $\frac{iC_{10}e^{2}v^{2}\left(s-M_{Z}^{2}\right)}{24\sqrt{2}\pi\Lambda^{4}c_{W}s_{W}}$\tabularnewline
\hline 
$\mu_{-\frac{1}{2}}^{-}\mu_{+\frac{1}{2}}^{+}\to Z_{0}\gamma_{\pm}$ & $\frac{iC_{10}e^{2}\sqrt{s}v^{2}e^{-i\phi}(1\mp\cos(\theta))\left(M_{Z}^{2}-s\right)}{4\sqrt{2}c_{W}M_{Z}s_{W}\Lambda^{4}}$ & $d_{-1,\mp1}^{1}(\phi,\theta)$ & $\frac{iC_{10}e^{2}\sqrt{s}v^{2}\left(M_{Z}^{2}-s\right)}{48\sqrt{2}\pi\Lambda^{4}c_{W}M_{Z}s_{W}}$\tabularnewline
\hline 
$\mu_{-\frac{1}{2}}^{-}\mu_{+\frac{1}{2}}^{+}\to Z_{\pm}\gamma_{\pm}$ & $\frac{\text{i\ensuremath{C_{10}}}e^{2}v^{2}e^{-i\phi}\sin(\theta)\left(M_{Z}^{2}-s\right)}{4c_{W}s_{W}\Lambda^{4}}$ & $d_{-1,0}^{1}(\phi,\theta)$ & $\frac{iC_{10}e^{2}v^{2}\left(s-M_{Z}^{2}\right)}{48\sqrt{2}\pi\Lambda^{4}c_{W}s_{W}}$\tabularnewline
\hline 
\end{tabular}

\caption{Helicity amplitudes and partial wave expansion coefficients for $\mathcal{O}_{BB}$}
\end{table}

\begin{table}[H]
\centering
\begin{tabular}{|c|c|c|c|}
\hline 
 & $\mathcal{M}$ & $d_{m_{1}m_{2}}^{J}(\phi,\theta)$ & $T_{J}$\tabularnewline
\hline 
\hline 
$\mu_{+\frac{1}{2}}^{-}\mu_{-\frac{1}{2}}^{+}\to Z_{0}\gamma_{\pm}$ & $\frac{\text{\ensuremath{C_{12}}}\sqrt{s}e^{i\phi}(\cos(\theta)\mp1)s_{W}^{2}\left(M_{Z}^{3}-sM_{Z}\right)}{\sqrt{2}\Lambda^{4}}$ & $d_{1,\mp1}^{1}(\phi,\theta)$ & $\pm\frac{C_{12}\sqrt{s}M_{Z}s_{W}^{2}\left(s-M_{Z}^{2}\right)}{12\sqrt{2}\pi\Lambda^{4}}$\tabularnewline
\hline 
$\mu_{+\frac{1}{2}}^{-}\mu_{-\frac{1}{2}}^{+}\to Z_{\pm}\gamma_{\pm}$ & $\frac{\pm\text{\ensuremath{C_{12}}}e^{i\phi}\sin(\theta)M_{Z}^{2}s_{W}^{2}\left(M_{Z}^{2}-s\right)}{\Lambda^{4}}$ & $d_{1,0}^{1}(\phi,\theta)$ & $\pm\frac{C_{12}M_{Z}^{2}s_{W}^{2}\left(M_{Z}^{2}-s\right)}{12\sqrt{2}\pi\Lambda^{4}}$\tabularnewline
\hline 
$\mu_{-\frac{1}{2}}^{-}\mu_{+\frac{1}{2}}^{+}\to Z_{0}\gamma_{\pm}$ & $\frac{C_{12}\sqrt{s}e^{-i\phi}(\cos(\theta)\pm1)M_{Z}s_{W}^{2}\left(s-M_{Z}^{2}\right)}{\sqrt{2}\Lambda^{4}}$ & $d_{-1,\mp1}^{1}(\phi,\theta)$ & $\pm\frac{C_{12}\sqrt{s}M_{Z}s_{W}^{2}\left(s-M_{Z}^{2}\right)}{12\sqrt{2}\pi\Lambda^{4}}$\tabularnewline
\hline 
$\mu_{-\frac{1}{2}}^{-}\mu_{+\frac{1}{2}}^{+}\to Z_{\pm}\gamma_{\pm}$ & $\frac{\text{\ensuremath{\pm}\ensuremath{C_{12}}}e^{-i\phi}\sin(\theta)M_{Z}^{2}s_{W}^{2}\left(s-M_{Z}^{2}\right)}{\Lambda^{4}}$ & $d_{-1,0}^{1}(\phi,\theta)$ & $\pm\frac{C_{12}M_{Z}^{2}s_{W}^{2}\left(M_{Z}^{2}-s\right)}{12\sqrt{2}\pi\Lambda^{4}}$\tabularnewline
\hline 
\end{tabular}

\caption{Helicity amplitudes and partial wave expansion coefficients for $\mathcal{O}_{G-}$}
\end{table}

\begin{table}[H]
\centering
\begin{tabular}{|c|c|c|c|}
\hline 
 & $\mathcal{M}$ & $d_{m_{1}m_{2}}^{J}(\phi,\theta)$ & $T_{J}$\tabularnewline
\hline 
\hline 
$\mu_{+\frac{1}{2}}^{-}\mu_{-\frac{1}{2}}^{+}\to Z_{0}\gamma_{\pm}$ & $\frac{iC_{14}\sqrt{s}e^{i\phi}(1\mp\cos(\theta))M_{Z}s_{W}^{2}\left(s-M_{Z}^{2}\right)}{2\sqrt{2}\Lambda^{4}}$ & $d_{1,\mp1}^{1}(\phi,\theta)$ & $\frac{iC_{14}\sqrt{s}M_{Z}s_{W}^{2}\left(s-M_{Z}^{2}\right)}{24\sqrt{2}\pi\Lambda^{4}}$\tabularnewline
\hline 
$\mu_{+\frac{1}{2}}^{-}\mu_{-\frac{1}{2}}^{+}\to Z_{\pm}\gamma_{\pm}$ & $\frac{i\text{\ensuremath{C_{14}}}e^{i\phi}\sin(\theta)M_{Z}^{2}s_{W}^{2}\left(M_{Z}^{2}-s\right)}{2\Lambda^{4}}$ & $d_{1,0}^{1}(\phi,\theta)$ & $\frac{iC_{14}M_{Z}^{2}s_{W}^{2}\left(M_{Z}^{2}-s\right)}{24\sqrt{2}\pi\Lambda^{4}}$\tabularnewline
\hline 
$\mu_{-\frac{1}{2}}^{-}\mu_{+\frac{1}{2}}^{+}\to Z_{0}\gamma_{\pm}$ & $\frac{iC_{14}\sqrt{s}e^{-i\phi}(1\pm\cos(\theta))M_{Z}s_{W}^{2}\left(s-M_{Z}^{2}\right)}{2\sqrt{2}\Lambda^{4}}$ & $d_{-1,\mp1}^{1}(\phi,\theta)$ & $\frac{iC_{14}\sqrt{s}M_{Z}s_{W}^{2}\left(s-M_{Z}^{2}\right)}{24\sqrt{2}\pi\Lambda^{4}}$\tabularnewline
\hline 
$\mu_{-\frac{1}{2}}^{-}\mu_{+\frac{1}{2}}^{+}\to Z_{\pm}\gamma_{\pm}$ & $\frac{i\text{\ensuremath{C_{14}}}e^{-i\phi}\sin(\theta)M_{Z}^{2}s_{W}^{2}\left(s-M_{Z}^{2}\right)}{2\Lambda^{4}}$ & $d_{-1,0}^{1}(\phi,\theta)$ & $\frac{iC_{14}M_{Z}^{2}s_{W}^{2}\left(M_{Z}^{2}-s\right)}{24\sqrt{2}\pi\Lambda^{4}}$\tabularnewline
\hline 
\end{tabular}

\caption{\protect\label{tab:A1-end}Helicity amplitudes and partial wave expansion
coefficients for $\widetilde{\mathcal{O}}_{G-}$}
\end{table}

\section{\label{sec:Bar-ZeeAmplitudes}Amplitudes of BarrZee diagrams}

The amplitudes of the Feynman diagrams in Figure~\ref{fig:EDM-diagrams}
are obtained with the help of \verb"FeynArts" and \verb"FeynCalc".
The amplitude corresponding to Figure~\ref{fig:EDM-diagrams}(a)
is:
\[
\mathcal{M}_{a}=0.
\]
The amplitude corresponding to Figure~\ref{fig:EDM-diagrams}(b)
is:
\begin{align*}
 & \mathcal{M}_{b}=\int\frac{d^{4}l}{(4\pi)^{4}}\frac{1}{el^{2}((l-k_{1})^{2}-m_{Z}^{2})((l-k_{1}-k_{2})^{2}-m_{e}^{2})}\varepsilon^{*\mu_{1}}(k_{1})g^{\mu_{2}\mu_{3}}g^{\mu_{4}\mu_{5}}\\
 & \bar{u}(k_{2},m_{e})\left(-\frac{ie\left(\left(2c_{W}^{2}-1\right)\gamma^{\mu_{5}}\left(\frac{1}{2}\left(1-\gamma^{5}\right)\right)-2s_{W}^{2}\gamma^{\mu_{5}}\left(\frac{1}{2}\left(\gamma^{5}+1\right)\right)\right)}{2c_{W}s_{W}}\right)\\
 & (-\slashed{l}+\slashed{k}_{1}+\slashed{k}_{2}+m_{e})\left(-ie\gamma^{\mu_{2}}\right)u(p_{1},m_{e})\\
 & s_{W}^{2}\biggl[C_{14}\left(-l^{4}g^{\mu_{1}\mu_{4}}k_{1}^{\mu_{3}}-l^{2}tg^{\mu_{1}\mu_{4}}k_{1}^{\mu_{3}}-l^{2}k_{1}^{\mu_{1}}k_{1}^{\mu_{3}}l^{\mu_{4}}+2l^{2}k_{1}^{\mu_{1}}k_{1}^{\mu_{3}}k_{1}^{\mu_{4}}-l^{2}g^{\mu_{3}\mu_{4}}k_{1}^{\mu_{1}}\left(k_{1}\cdot l\right)+\right.\\
 & l^{2}g^{\mu_{1}\mu_{4}}l^{\mu_{3}}\left(k_{1}\cdot l\right)+2l^{2}g^{\mu_{1}\mu_{4}}k_{1}^{\mu_{3}}\left(k_{1}\cdot l\right)+2g^{\mu_{3}\mu_{4}}k_{1}^{\mu_{1}}\left(k_{1}\cdot l\right){}^{2}-2g^{\mu_{1}\mu_{4}}l^{\mu_{3}}\left(k_{1}\cdot l\right){}^{2}+\\
 & tk_{1}^{\mu_{1}}k_{1}^{\mu_{3}}l^{\mu_{4}}-g^{\mu_{1}\mu_{3}}\left(t-l^{2}\right)\left(l^{\mu_{4}}\left(t-k_{1}\cdot l\right)+k_{1}^{\mu_{4}}\left(l^{2}-k_{1}\cdot l\right)\right)-tg^{\mu_{3}\mu_{4}}k_{1}^{\mu_{1}}\left(k_{1}\cdot l\right)+\\
 & tg^{\mu_{1}\mu_{4}}l^{\mu_{3}}\left(k_{1}\cdot l\right)+2k_{1}^{\mu_{1}}l^{\mu_{3}}l^{\mu_{4}}\left(k_{1}\cdot l\right)-2k_{1}^{\mu_{1}}k_{1}^{\mu_{3}}l^{\mu_{4}}\left(k_{1}\cdot l\right)-2k_{1}^{\mu_{1}}k_{1}^{\mu_{4}}l^{\mu_{3}}\left(k_{1}\cdot l\right)+\\
 & \left.l^{\mu_{1}}\left(tg^{\mu_{3}\mu_{4}}\left(-2\left(k_{1}\cdot l\right)+l^{2}+t\right)-k_{1}^{\mu_{3}}\left(l^{2}+t\right)\left(k_{1}^{\mu_{4}}-l^{\mu_{4}}\right)+l^{\mu_{3}}\left(k_{1}^{\mu_{4}}\left(2k_{1}\cdot l-l^{2}+t\right)-2tl^{\mu_{4}}\right)\right)\right)+\\
 & C_{13}\left(t-l^{2}\right)\left(-l^{2}g^{\mu_{1}\mu_{4}}k_{1}^{\mu_{3}}+g^{\mu_{3}\mu_{4}}k_{1}^{\mu_{1}}\left(k_{1}\cdot l\right)+g^{\mu_{1}\mu_{4}}l^{\mu_{3}}\left(k_{1}\cdot l\right)+\right.\\
 & \left.l^{\mu_{1}}\left(-tg^{\mu_{3}\mu_{4}}-k_{1}^{\mu_{4}}l^{\mu_{3}}+k_{1}^{\mu_{3}}\left(k_{1}^{\mu_{4}}+l^{\mu_{4}}\right)\right)+g^{\mu_{1}\mu_{3}}\left(l^{\mu_{4}}\left(t-k_{1}\cdot l\right)+k_{1}^{\mu_{4}}\left(l^{2}-k_{1}\cdot l\right)\right)-k_{1}^{\mu_{1}}k_{1}^{\mu_{3}}l^{\mu_{4}}\right)\biggr].
\end{align*}
The amplitude corresponding to Figure~\ref{fig:EDM-diagrams}(c)
is:
\begin{align*}
 & \mathcal{M}_{c}=\int\frac{d^{4}l}{(4\pi)^{4}}\frac{1}{e(l^{2}-m_{Z}^{2})(l-k_{1})^{2}((l-k_{1}-k_{2})^{2}-m_{e}^{2})}\varepsilon^{*\mu_{1}}(k_{1})g^{\mu_{2}\mu_{3}}g^{\mu_{4}\mu_{5}}\\
 & \bar{u}(k_{2},m_{e})(-ie\gamma^{\mu_{5}})(-\slashed{l}+\slashed{k}_{1}+\slashed{k}_{2}+m_{e})\\
 & \left(-\frac{ie\left(\left(2c_{W}^{2}-1\right)\gamma^{\mu_{2}}\left(\frac{1}{2}\left(1-\gamma^{5}\right)\right)-2s_{W}^{2}\gamma^{\mu_{2}}\left(\frac{1}{2}\left(\gamma^{5}+1\right)\right)\right)}{2c_{W}s_{W}}\right)u(p_{1},m_{e})\\
 & s_{W}^{2}\biggl[C_{14}\left(2g^{\mu_{1}\mu_{4}}l^{\mu_{3}}\left(k_{1}\cdot l\right){}^{2}+l^{4}g^{\mu_{1}\mu_{4}}k_{1}^{\mu_{3}}-l^{4}g^{\mu_{1}\mu_{3}}k_{1}^{\mu_{4}}-l^{2}g^{\mu_{1}\mu_{4}}l^{\mu_{3}}\left(k_{1}\cdot l\right)-2l^{2}g^{\mu_{1}\mu_{4}}k_{1}^{\mu_{3}}\left(k_{1}\cdot l\right)+\right.\\
 & l^{2}g^{\mu_{1}\mu_{3}}l^{\mu_{4}}\left(k_{1}\cdot l\right)+l^{2}g^{\mu_{1}\mu_{3}}k_{1}^{\mu_{4}}\left(k_{1}\cdot l\right)+k_{1}^{\mu_{1}}\left(l^{2}\left(g^{\mu_{3}\mu_{4}}\left(k_{1}\cdot l\right)+k_{1}^{\mu_{3}}l^{\mu_{4}}\right)-2l^{\mu_{3}}l^{\mu_{4}}\left(k_{1}\cdot l\right)\right)+\\
 & \left.l^{\mu_{1}}\left(l^{\mu_{3}}\left(k_{1}^{\mu_{4}}\left(l^{2}-2\left(k_{1}\cdot l\right)\right)+2tl^{\mu_{4}}\right)-l^{2}\left(tg^{\mu_{3}\mu_{4}}+k_{1}^{\mu_{3}}\left(l^{\mu_{4}}-k_{1}^{\mu_{4}}\right)\right)\right)-l^{2}tg^{\mu_{1}\mu_{3}}l^{\mu_{4}}\right)-\\
 & C_{13}\left(2\left(k_{1}\cdot l\right)-l^{2}\right)\left(-l^{2}g^{\mu_{1}\mu_{4}}k_{1}^{\mu_{3}}+g^{\mu_{3}\mu_{4}}k_{1}^{\mu_{1}}\left(k_{1}\cdot l\right)+g^{\mu_{1}\mu_{4}}l^{\mu_{3}}\left(k_{1}\cdot l\right)+\right.\\
 & \left.l^{\mu_{1}}\left(-tg^{\mu_{3}\mu_{4}}-k_{1}^{\mu_{4}}l^{\mu_{3}}+k_{1}^{\mu_{3}}\left(k_{1}^{\mu_{4}}+l^{\mu_{4}}\right)\right)+g^{\mu_{1}\mu_{3}}\left(l^{\mu_{4}}\left(t-k_{1}\cdot l\right)+k_{1}^{\mu_{4}}\left(l^{2}-k_{1}\cdot l\right)\right)-k_{1}^{\mu_{1}}k_{1}^{\mu_{3}}l^{\mu_{4}}\right)\biggr].
\end{align*}
The amplitude corresponding to Figure~\ref{fig:EDM-diagrams}(d)
is:
\begin{align*}
 & \mathcal{M}_{d}=\int\frac{d^{4}l}{(4\pi)^{4}}\frac{1}{e(l^{2}-m_{Z}^{2})((l-k_{1})-m_{Z}^{2})^{2}((l-k_{1}-k_{2})^{2}-m_{e}^{2})}\varepsilon^{*\mu_{1}}(k_{1})g^{\mu_{2}\mu_{3}}g^{\mu_{4}\mu_{5}}\\
 & \bar{u}(k_{2},m_{e})\left(-\frac{ie\left(\left(2c_{W}^{2}-1\right)\gamma^{\mu_{5}}\left(\frac{1}{2}\left(1-\gamma^{5}\right)\right)-2s_{W}^{2}\gamma^{\mu_{5}}\left(\frac{1}{2}\left(\gamma^{5}+1\right)\right)\right)}{2c_{W}s_{W}}\right)\\
 & (-\slashed{l}+\slashed{k}_{1}+\slashed{k}_{2}+m_{e})\left(-\frac{ie\left(\left(2c_{W}^{2}-1\right)\gamma^{\mu_{2}}\left(\frac{1}{2}\left(1-\gamma^{5}\right)\right)-2s_{W}^{2}\gamma^{\mu_{2}}\left(\frac{1}{2}\left(\gamma^{5}+1\right)\right)\right)}{2c_{W}s_{W}}\right)u(p_{1},m_{e})\\
 & s_{W}c_{W}\biggl[C_{14}\left(2g^{\mu_{3}\mu_{4}}k_{1}^{\mu_{1}}\left(k_{1}\cdot l\right){}^{2}-l^{2}tg^{\mu_{1}\mu_{4}}k_{1}^{\mu_{3}}-tg^{\mu_{3}\mu_{4}}k_{1}^{\mu_{1}}\left(k_{1}\cdot l\right)+tg^{\mu_{1}\mu_{4}}l^{\mu_{3}}\left(k_{1}\cdot l\right)+\right.\\
 & tl^{\mu_{1}}\left(g^{\mu_{3}\mu_{4}}\left(t-2\left(k_{1}\cdot l\right)\right)+k_{1}^{\mu_{4}}l^{\mu_{3}}+k_{1}^{\mu_{3}}\left(l^{\mu_{4}}-k_{1}^{\mu_{4}}\right)\right)+tg^{\mu_{1}\mu_{3}}\left(l^{\mu_{4}}\left(k_{1}\cdot l-t\right)+k_{1}^{\mu_{4}}\left(k_{1}\cdot l-l^{2}\right)\right)+\\
 & \left.-2k_{1}^{\mu_{4}}k_{1}^{\mu_{1}}l^{\mu_{3}}\left(k_{1}\cdot l\right)-2k_{1}^{\mu_{3}}k_{1}^{\mu_{1}}l^{\mu_{4}}\left(k_{1}\cdot l\right)+tk_{1}^{\mu_{3}}k_{1}^{\mu_{1}}l^{\mu_{4}}+2l^{2}k_{1}^{\mu_{3}}k_{1}^{\mu_{4}}k_{1}^{\mu_{1}}\right)-\\
 & C_{13}\left(t-2\left(k_{1}\cdot l\right)\right)\left(l^{2}g^{\mu_{1}\mu_{4}}k_{1}^{\mu_{3}}-g^{\mu_{3}\mu_{4}}k_{1}^{\mu_{1}}\left(k_{1}\cdot l\right)-g^{\mu_{1}\mu_{4}}l^{\mu_{3}}\left(k_{1}\cdot l\right)+\right.\\
 & \left.l^{\mu_{1}}\left(tg^{\mu_{3}\mu_{4}}+k_{1}^{\mu_{4}}l^{\mu_{3}}-k_{1}^{\mu_{3}}\left(k_{1}^{\mu_{4}}+l^{\mu_{4}}\right)\right)+g^{\mu_{1}\mu_{3}}\left(l^{\mu_{4}}\left(k_{1}\cdot l-t\right)+k_{1}^{\mu_{4}}\left(k_{1}\cdot l-l^{2}\right)\right)+k_{1}^{\mu_{1}}k_{1}^{\mu_{3}}l^{\mu_{4}}\right)\biggr].
\end{align*}
The amplitude corresponding to Figure~\ref{fig:EDM-diagrams}(e)
is:
\begin{align*}
 & \mathcal{M}_{e}=\int\frac{d^{4}l}{(4\pi)^{4}}\frac{1}{e(l^{2}-m_{W}^{2})((l-k_{1})-m_{W}^{2})^{2}(l-k_{1}-k_{2})^{2}}\varepsilon^{*\mu_{1}}(k_{1})g^{\mu_{2}\mu_{3}}g^{\mu_{4}\mu_{5}}\\
 & \bar{u}(k_{2},m_{e})\left(\frac{ie\gamma^{\mu_{5}}.\left(\frac{1}{2}\left(1-\gamma^{5}\right)\right)}{\sqrt{2}s_{W}}\right)(-\slashed{l}+\slashed{k}_{1}+\slashed{k}_{2})\left(\frac{ie\gamma^{\mu_{2}}.\left(\frac{1}{2}\left(1-\gamma^{5}\right)\right)}{\sqrt{2}s_{W}}\right)u(p_{1},m_{e})\\
 & s_{W}c_{W}\biggl[C_{14}\left(2g^{\mu_{3}\mu_{4}}k_{1}^{\mu_{1}}\left(k_{1}\cdot l\right){}^{2}-l^{2}tg^{\mu_{1}\mu_{4}}k_{1}^{\mu_{3}}-tg^{\mu_{3}\mu_{4}}k_{1}^{\mu_{1}}\left(k_{1}\cdot l\right)+tg^{\mu_{1}\mu_{4}}l^{\mu_{3}}\left(k_{1}\cdot l\right)+\right.\\
 & tl^{\mu_{1}}\left(g^{\mu_{3}\mu_{4}}\left(t-2\left(k_{1}\cdot l\right)\right)+k_{1}^{\mu_{4}}l^{\mu_{3}}+k_{1}^{\mu_{3}}\left(l^{\mu_{4}}-k_{1}^{\mu_{4}}\right)\right)+tg^{\mu_{1}\mu_{3}}\left(l^{\mu_{4}}\left(k_{1}\cdot l-t\right)+k_{1}^{\mu_{4}}\left(k_{1}\cdot l-l^{2}\right)\right)+\\
 & \left.-2k_{1}^{\mu_{4}}k_{1}^{\mu_{1}}l^{\mu_{3}}\left(k_{1}\cdot l\right)-2k_{1}^{\mu_{3}}k_{1}^{\mu_{1}}l^{\mu_{4}}\left(k_{1}\cdot l\right)+tk_{1}^{\mu_{3}}k_{1}^{\mu_{1}}l^{\mu_{4}}+2l^{2}k_{1}^{\mu_{3}}k_{1}^{\mu_{4}}k_{1}^{\mu_{1}}\right)-\\
 & C_{13}\left(t-2\left(k_{1}\cdot l\right)\right)\left(l^{2}g^{\mu_{1}\mu_{4}}k_{1}^{\mu_{3}}-g^{\mu_{3}\mu_{4}}k_{1}^{\mu_{1}}\left(k_{1}\cdot l\right)-g^{\mu_{1}\mu_{4}}l^{\mu_{3}}\left(k_{1}\cdot l\right)+\right.\\
 & \left.l^{\mu_{1}}\left(tg^{\mu_{3}\mu_{4}}+k_{1}^{\mu_{4}}l^{\mu_{3}}-k_{1}^{\mu_{3}}\left(k_{1}^{\mu_{4}}+l^{\mu_{4}}\right)\right)+g^{\mu_{1}\mu_{3}}\left(l^{\mu_{4}}\left(k_{1}\cdot l-t\right)+k_{1}^{\mu_{4}}\left(k_{1}\cdot l-l^{2}\right)\right)+k_{1}^{\mu_{1}}k_{1}^{\mu_{3}}l^{\mu_{4}}\right)\biggr].
\end{align*}
In the above amplitudes, $p_{1}$, $k_{2}$, and $k_{1}$ are the
4-momentum of incoming electron, outgoing electron and photon respectively,
and $t=k_{1}^{2}$. The loop momentum is denoted by $l$. By removing
the external spinors and polarization vectors from the above amplitudes,
we identify the remaining structure as the vertex function $M_{\mu}$.
We then apply the projection operator to $M_{\mu}$ to extract the
EDM form factor $G_{2}(t)$. 

The contribution of operator $\widetilde{\mathcal{O}}_{G+}$ to $G_{2}(t)$
is given by:
\begin{equation}
G_{2}^{\widetilde{\mathcal{O}}_{G+}}(t)=G_{2}^{\widetilde{\mathcal{O}}_{G+},b}(t)+G_{2}^{\widetilde{\mathcal{O}}_{G+},c}(t)+G_{2}^{\widetilde{\mathcal{O}}_{G+},d}(t)+G_{2}^{\widetilde{\mathcal{O}}_{G+},e}(t),\label{eq:EDM-G3}
\end{equation}
with
\begin{align*}
 & G_{2}^{\widetilde{\mathcal{O}}_{G+},b}(t)=\int\frac{d^{4}l}{(2\pi)^{4}}\frac{1}{k_{1}^{2}c_{W}s_{W}\left(k_{1}^{2}-4m_{e}^{2}\right)l^{2}\left[\left(l-k_{1}\right)^{2}-m_{Z}^{2}\right]\left[\left(l-k_{1}-k_{2}\right)^{2}-m_{e}^{2}\right]}\\
 & \biggl[e\left(c_{W}^{2}-1\right)m_{e}^{2}(2\left(l\cdot k_{1}\right)^{2}\left(l^{2}\left(2C_{13}m_{e}^{2}+C_{13}k_{1}^{2}\right)-C_{13}k_{1}^{2}\left(2m_{e}^{2}+k_{1}^{2}\right)\right)\\
 & -C_{13}k_{1}^{2}\left(k_{1}^{2}-l^{2}\right)\left(4\left(l\cdot k_{2}\right)^{2}+l^{2}\left(k_{1}^{2}-4m_{e}^{2}\right)\right)+8k_{1}^{2}\left(C_{13}l^{2}-C_{13}k_{1}^{2}\right)\left(l\cdot k_{2}\right)\left(l\cdot k_{1}\right)\biggr],
\end{align*}
\begin{align*}
 & G_{2}^{\widetilde{\mathcal{O}}_{G+},c}(t)=\int\frac{d^{4}l}{(2\pi)^{4}}\frac{1}{k_{1}^{2}c_{W}s_{W}\left(k_{1}^{2}-4m_{e}^{2}\right)(l^{2}-m_{Z}^{2})\left(l-k_{1}\right)^{2}\left[\left(l-k_{1}-k_{2}\right)^{2}-m_{e}^{2}\right]}\\
 & \biggl[e(c_{W}^{2}-1)m_{e}^{2}\left(4\left(C_{13}k_{1}^{2}-2C_{13}m_{e}^{2}\right)\left(k_{1}\cdot l\right){}^{3}-2\left(k_{1}\cdot l\right){}^{2}\left(l^{2}\left(C_{13}k_{1}^{2}-2C_{13}m_{e}^{2}\right)+2C_{13}k_{1}^{4}\right)+\right.\\
 & k_{1}^{2}l^{2}\left(4C_{13}\left(k_{2}\cdot l\right){}^{2}+C_{13}l^{2}\left(k_{1}^{2}-4m_{e}^{2}\right)+4C_{13}k_{1}^{2}\left(k_{2}\cdot l\right)\right)+\\
 & 4k_{1}^{2}\left(k_{1}\cdot l\right)\left(-2C_{13}\left(k_{2}\cdot l\right){}^{2}+2C_{13}l^{2}m_{e}^{2}-2C_{13}k_{1}^{2}\left(k_{2}\cdot l\right)\right)\biggr],
\end{align*}
\begin{align*}
 & G_{2}^{\widetilde{\mathcal{O}}_{G+},d}(t)=\int\frac{d^{4}l}{(2\pi)^{4}}\frac{1}{2c_{W}s_{W}\left(k_{1}^{2}-4m_{e}^{2}\right)(l^{2}-m_{Z}^{2})\left[\left(l-k_{1}\right)^{2}-m_{Z}^{2}\right]\left[\left(l-k_{1}-k_{2}\right)^{2}-m_{e}^{2}\right]}\\
 & \biggl[em_{e}^{2}(-4s_{W}^{2}+1)\left(-4C_{13}\left(k_{1}\cdot l\right){}^{3}+\right.\\
 & \left.4\left(C_{13}k_{1}^{2}-2C_{13}\left(k_{2}\cdot l\right)\right)\left(k_{1}\cdot l\right){}^{2}-2C_{13}k_{1}^{4}\left(k_{2}\cdot l\right)+k_{1}^{2}\left(k_{1}\cdot l\right)\left(8C_{13}\left(k_{2}\cdot l\right)-C_{13}k_{1}^{2}\right)\right)\biggr],
\end{align*}
\begin{align*}
 & G_{2}^{\widetilde{\mathcal{O}}_{G+},e}(t)=\int\frac{d^{4}l}{(2\pi)^{4}}\frac{1}{2k_{1}^{2}s_{W}^{2}\left(k_{1}^{2}-4m_{e}^{2}\right)(l^{2}-m_{W}^{2})\left[\left(l-k_{1}\right)^{2}-m_{W}^{2}\right]\left(l-k_{1}-k_{2}\right)^{2}}\\
 & \biggl[em_{e}^{2}\left(-8C_{13}k_{1}^{2}+8\left(k_{1}\cdot l\right){}^{2}\left(-2C_{13}k_{1}^{2}c_{W}s_{W}\left(k_{2}\cdot l\right)+C_{13}k_{1}^{4}c_{W}s_{W}+ie^{2}\right)\right.\\
 & c_{W}s_{W}\left(k_{1}\cdot l\right){}^{3}+2(k_{1}\cdot l)\left(4\left(k_{2}\cdot l\right)\left(2C_{13}k_{1}^{4}c_{W}s_{W}+2ie^{2}\right)+\right.\\
 & \left.C_{13}k_{1}^{6}\left(-c_{W}\right)s_{W}-ie^{2}k_{1}^{2}-4ie^{2}m_{e}^{2}\right)-\\
 & \left.k_{1}^{2}\left(4\left(k_{2}\cdot l\right)\left(C_{13}k_{1}^{4}c_{W}s_{W}+2ie^{2}\right)+ie^{2}\left(k_{1}^{2}-4m_{e}^{2}\right)\right)\right)\biggr].
\end{align*}

The contribution of operator $\widetilde{\mathcal{O}}_{G-}$ to $G_{2}(t)$
is given by:
\begin{equation}
G_{2}^{\widetilde{\mathcal{O}}_{G-}}(t)=G_{2}^{\widetilde{\mathcal{O}}_{G-},b}(t)+G_{2}^{\widetilde{\mathcal{O}}_{G-},c}(t)+G_{2}^{\widetilde{\mathcal{O}}_{G-},d}(t)+G_{2}^{\widetilde{\mathcal{O}}_{G-},e}(t),\label{eq:EDM-G4}
\end{equation}
with
\begin{align*}
 & G_{2}^{\widetilde{\mathcal{O}}_{G-},b}(t)=\int\frac{d^{4}l}{(2\pi)^{4}}\frac{1}{k_{1}^{2}c_{W}s_{W}\left(k_{1}^{2}-4m_{e}^{2}\right)l^{2}\left[\left(l-k_{1}\right)^{2}-m_{Z}^{2}\right]\left[\left(l-k_{1}-k_{2}\right)^{2}-m_{e}^{2}\right]}\\
 & \biggl[e(c_{W}^{2}-1)m_{e}^{2}\left(C_{14}k_{1}^{2}\left(k_{1}^{2}-l^{2}\right)\left(4\left(k_{2}\cdot l\right){}^{2}+l^{2}\left(k_{1}^{2}-4m_{e}^{2}\right)\right)+2\left(k_{1}\cdot l\right){}^{2}\right.\\
 & l^{2}\left(C_{14}k_{1}^{2}-2C_{14}m_{e}^{2}\right)-4C_{14}k_{1}^{2}\left(k_{2}\cdot l\right)+C_{14}k_{1}^{2}\left(2m_{e}^{2}+k_{1}^{2}\right)-\\
 & \left.4C_{14}k_{1}^{2}\left(k_{1}\cdot l\right){}^{3}+8C_{14}k_{1}^{4}\left(k_{1}\cdot l\right)\left(k_{2}\cdot l\right)\right)\biggr],
\end{align*}
\begin{align*}
 & G_{2}^{\widetilde{\mathcal{O}}_{G-},c}(t)=\int\frac{d^{4}l}{(2\pi)^{4}}\frac{1}{k_{1}^{2}c_{W}s_{W}\left(k_{1}^{2}-4m_{e}^{2}\right)(l^{2}-m_{Z}^{2})\left(l-k_{1}\right)^{2}\left[\left(l-k_{1}-k_{2}\right)^{2}-m_{e}^{2}\right]}\\
 & \biggl[e(c_{W}^{2}-1)m_{e}^{2}\left(8C_{14}m_{e}^{2}\left(k_{1}\cdot l\right){}^{3}-\right.\\
 & 2\left(k_{1}\cdot l\right)^{2}\left(l^{2}\left(2C_{14}m_{e}^{2}+C_{14}k_{1}^{2}\right)-4C_{14}k_{1}^{2}\left(k_{2}\cdot l\right)\right)+4k_{1}^{2}\left(k_{1}\cdot l\right)\\
 & 2C_{14}\left(k_{2}\cdot l\right){}^{2}+l^{2}\left(C_{14}k_{1}^{2}-2C_{14}m_{e}^{2}\right)-2C_{14}l^{2}\left(k_{2}\cdot l\right)+\\
 & \left.k_{1}^{2}l^{2}\left(-4C_{14}\left(k_{2}\cdot l\right){}^{2}-C_{14}l^{2}\left(k_{1}^{2}-4m_{e}^{2}\right)+4C_{14}k_{1}^{2}\left(k_{2}\cdot l\right)\right)\right)\biggr],
\end{align*}
\begin{align*}
 & G_{2}^{\widetilde{\mathcal{O}}_{G-},d}(t)=\int\frac{d^{4}l}{(2\pi)^{4}}\frac{1}{2c_{W}s_{W}\left(k_{1}^{2}-4m_{e}^{2}\right)(l^{2}-m_{Z}^{2})\left[\left(l-k_{1}\right)^{2}-m_{Z}^{2}\right]\left[\left(l-k_{1}-k_{2}\right)^{2}-m_{e}^{2}\right]}\\
 & \biggl[em_{e}^{2}(-4s_{W}^{2}+1)\left(4\left(k_{1}\cdot l\right){}^{2}\left(2C_{14}\left(k_{2}\cdot l\right)-C_{14}m_{e}^{2}-C_{14}k_{1}^{2}\right)-\right.\\
 & k_{1}^{2}\left(4C_{14}\left(k_{2}\cdot l\right){}^{2}+C_{14}l^{2}\left(k_{1}^{2}-4m_{e}^{2}\right)-2C_{14}k_{1}^{2}\left(k_{2}\cdot l\right)\right)+\\
 & \left.4C_{14}\left(k_{1}\cdot l\right){}^{3}+k_{1}^{2}\left(k_{1}\cdot l\right)\left(C_{14}k_{1}^{2}-12C_{14}\left(k_{2}\cdot l\right)\right)\right)\biggr],
\end{align*}
\begin{align*}
 & G_{2}^{\widetilde{\mathcal{O}}_{G-},e}(t)=\int\frac{d^{4}l}{(2\pi)^{4}}\frac{1}{2k_{1}^{2}s_{W}^{2}\left(k_{1}^{2}-4m_{e}^{2}\right)(l^{2}-m_{W}^{2})\left[\left(l-k_{1}\right)^{2}-m_{W}^{2}\right]\left(l-k_{1}-k_{2}\right)^{2}}\\
 & \biggl[em_{e}^{2}\left(8\left(\left(k_{1}\cdot l\right)\right){}^{2}\left(2C_{14}k_{1}^{2}c_{W}s_{W}\left(k_{2}\cdot l\right)-C_{14}k_{1}^{2}c_{W}m_{e}^{2}s_{W}+C_{14}k_{1}^{4}\left(-c_{W}\right)s_{W}+ie^{2}\right)-\right.\\
 & k_{1}^{2}\left(8C_{14}k_{1}^{2}c_{W}s_{W}\left(k_{2}\cdot l\right){}^{2}+\left(k_{1}^{2}-4m_{e}^{2}\right)\left(2C_{14}k_{1}^{2}l^{2}c_{W}s_{W}+ie^{2}\right)+\right.\\
 & \left.4\left(k_{2}\cdot l\right)\left(-C_{14}k_{1}^{4}c_{W}s_{W}+2ie^{2}\right)\right)+8C_{14}k_{1}^{2}c_{W}s_{W}\left(k_{1}\cdot l\right){}^{3}+\\
 & \left.2\left(k_{1}\cdot l\right)\left(4\left(k_{2}\cdot l\right)\left(-3C_{14}k_{1}^{4}c_{W}s_{W}+2ie^{2}\right)+C_{14}k_{1}^{6}c_{W}s_{W}-ie^{2}k_{1}^{2}-4ie^{2}m_{e}^{2}\right)\right)\biggr].
\end{align*}

\section{\label{sec:PV-reduction}Passarino-Veltman functions used
in this work}

\subsection{Passarino-Veltman functions contributing to the EDM form factors}

The Passarino-Veltman functions $A_{0}$, $B_{0}$, $C_{0}$ are the
basic scalar integrals that appear after reducing tensor loop integrals.
Take dimension $d=4-\epsilon$ and in dimensional regularization~(DR),
the functions $A_{0}$, $B_{0}$, $C_{0}$ are:
\begin{equation}
\begin{split} & A_{0}(m^{2})=\frac{(2\pi\mu)^{4-d}}{i\pi^{2}}\int d^{d}k\frac{1}{k^{2}-m^{2}},\\
 & B_{0}(p^{2},m_{0}^{2},m_{1}^{2})=\frac{(2\pi\mu)^{4-d}}{i\pi^{2}}\int d^{d}k\frac{1}{(k^{2}-m_{0}^{2})((k+p)^{2}-m_{1}^{2})},\\
 & C_{0}(p_{1}^{2},p_{2}^{2},(p_{1}+p_{2})^{2},m_{0}^{2},m_{1}^{2},m_{2}^{2})\\
 & =\frac{(2\pi\mu)^{4-d}}{i\pi^{2}}\int d^{d}k\frac{1}{(k^{2}-m_{0}^{2})((k+p_{1})^{2}-m_{1}^{2})((k+p_{1}+p_{2})^{2}-m_{2}^{2})}.
\end{split}
\label{eq.definitionPAVE}
\end{equation}

In this work, the expressions involving the $A_{0}$, $B_{0}$, $C_{0}$
functions are Eq.~(\ref{eq:EDM-G3-AfterTrace}) and Eq.~(\ref{eq:EDM-G4-AfterTrace}).
The analytical results of the $A_{0}$, $B_{0}$, $C_{0}$ functions
appearing Eq.~(\ref{eq:EDM-G3-AfterTrace}) and Eq.~(\ref{eq:EDM-G4-AfterTrace})
will be discussed below. We employ both DR and cutoff regularization,
and explicitly present the matching between the two schemes.

\subsection{\texorpdfstring{$A_{0}$}{A0} function}

In DR, $A_{0}(m^{2})$ is given by:
\begin{equation}
A_{0}^{\mathrm{DR}}(m^{2})=m^{2}\left(\Delta+1-\ln\frac{m^{2}}{\mu^{2}}\right).
\end{equation}
with the divergent term:
\begin{equation}
\Delta\equiv\frac{2}{\epsilon}-\gamma_{E}+\log(4\pi).\label{eq.Delta}
\end{equation}

Start from Eq.~(\ref{eq.definitionPAVE}) and perform the computation
in $d=4$ with a hard cutoff $|k_{E}|<\Lambda_{c}$ in Euclidean space,
the integration becomes:
\begin{align*}
A_{0}^{\mathrm{cutoff}}(m^{2}) & =\;-\frac{1}{\pi^{2}}\int_{|k_{E}|\le\Lambda_{c}}\frac{d^{4}k_{E}}{k_{E}^{2}+m^{2}}\\
 & =-\frac{1}{\pi^{2}}2\pi^{2}\int_{0}^{\Lambda_{c}}\frac{k^{3}\,dk}{k^{2}+m^{2}}\\
 & =-\bigg[\Lambda_{c}^{2}-m^{2}\log\!\Big(1+\frac{\Lambda_{c}^{2}}{m^{2}}\Big)\bigg].
\end{align*}
For $\Lambda_{c}\gg m$, we have $\Lambda_{c}^{2}-m^{2}\log\Big(1+\Lambda_{c}^{2}/m^{2}\Big)=\Lambda_{c}^{2}-m^{2}\log\left(\Lambda_{c}^{2}/m^{2}\right)+O\left(m^{3}\right)$,
then: 
\begin{align*}
A_{0}^{\mathrm{cutoff}}(m^{2}) & =\;m^{2}\log\left(\frac{\Lambda_{c}^{2}}{m^{2}}\right)-\Lambda_{c}^{2}.
\end{align*}
For convenience, following common practice in effective field theory
calculations, one may choose the simple scheme: 
\begin{equation}
\Delta=-1\quad\text{and}\quad\Lambda_{c}=\mu,
\end{equation}
which eliminates scheme-dependent constants and aligns the logarithmic
contributions in both regularizations. 

\subsection{\texorpdfstring{$B_{0}$}{B0} functions}

The $B_{0}(p^{2},m_{1}^{2},m_{2}^{2})$ function is one of the Passarino-Veltman
scalar functions that comes from evaluating two-point loop integrals.
In dimensional regularization the well known compact formula is:
\begin{equation}
\begin{split} B_{0}^{\mathrm{DR}}(p^{2},m_{1}^{2},m_{2}^{2})  &  =\Delta-\int_{0}^{1}dx\log\frac{m_{1}^{2}x+m_{2}^{2}(1-x)-p^{2}x(1-x)}{\mu^{2}}\\
  &  =\Delta-\int_{0}^{1}dx\log\frac{Q(x)}{\mu^{2}}, 
\end{split}
\label{eq:bDR}
\end{equation}
in which we define $Q(x)=m_{1}^{2}x+m_{2}^{2}(1-x)-p^{2}x(1-x)$.

We perform the integration in $d=4$ with a hard cutoff $|k_{E}|<\Lambda_{c}$:
\begin{equation}
\begin{split}B_{0}^{\mathrm{cutoff}}(p^{2},m_{1}^{2},m_{2}^{2}) & =\int\frac{d^{4}k}{i\pi^{2}}\frac{1}{(k^{2}-m_{1}^{2}+i\epsilon)((k+p)^{2}-m_{2}^{2}+i\epsilon)}\\
 & =\int_{0}^{1}dx\int\frac{d^{4}k}{i\pi^{2}}\frac{1}{\big[k^{2}-Q(x)+i\epsilon\big]^{2}}\\
 & =\int_{0}^{1}dx\int_{|k_{E}|\le\Lambda_{c}}\frac{d^{4}k_{E}}{\pi^{2}}\frac{1}{(k_{E}^{2}+Q(x))^{2}}\\
 & =\int_{0}^{1}dx\left[\frac{Q(x)}{\Lambda_{c}^{2}+Q(x)}+\log\left(\frac{\Lambda_{c}^{2}+Q(x)}{Q(x)}\right)-1\right].
\end{split}
\end{equation}
Here $Q(x)=m_{1}^{2}x+m_{2}^{2}(1-x)-p^{2}x(1-x)$. For $\Lambda_{c}^{2}\gg Q(x)$,
it reduces to:
\begin{equation}
\begin{split}B_{0}^{\mathrm{cutoff}}(p^{2},m_{1}^{2},m_{2}^{2}) & =\int_{0}^{1}dx\left[\log\frac{\Lambda_{c}^{2}}{Q(x)}-1\right]\\
 & =-1+\int_{0}^{1}dx\log\frac{\Lambda_{c}^{2}}{Q(x)}\\
 & =-1-\int_{0}^{1}dx\log\frac{Q(x)}{\Lambda_{c}^{2}}.
\end{split}
\label{eq:Bcutoff}
\end{equation}
By choosing the simple scheme $\Delta=-1$ and $\Lambda_{c}=\mu$,
the logarithmic contributions in dimensional regularization and cutoff
regularization align directly, making the matching procedure straightforward.

Expressions of $B_{0}$ functions for both dimensional regularization
and cutoff can be calculated from Eq.~(\ref{eq:bDR}) and Eq.~(\ref{eq:Bcutoff})
respectively. Here we list the expressions of $B_{0}$ functions used
in this work:
\begin{equation}
\begin{split}B_{0}(0,m_{Z}^{2},m_{Z}^{2}) & =\text{\ensuremath{\Delta}}-\log\left(\frac{m_{Z}^{2}}{\mu^{2}}\right),\\
B_{0}(0,0,m_{Z}^{2}) & =\Delta+1-\log\left(\frac{m_{Z}^{2}}{\mu^{2}}\right),\\
B_{0}(m_{e}^{2},0,m_{e}^{2}) & =\Delta+2-\log\left(\frac{m_{e}^{2}}{\mu^{2}}\right),\\
B_{0}(m_{e}^{2},m_{e}^{2},m_{Z}^{2}) & =\Delta+2-\frac{m_{Z}^{2}\log\left(\frac{m_{Z}^{2}}{m_{e}^{2}}\right)+m_{Z}\sqrt{m_{Z}^{2}-4m_{e}^{2}}\log\left(\frac{m_{Z}-\sqrt{m_{Z}^{2}-4m_{e}^{2}}}{\sqrt{m_{Z}^{2}-4m_{e}^{2}}+m_{Z}}\right)}{2m_{e}^{2}}\\
 & -\log\left(\frac{m_{e}^{2}}{\mu^{2}}\right).
\end{split}
\label{eq:b0functions}
\end{equation}
In DR, $\Delta$ is the one in Eq.~(\ref{eq.Delta}). In cutoff regularization,
one can just replace $\Delta\to-1$ and $\mu\to\Lambda_{c}$.

\subsection{\texorpdfstring{$C_{0}$}{C0} functions}

The $C_{0}$ function (three-point scalar function) is finite in $d=4$,
so regularization and cutoff are not needed. We list the analytical
results of $C_{0}$ functions used in this work:
\begin{equation}
\begin{split} & C_{0}(0,m_{e}^{2},m_{e}^{2},0,m_{Z}^{2},m_{e}^{2})\\
 & =-\frac{\log\left(\frac{m_{Z}^{2}}{m_{e}^{2}}\right)}{2m_{e}^{2}}-\frac{i\sqrt{4m_{e}^{2}-m_{Z}^{2}}\log\left(\frac{-im_{Z}\sqrt{4m_{e}^{2}-m_{Z}^{2}}-2m_{e}^{2}+m_{Z}^{2}}{2m_{e}^{2}}\right)}{2m_{e}^{2}m_{Z}},
\end{split}
\end{equation}
and
\begin{equation}
\begin{split} & C_{0}(0,m_{e}^{2},m_{e}^{2},m_{Z}^{2},m_{Z}^{2},m_{e}^{2})\\
= & \frac{\log\left(\frac{m_{e}^{2}}{m_{Z}^{2}}\right)}{2m_{e}^{2}}-\frac{\log\left(\frac{m_{Z}\sqrt{m_{Z}^{2}-4m_{e}^{2}}-2m_{e}^{2}+m_{Z}^{2}}{2m_{e}^{2}}\right)}{m_{Z}\sqrt{m_{Z}^{2}-4m_{e}^{2}}}+\frac{m_{Z}\log\left(\frac{m_{Z}\sqrt{m_{Z}^{2}-4m_{e}^{2}}-2m_{e}^{2}+m_{Z}^{2}}{2m_{e}^{2}}\right)}{2m_{e}^{2}\sqrt{m_{Z}^{2}-4m_{e}^{2}}}.
\end{split}
\end{equation}

\bibliographystyle{JHEP}
\bibliography{muonzgamma}

\end{document}